\documentclass[preprint2]{aastex}

\usepackage{amsmath} 
\usepackage{natbib}
\usepackage{subfigure}

\shorttitle{Transit Search from Antarctica and Chile}
\shortauthors{Fruth et al.}

\begin{document}
\title{Transit Search from Antarctica and Chile -- \\Comparison and Combination}

\author{
T.~Fruth\altaffilmark{1}, J.~Cabrera\altaffilmark{1}, Sz.~Csizmadia\altaffilmark{1}, C.~Dreyer\altaffilmark{1,2}, P.~Eigm\"uller\altaffilmark{1}, A.~Erikson\altaffilmark{1}, P.~Kabath\altaffilmark{1,3}, T.~Pasternacki\altaffilmark{1}, H.~Rauer\altaffilmark{1,2}, R.~Titz-Weider\altaffilmark{1}, \\
L.~Abe\altaffilmark{4}, A.~Agabi\altaffilmark{4}, I.~Gon\c{c}alves\altaffilmark{4}, T.~Guillot\altaffilmark{4}, D.~M\'ekarnia\altaffilmark{4}, J.-P.~Rivet\altaffilmark{4}, N.~Crouzet\altaffilmark{5}, \\
R.~Chini\altaffilmark{6,7}, R.~Lemke\altaffilmark{6}, and M.~Murphy\altaffilmark{8} 
}

\altaffiltext{1}{Institut f\"ur Planetenforschung, Deutsches Zentrum f\"ur Luft- und Raumfahrt, Rutherfordstr.~2, 12489~Berlin, Germany.}
\altaffiltext{2}{Zentrum f\"ur Astronomie und Astrophysik, Technische Universit\"at Berlin, 10623~Berlin, Germany.}
\altaffiltext{3}{European Southern Observatory, Alonso de C\'ordova~3107, Vitacura, Casilla~19001, Santiago~19, Chile.}
\altaffiltext{4}{Laboratoire Lagrange, UMR7293, Universit\'e de Nice Sophia-Antipolis, CNRS, Observatoire de la C\^ote d'Azur, BP4229, 06304~Nice Cedex~4, France.}
\altaffiltext{5}{Space Telescope Science Institute, Baltimore, MD~21218, USA.}
\altaffiltext{6}{Astronomisches Institut, Ruhr-Universit\"at Bochum, 44780~Bochum, Germany.}
\altaffiltext{7}{Instituto de Astronom\'{\i}a, Universidad Cat\'olica del Norte, Avenida Angamos 0610, Antofagasta, Chile.}
\altaffiltext{8}{Departamento de F\'isica, Universidad Cat\'olica del Norte, Avenida Angamos 0610, Antofagasta, Chile.}

\email{thomas.fruth@dlr.de}

\begin{abstract}
Observing sites at the East-Antarctic plateau are considered to provide exceptional conditions for astronomy. The aim of this work is to assess its potential for detecting transiting extrasolar planets through a comparison and combination of photometric data from Antarctica with time series from a midlatitude site.

During 2010, the two small aperture telescopes ASTEP\,400 (Dome~C) and BEST\,II (Chile) together performed an observing campaign of two target fields and the transiting planet \mbox{WASP-18b}. For the latter, a bright star, Dome~C appears to yield an advantageous signal-to-noise ratio. For field surveys, both Dome~C and Chile appear to be of comparable photometric quality. However, within two weeks, observations at Dome~C yield a transit detection efficiency that typically requires a whole observing season in Chile. For the first time, data from Antarctica and Chile have been combined to extent the observational duty cycle. This approach is both feasible in practice and favorable for transit search, as it increases the detection yield by 12--18\%.
\end{abstract}

\keywords{Astronomical Instrumentation, Extrasolar Planets}

\section{Introduction}
To obtain a better understanding of our Universe, improved observing conditions have constantly been sought by astronomers. While the limiting factors can be many and diverse, the selection of an observing site particularly impacts the quality of the astronomical data recorded. Among the most important constraints are the fraction of clear skies, the level of atmospheric seeing and scintillation, the accessibility of a wide spectral range, and a low sky brightness due to emission, scattered light, and light pollution. 

Optimal conditions are generally achieved high above the atmosphere, i.e., using air-borne or space observatories. However, such projects are limited by extensive costs, and technical considerations impose further constraints. Therefore, the search for excellent observational sites on ground is being pursued with unwaned interest. Over the last few decades, high-altitude sites such as in the Chilean Atacama desert or the mountain tops of Hawaii have generally been recognized to provide the best observing conditions for large ground-based observatories. 

In recent times, Antarctica is expected to provide a number of advantages for astronomy \citep[see, e.g.,][]{Indermuehle2005,Saunders2009,Burton2010,Fossat2010}, and therefore sites at the East Antarctic plateau such as Dome~C are currently being considered for future large-scale observatories \citep{Burton2005,Cui2010,Ichikawa2010,Abe2013a}. In particular, time series observations are considered to benefit from a high duty cycle and low photometric noise. These are of key importance for detecting and characterizing transiting extrasolar planets.

First, more and/or smaller planets are expected to be found at Dome~C due to an increased photometric precision \citep{Rauer2010a}. Two conditions are considered important in this respect: Less systematic noise due to stable environmental conditions (in particular, the lack of day/night temperature variations; \citealt{Pont2005a}), and less scintillation noise due to a low level of atmospheric turbulence. The latter is expected to be 2--4~times smaller at Dome~C compared to temperate sites \citep{Kenyon2006b}.

Second, the Antarctic winter allows for an almost continuous time series to be obtained \citep{Caldwell2004,Pont2005a}, although the total amount of usable dark time is not increased compared to midlatitude sites \citep{Kenyon2006a}. However, observations can cover large planetary orbits better than temperate sites with diurnal interruptions. \citet{Rauer2008} showed that planets with orbital periods of up to two weeks are covered well within one observing season at Dome C; in contrast, a similar performance with midlatitude sites could only be achieved if three stations were combined into a network. While \citet{Rauer2008} relied on observing times modeled from weather data and astronomical dark time, \citet{Crouzet2010} obtained statistics directly from the 10\,cm Antarctic Search for Transiting ExoPlanets (ASTEP)~South telescope. They estimated the transit yield and compared it to an analogous instrumental setup at La Silla: The ASTEP~South 2008 campaign is expected to yield a number of planet detections comparable to a modeled observing season at La Silla. However, if the ASTEP~South observations were extended over the whole winter season, the expected yield would be larger at Dome~C. In addition, \citet{Abe2013} recently obtained an unprecedented ground-based duty cycle while monitoring the transiting planet WASP-19b from Dome~C.

While these previous studies indicate an advantage for transit search at Dome~C, this still needs to be confirmed on the basis of extensive photometric data. For example, the study of \citet{Kenyon2006b} derived the scintillation noise from measurements of atmospheric turbulence profiles above Dome~C; however, this will only yield an advantage if it forms the dominant component in the noise budget for bright stars. \citet{Crouzet2010} used observing statistics from Dome~C, but modeled the photometric quality in Antarctica and Chile from instrument characteristics.

This study aims to address two open questions: First, whether transiting planets can be better photometrically characterized or detected from Antarctica in comparison to midlatitude sites. Second, if a transit survey from Antarctica together with a midlatitude site is feasible and promising in practice. In order to obtain a first comparison based on photometric data, a coordinated survey has been performed both from Dome~C and Chile. 

The paper is outlined as follows. It first introduces the instruments used (Section~\ref{sec:telescopes}), the observations (Section~\ref{sec:obs}), and their analysis (Section~\ref{sec:analysis}). Section~\ref{sec:phot} presents the scientific results regarding the photometric quality (i.e., limiting the radius of detectable transiting planets), while Section~\ref{sec:dc} focuses on the observational phase coverage (limiting the orbital period found by transit surveys). Section~\ref{sec:summary} summarizes the paper.

\section{Telescopes}\label{sec:telescopes}
The most important specifications of ASTEP\,400 and BEST\,II are summarized in Table~\ref{tab:telescopes} and described in the following text.

\subsection{ASTEP\,400}
The ASTEP project comprises two small telescopes at Dome~C, Antarctica: ASTEP~South \citep[not considered here]{Crouzet2010} and ASTEP\,400 \citep{Fressin2006,Daban2010,Crouzet2011}. Their main scientific objectives are first, to assess the photometric quality of Dome~C and second, to search for transiting planets \citep{Fressin2006}. The two ASTEP telescopes are operated by an international consortium under the responsibility of the Observatoire de la C\^ote d'Azur. 

The ASTEP\,400 telescope (simply refered to as ``ASTEP'' in the following text) was installed at Dome~C during the summer campaign 2009/2010 and achieved first light on 25th March 2010. It has an aperture of 40\,cm and is being operated on an Astrophysics AP3600 mount that was modified to operate down to $-80\degr$C. A thermalized enclosure contains two CCDs, correction lenses, and a dichroic mirror. The latter is used to forward the blue part ($\lambda\lesssim 550$\,nm) of the light beam to the guiding camera (SBIG~ST402M), while the red part ($\lambda\gtrsim 550$\,nm) is reflected to the main focus with the science camera (FLI ProLine 16801); its sensitivity peaks at $\lambda\approx 650$\,nm \citep{Abe2013}. The 4k~$\times$~4k~Pixel CCD covers a FOV of $1\fdg 0\times1\fdg 0$, thus providing an angular resolution of $0\farcs 9/\mbox{Px}$. 

\begin{table}[t]\footnotesize\centering
\caption{Instrument Specifications \label{tab:telescopes}}
\begin{tabular}{lcc}
\tableline\tableline
 & ASTEP\,400 & BEST\,II \\
\tableline
FOV         & $1\fdg 0\times1\fdg 0$ & $1\fdg 7\times 1\fdg 7$ \\
Aperture    & 400\,mm & 250\,mm \\
CCD         & FLI ProLine 16801 & FLI IMG-16801E2 \\
CCD Size    & 4k~$\times$~4k~Px & 4k~$\times$~4k~Px \\
Pixel Scale & $0\farcs 9/\mbox{Px}$ & $1\farcs 5$/Px \\
Gain $g$    & 1.53\,e$^-$/ADU & 1.98\,e$^-$/ADU\\
Overhead\tablenotemark{\dagger}    & 20.8\,s & 145\,s \\
\tableline
\end{tabular}
\begin{flushleft}\footnotesize
$^\dagger$ Average time between two exposures, including readout and processing time.
\end{flushleft}
\end{table} 

\subsection{BEST\,II}
The Berlin Exoplanet Search Telescope~II \citep[BEST\,II;][]{Kabath2009} is a small aperture telescope operated by the Institute of Planetary Research of the German Aerospace Center (DLR), Berlin. Its prime scientific focus is to support the CoRoT space mission \citep{Baglin2006} with photometric follow-up observations of transiting planetary candidates \citep{Deeg2009,Rauer2010,Csizmadia2011}. In addition, surveyed target fields are regularly analyzed for stellar variability \citep{Karoff2007,Kabath2007,Kabath2008,Kabath2009,Kabath2009a,Pasternacki2011,Fruth2012,Fruth2013,Klagyivik2013}.

BEST\,II is located in the Chilean Atacama desert at the Observatorio Cerro Armazones, i.e., in immediate neighborhood of the future site for the European Extremely Large Telescope \mbox{(E-ELT)}, and operated in robotic mode. It consists of a 25\,cm-aperture Baker-Ritchey-Chr\'etien reflector and a 14\,cm-aperture guiding refractor. The main instrument is a 4k$\times$4k CCD (FLI IMG-16801E2), i.e., BEST\,II and ASTEP\,400 use the same CCD chip. BEST\,II has a pixel scale of $1\farcs 5$/Px and a FOV of \mbox{$1\fdg 7\times 1\fdg 7$}. Observations are obtained without any filter, and the CCD sensitivity also peaks at $\lambda\approx 650$\,nm.

\section{Observations}\label{sec:obs}
The performance of a transit search at Dome~C can best be evaluated using a photometric survey. For that, five target fields, named Exo1--Exo5, have been selected by the ASTEP team for transit search from Antarctica in 2010. BEST\,II joined the campaign for two target fields, which are described in Section~\ref{sec:obs:e23}. In addition, the planet WASP-18b was observed with both telescopes in order to compare the photometric quality using a known transit signal; its observations are covered in Section~\ref{sec:obs:wasp18}. 

\begin{table*}[p]\centering\footnotesize
\caption{Details of Exo2 and Exo3 Observations. \label{tab:obs:fieldstats}}
\begin{tabular}[htbp]{lrrrcrrr}
\tableline\tableline
 & \multicolumn{3}{c}{\dotfill$N_{\textrm{Frames}}$\dotfill} && \multicolumn{3}{c}{\dotfill$N_{\textrm{Nights}}$\dotfill} \\
                        & ASTEP   & BEST\,II & total       && ASTEP           & BEST\,II       & together\tablenotemark{\dagger} \\\cline{2-8}\\[-.95em]
 Exo2                   & 5{,}895 & 391      & 6{,}286     && 16 \phantom{0}  & 6  \phantom{0} & 18 \phantom{0}  \\
 Exo3                   & 3{,}418 & 437      & 3{,}855     && 16 \phantom{0}  & 12 \phantom{0} & 26 \phantom{0}  \\
 total (both fields)    & 9{,}313 & 828      & 10{,}141    && 25 \phantom{0}  & 18 \phantom{0} & 37 \phantom{0}  \\
\tableline
 & \multicolumn{3}{c}{\dotfill$N_\star$\dotfill} && \multicolumn{3}{c}{\dotfill$N_\star$ with $\sigma\leq 0.01$\,mag\dotfill} \\
                        & ASTEP    & BEST\,II  & both\tablenotemark{\ddagger}  && ASTEP   &  BEST\,II & both\tablenotemark{\ddagger} \\\cline{2-8}\\[-.95em]
 Exo2                   & 37{,}619 &  90{,}330 & 9{,}124  && 2{,}318 &  8{,}229  & 745  \\
 Exo3                   & 57{,}346 & 134{,}222 & 49{,}698 && 1{,}838 &  6{,}436  & 1{,}779 \\
 total (both fields)    & 94{,}965 & 224{,}552 & 58{,}822 && 4{,}156 & 14{,}665  & 2{,}524 \\
\tableline
\end{tabular}\vspace{-3mm}
\tablenotetext{\dagger}{number of nights with ASTEP and/or BEST\,II observations}
\tablenotetext{\ddagger}{number of stars observed with ASTEP and BEST\,II}
\tablecomments{The table gives the number of frames ($N_{\textrm{Frames}}$), nights ($N_{\textrm{Nights}}$), and light curves ($N_\star$) obtained with each telescope in each target field. The latter quantity is  given both as the total and the number of high-precision ($\sigma\leq 0.01$\,mag) light curves.}
\end{table*}

\begin{figure*}[p]\centering
\subfigure[Exo2]{
  \begin{minipage}[b]{0.25\linewidth}
   \hspace{5mm}\includegraphics[width=\linewidth]{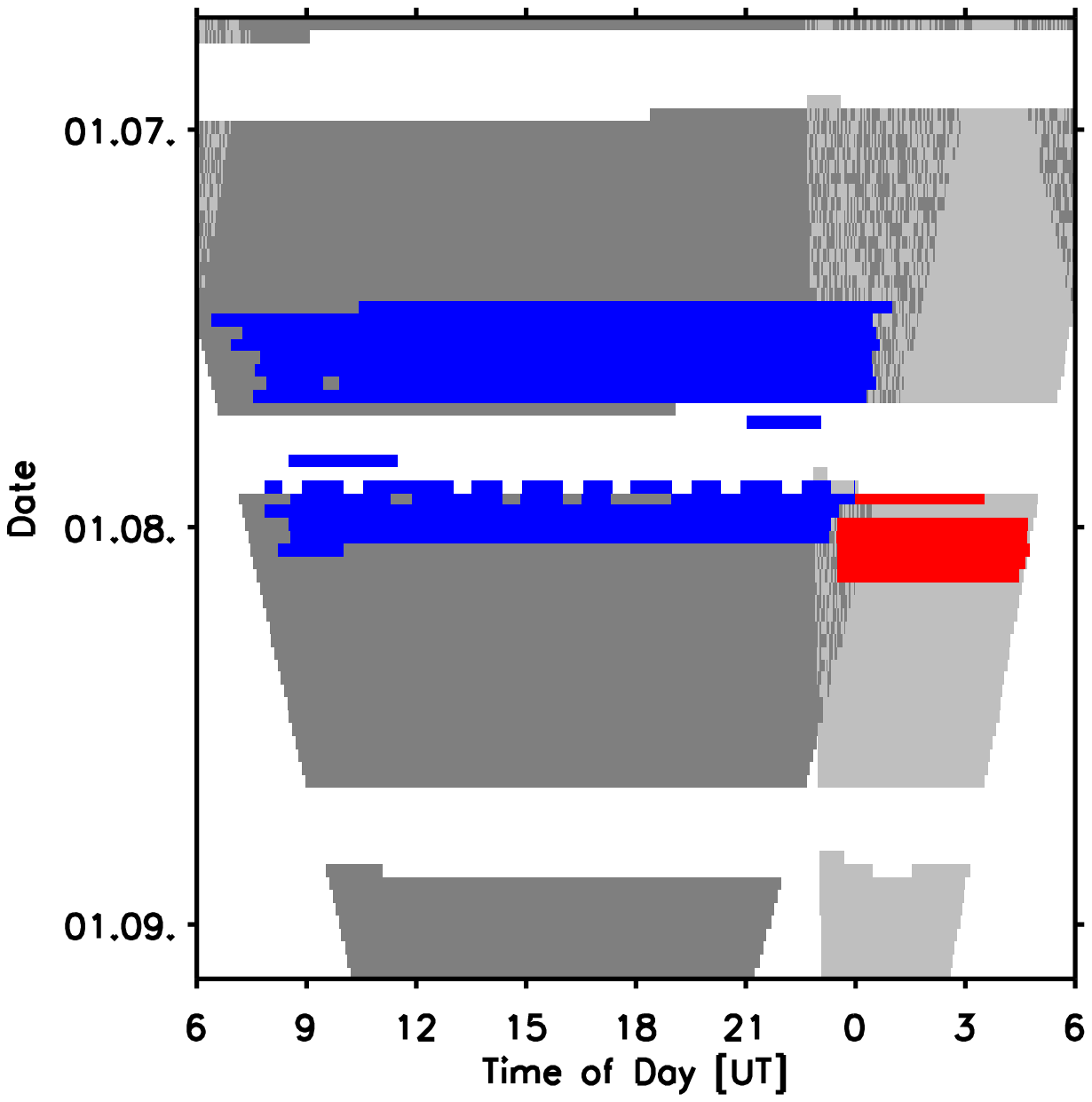}\vspace{6mm}
  \end{minipage}}\hspace{10mm}
\subfigure[Exo3]{
  \begin{minipage}[b]{0.25\linewidth}
   \hspace{5mm}\includegraphics[width=\linewidth]{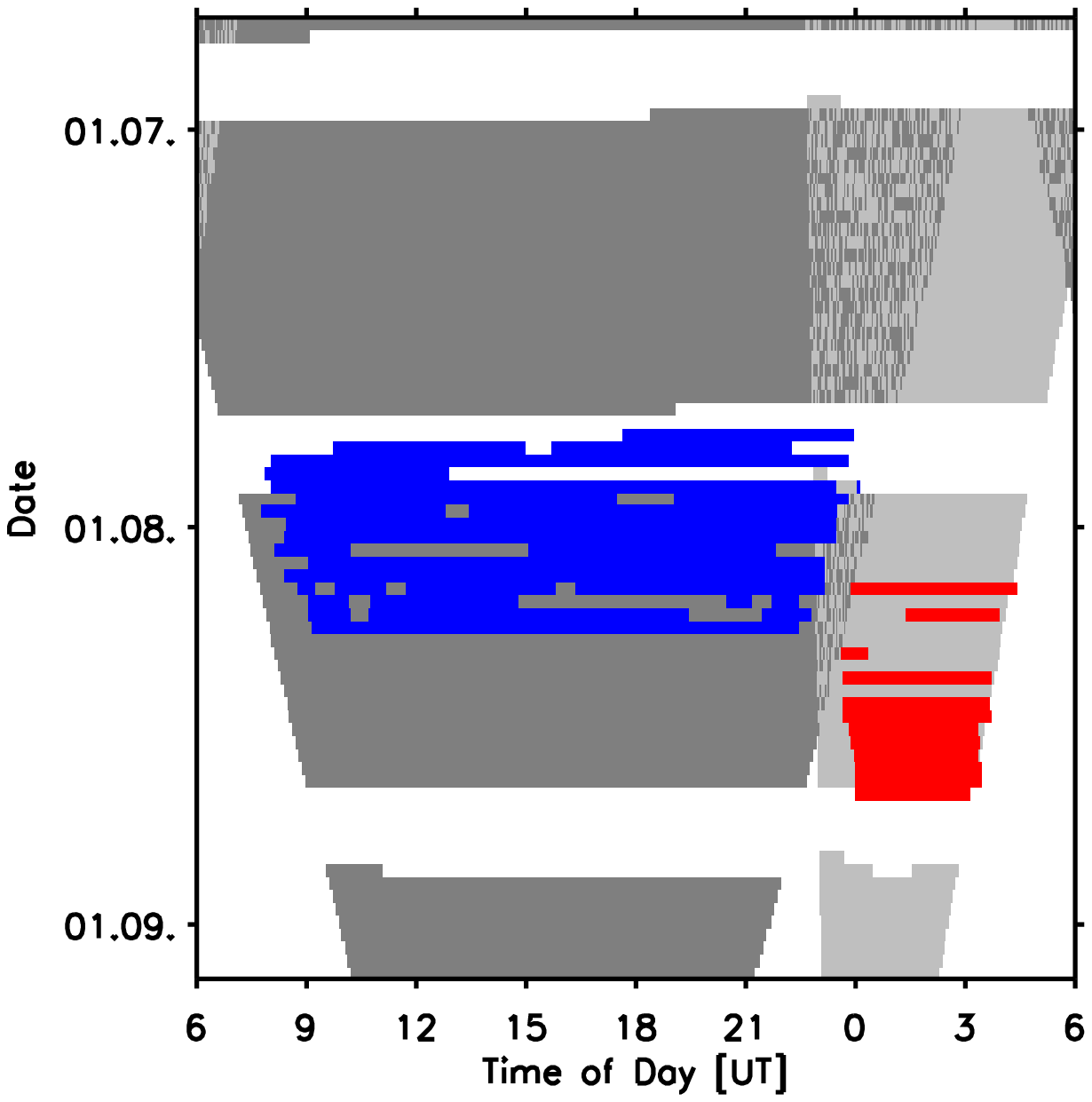}\vspace{6mm}
  \end{minipage}}\vspace*{-2mm}   
\caption{Joint BEST\,II/ASTEP field observations in 2010. Times of observations are shown for the fields (a) Exo2 and (b) Exo3. ASTEP time series are marked blue, BEST\,II observations red. For comparison, gray shaded areas indicate the maximum astronomical visibility of each respective field and site (i.e., target at least $30\degr$ above the horizon, Sun below~$-8\degr$, Moon phase $\varphi\leq 0.9$).}
\label{fig:obs:exo23}
\end{figure*}

\begin{figure*}[p]\centering
  \subfigure[Exo2]{\includegraphics[width=0.456\linewidth]{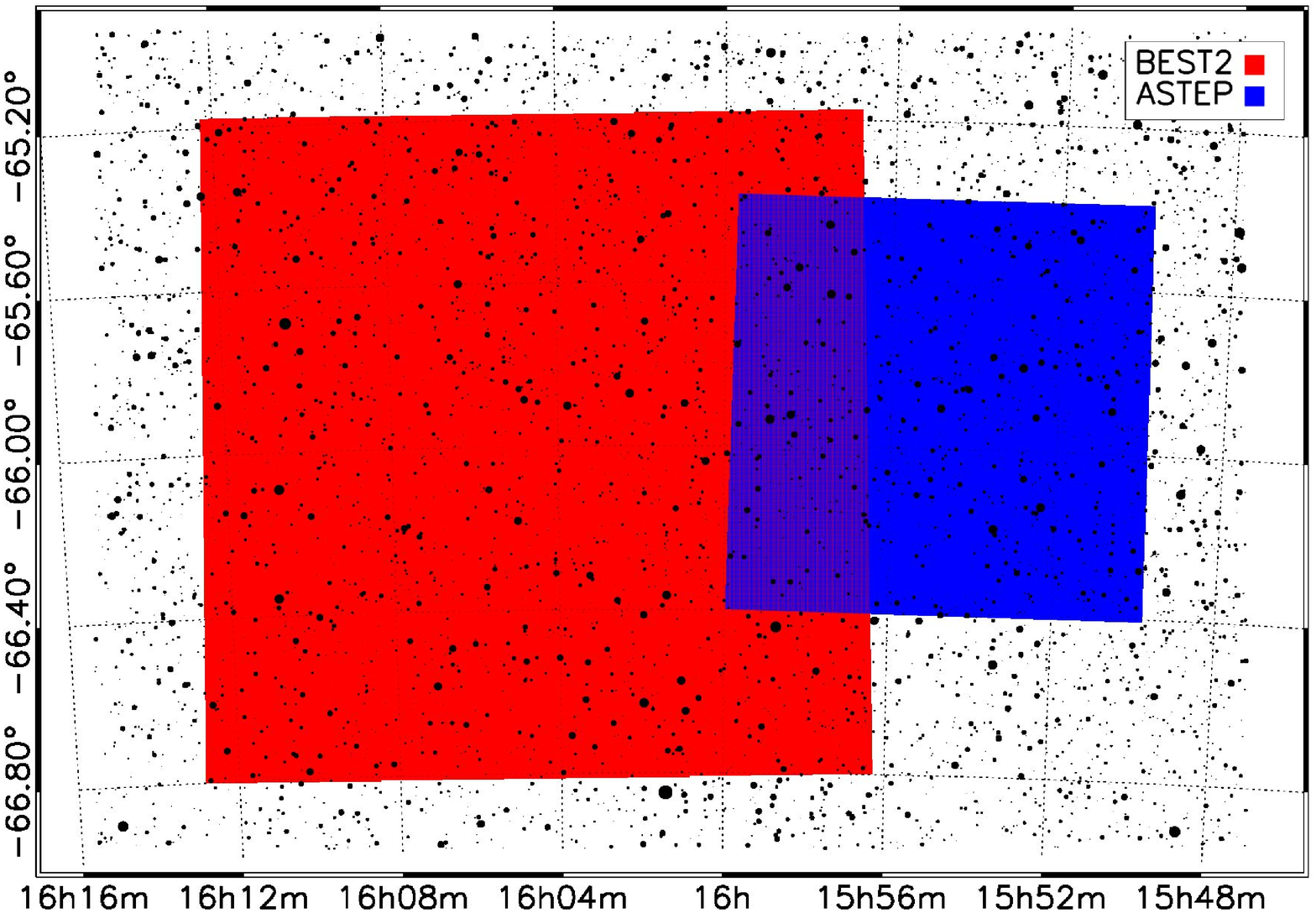}}
  \subfigure[Exo3]{\includegraphics[width=0.33\linewidth]{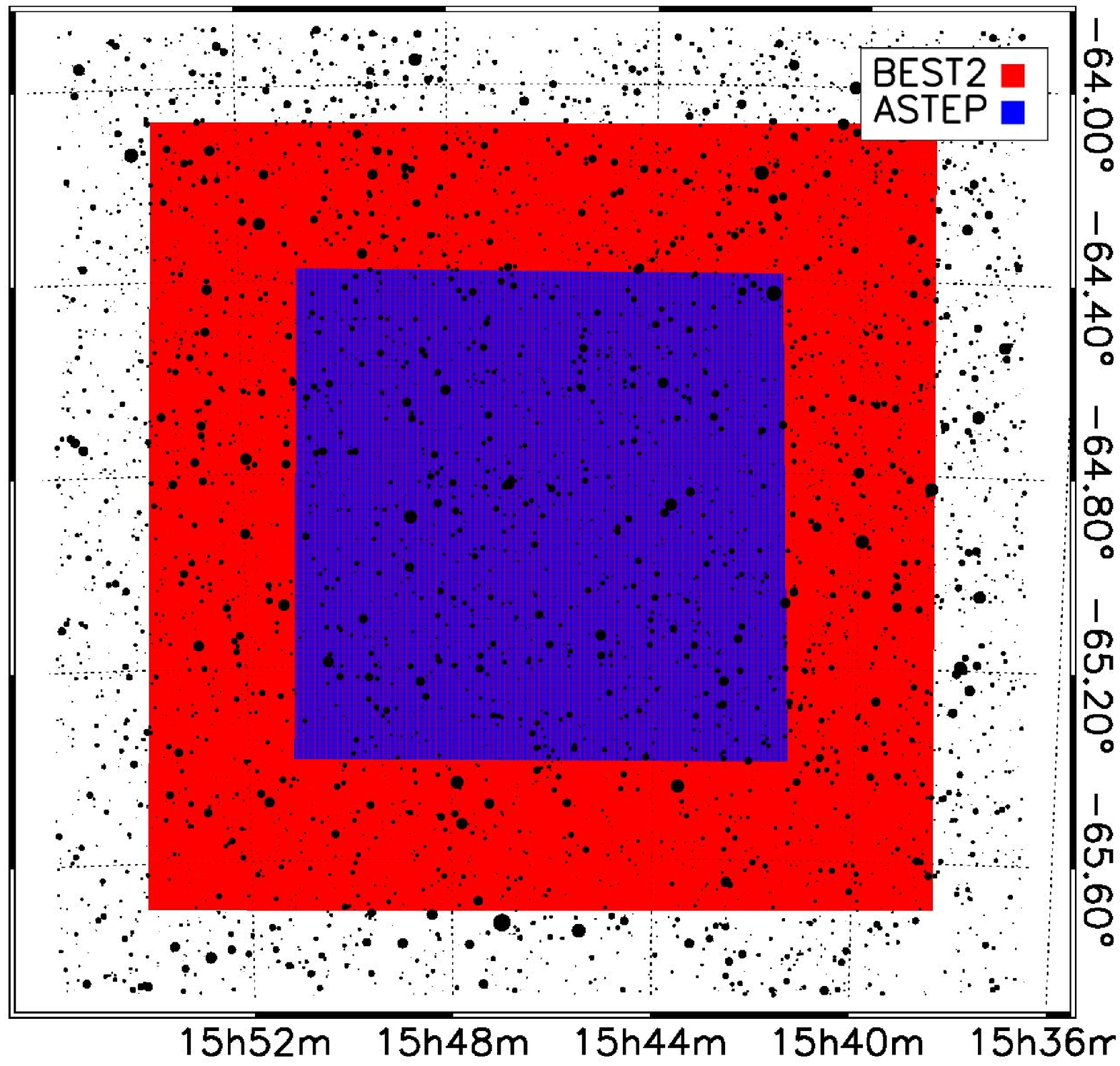}}
\caption{Sky position and orientation of the fields (a) Exo2 and (b) Exo3. The FOV of BEST\,II ($1\fdg 7\times1\fdg 7$) is marked red, while the ASTEP FOV ($1\fdg 0\times1\fdg 0$) is shown in blue.} \label{fig:obs:field-at-sky}
\end{figure*}

\subsection{Target Fields Exo2 and Exo3}\label{sec:obs:e23}
A first target field, Exo2, was observed with ASTEP from 14th July until 2nd August 2010 for a total of 16~nights, while BEST\,II pointed at the field for six nights between 29th July and 4th August 2010. Observations with both telescopes have been obtained during four nights, but overlap only for 28~minutes on 29th July. The second field, Exo3, was observed with a slightly larger timing offset between the two telescopes: ASTEP observed the field for 16~nights between 24th July and 8th August 2010, BEST\,II for 12~nights during 5th to 21st August 2010. The data contain two nights with observations from both sites without overlap. Table~\ref{tab:obs:fieldstats} compares the number of frames and nights obtained with each telescope, and Figure~\ref{fig:obs:exo23} shows the corresponding time series.

BEST\,II can cover the FOV of ASTEP in a single pointing: Figure~\ref{fig:obs:field-at-sky} shows a sky map with the relative positions and orientations of both fields for each telescope. 
The ASTEP center coordinates for the target field Exo3, 
\begin{center}
  $(\alpha,\delta)_\textrm{Exo3} = (15^h 46^m 11\fs 042, -64\degr 53' 32\farcs 52 )$, 
\end{center}
coincide with the BEST\,II observations. However, the Exo2 field was observed with ASTEP at a different pointing than initially announced: BEST\,II observed at coordinates 
\begin{center}
 $(\alpha,\delta)_\textrm{Exo2}^\textrm{BEST\,II} = (16^h 04^m 32\fs 414, -65\degr 50' 35\farcs 31 )$, 
\end{center}
which are offset by $1\fdg 11$ from the final ASTEP pointing at
\begin{center}
 $(\alpha,\delta)_\textrm{Exo2}^\textrm{ASTEP} = (15^h 54^m 48\fs 499, -65\degr 54' 04\farcs 35 )$,
\end{center}
i.e., BEST\,II observations only cover $\sim$\,35\% of the Exo2 field.

Both fields have been observed at airmasses of $X=1.0$--1.3 with ASTEP (circumpolar) and 1.3--2.0 with BEST\,II. This translates into slightly worse observing conditions in Chile: The scintillation index $\sigma_I$ reaches at most 1.5 of its zenith value $\sigma_I^0$ at Dome~C, but varies from 1.5--$2.8\,\sigma_I^0$ in Chile \citep[with $\sigma_I\propto X^{1.5}$,][]{Young1967}.

\subsection{WASP-18b}\label{sec:obs:wasp18}
The transiting hot Jupiter WASP-18b \citep{Hellier2009} was monitored intensively with ASTEP during its first observing season to test the quality of the instrument. With a depth of~$\sim$\,1\% and a magnitude of $V=9.3$\,mag, the primary transit should be well visible. In addition, the target was selected to test whether ASTEP can measure phase variations and secondary eclipses: With a period of 0.94\,days, WASP-18b is placed among the fastest orbiting exoplanets known, and thus highly irradiated. In the optical, phase variations due to reflection are expected in the order of a few 100\,ppm.

During the southern winter 2010, ASTEP observed WASP-18b for 66~nights (including 34 contiguous nights from 8th June to 11th July). In order to compare with measurements from a midlatitude site, BEST\,II also monitored WASP-18b for 19~nights between 12th August and 7th December 2010.

\section{Data Analysis}\label{sec:analysis}

\subsection{Data Reduction}\label{sec:datared} 
The scientific images from each telescope have been calibrated and reduced using standard photometric procedures. In order to allow for a homogeneous comparison, the custom-built BEST/BEST\,II pipeline was adapted to work with ASTEP data. It is described in detail in previous publications \citep[e.g.,][]{Kabath2009,Kabath2009a,Pasternacki2011,Fruth2012,Fruth2013}.

The processing steps include raw image calibration (bias, dark, flat) on a nightly basis, image subtraction \citep{Alard1998,Alard2000}, simple unit-weight aperture photometry with background subtraction, noise reduction by mean light curve subtraction plus removal of higher order systematics \citep{Tamuz2005}, astrometric matching \citep{Pal2006} with the UCAC3 catalog \citep{Zacharias2010}, and a zero-order magnitude calibration to the R2MAG band of UCAC3.

\subparagraph{Exo2 and Exo3.} Table~\ref{tab:obs:fieldstats} lists the number of light curves in each data set. The ASTEP data set contains 57{,}346 light curves on Exo3, and 37{,}619 on Exo2. Due to the larger FOV, the BEST\,II data include more light curves than ASTEP: 134{,}222 for Exo3, and 90{,}330 for the (slightly different) Exo2 pointing.

\subparagraph{WASP-18b.} 
As part of this study, only one night of ASTEP data covering a WASP-18b transit was reduced as a test case for the joint data analysis. In addition, all nights of BEST\,II observations on WASP-18b were reduced using the same pipeline. Image subtraction was not applied to either data set, since it proved not to increase the photometric quality for this uncrowded field. The BEST\,II observations on WASP-18b cover six full and three partial transits of WASP-18b, but only include one photometric transit event of WASP-18b. The most important photometric parameters (such as the aperture radius) were optimized in both reductions with respect to the transit to signal-to-noise ratio (S/N).

\subsection{Data Combination}\label{sec:datacombi} 
As \citet{Rauer2008} showed, the combination of data from Antarctica and Chile can potentially extend the observational phase coverage significantly. In order to assess whether the duty cycle of ASTEP could reasonably be extended with BEST\,II time series, light curves from both telescopes have been combined. This procedure is described in the following text, while Appendix~\ref{sec:app:datacombi} presents a case-by-case comparison of photometric parameters. The feasibility of a transit search in joint data will be addressed in Section~\ref{sec:dc:joint}. 

\begin{figure}[tc]\centering
  \includegraphics[width=.9\linewidth]{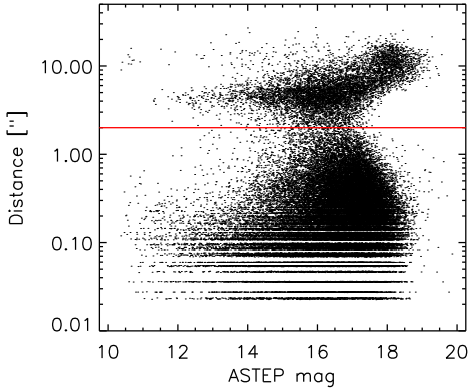}  
  \caption{Distance $d$ to nearest BEST\,II object for each star in the Exo3 data set. The red line denotes the limit of $2''$ that is used for matching.}
  \label{fig:e23:dist-mag}
\end{figure}

The first step comprises a light curve match using the equatorial coordinates of each reduction. For each ASTEP light curve, the angular distance~$d$ to the nearest BEST\,II star has been calculated; it is plotted versus the magnitude in Figure~\ref{fig:e23:dist-mag} for field Exo3 as an example. The figure shows a clear distinction between stars that are matched within~\mbox{$d\ll 2''$} and those that do not have a counterpart, i.e., with~$d\gg 2''$. Therefore, the limit~\mbox{$d=2''$} is used as the criterion for a successful match. In target field Exo2, 9{,}124 stars meet this criterion (i.e., 10.1\% of all BEST\,II, and 24.3\% of all ASTEP stars; see Table~\ref{tab:obs:fieldstats}). For field Exo3, the overlap is much better; here, 49{,}698 stars are matched (37.0\% of BEST\,II, and 86.7\% of all ASTEP stars). In total, joint observations are obtained for 58{,}822 field stars. 

Since the BEST\,II/ASTEP observing strategy does not aim at a precise calibration for absolute magnitudes, two light curves of the same star may show significantly different base levels in two reductions (Figure~\ref{fig:app:best2astep:mag-rms} in Appendix~\ref{sec:app:datacombi}). In a second step, each matched light curve $i$ is thus adjusted using a mean magnitude shift, i.e., the value
\begin{equation}\label{eq:Deltam}
  \Delta m_i=\overline{m}_i^B-\overline{m}_i^A, 
\end{equation}
is subtracted from all BEST\,II measurements (with $\overline{m}_i^{A/B}$ denoting the mean magnitude of the ASTEP/BEST\,II time series). 

BEST\,II and ASTEP use the same CCD chip and both observed the fields Exo2 and Exo3 in white light. Thus, the photometric systems are expected to be very similar. However, a dichroic mirror in the optical path of ASTEP\,400 only reflects wavelengths longward of $\sim$\,550\,nm to the scientific instrument, while the whole spectrum is used with BEST\,II. Thus, blue stars are expected to appear somewhat brighter when being observed with BEST\,II. In Appendix~\ref{sec:app:photsys}, the correlation between $\Delta m$ and catalog colors is evaluated quantitatively; it confirms the expected color dependency between both optical designs. However, the differential amplitude is rather low, so that its influence upon the combined data is considered negligible.

\subsection{Photometric Noise}\label{sec:noise} 
\begin{figure*}[t]\centering\vspace{-2mm}
  \subfigure[Exo2]{
   \includegraphics[width=.5\linewidth]{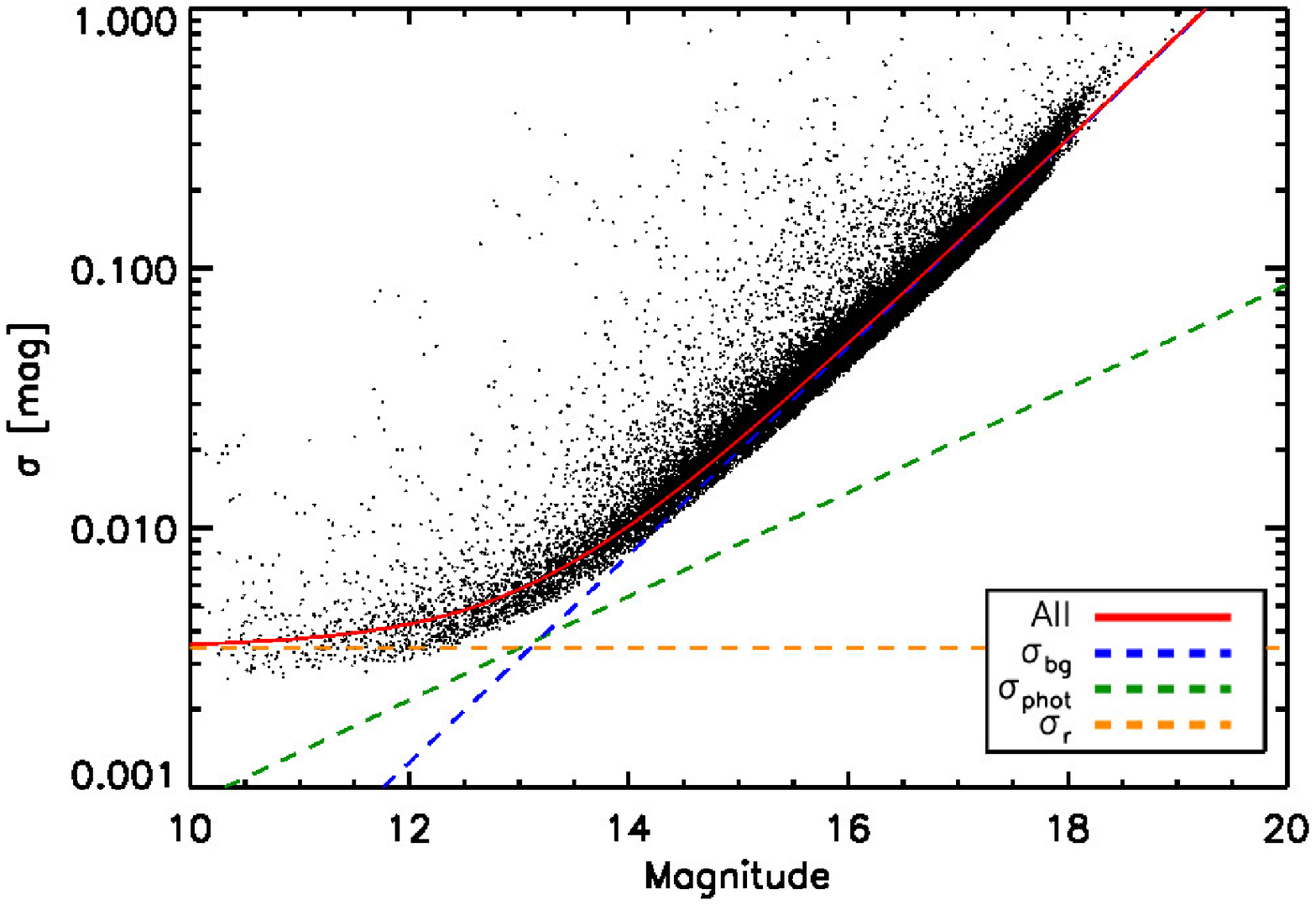}
   \includegraphics[width=.5\linewidth]{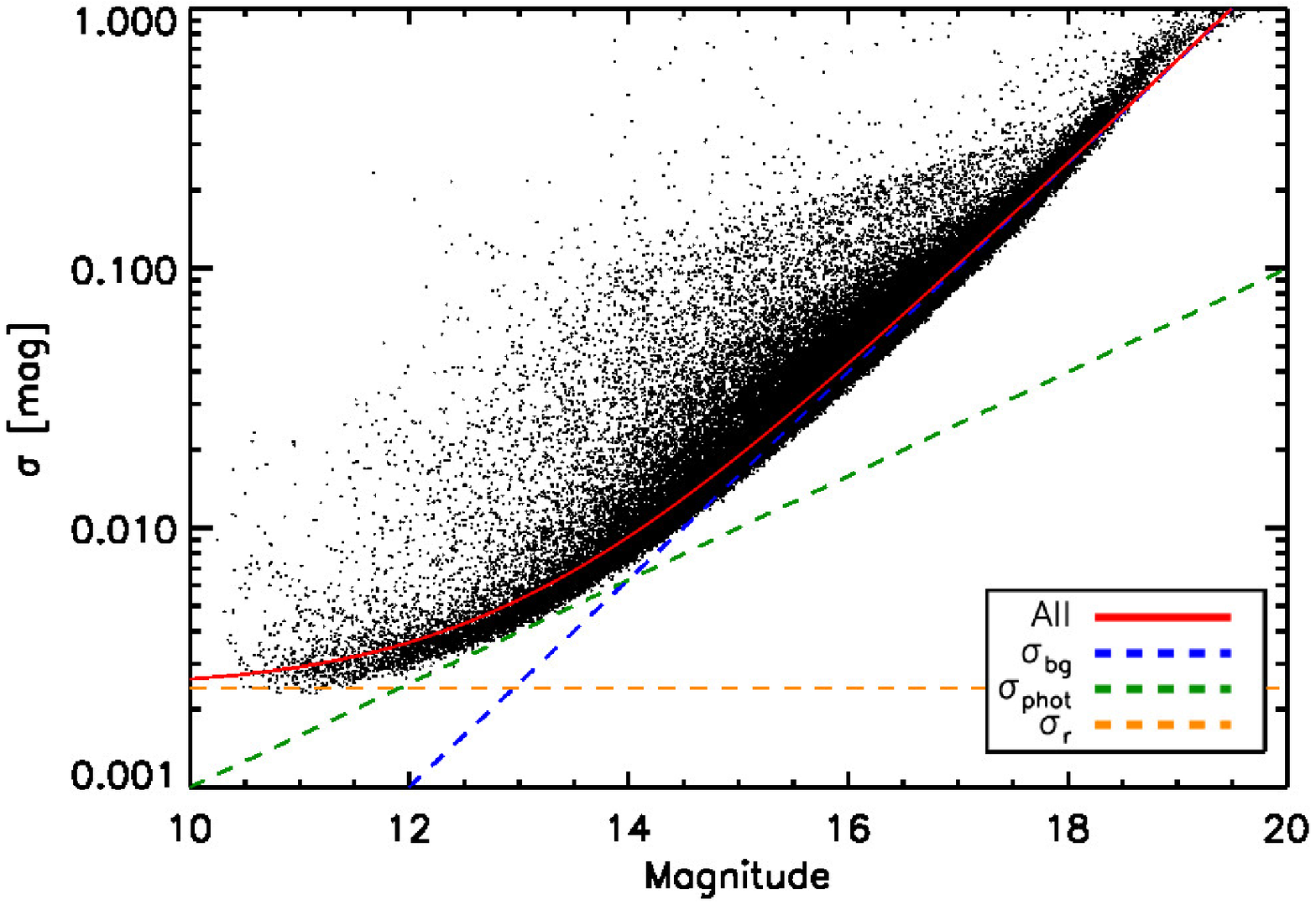}
  }  
  \subfigure[Exo3]{
   \includegraphics[width=.5\linewidth]{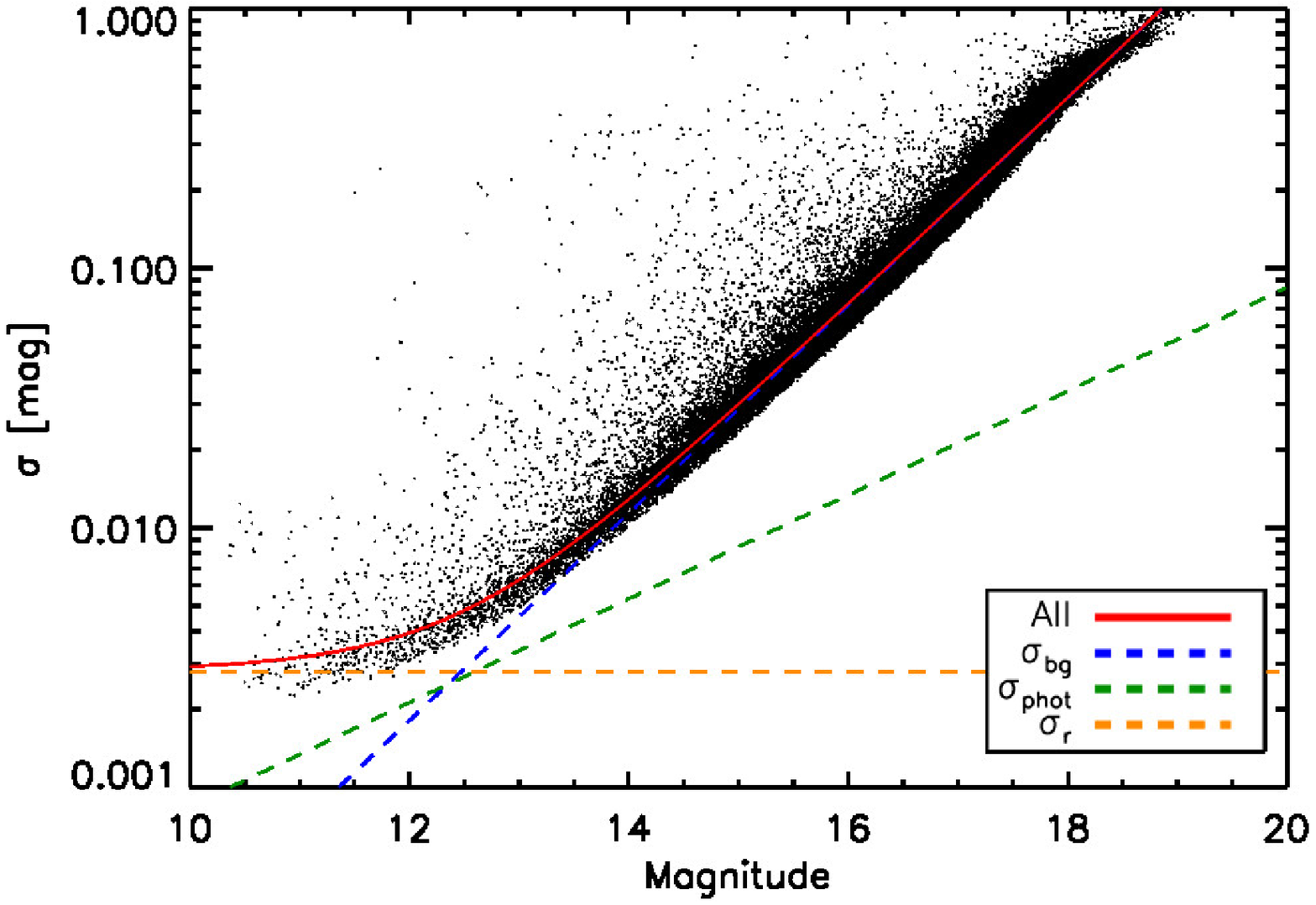}
   \includegraphics[width=.5\linewidth]{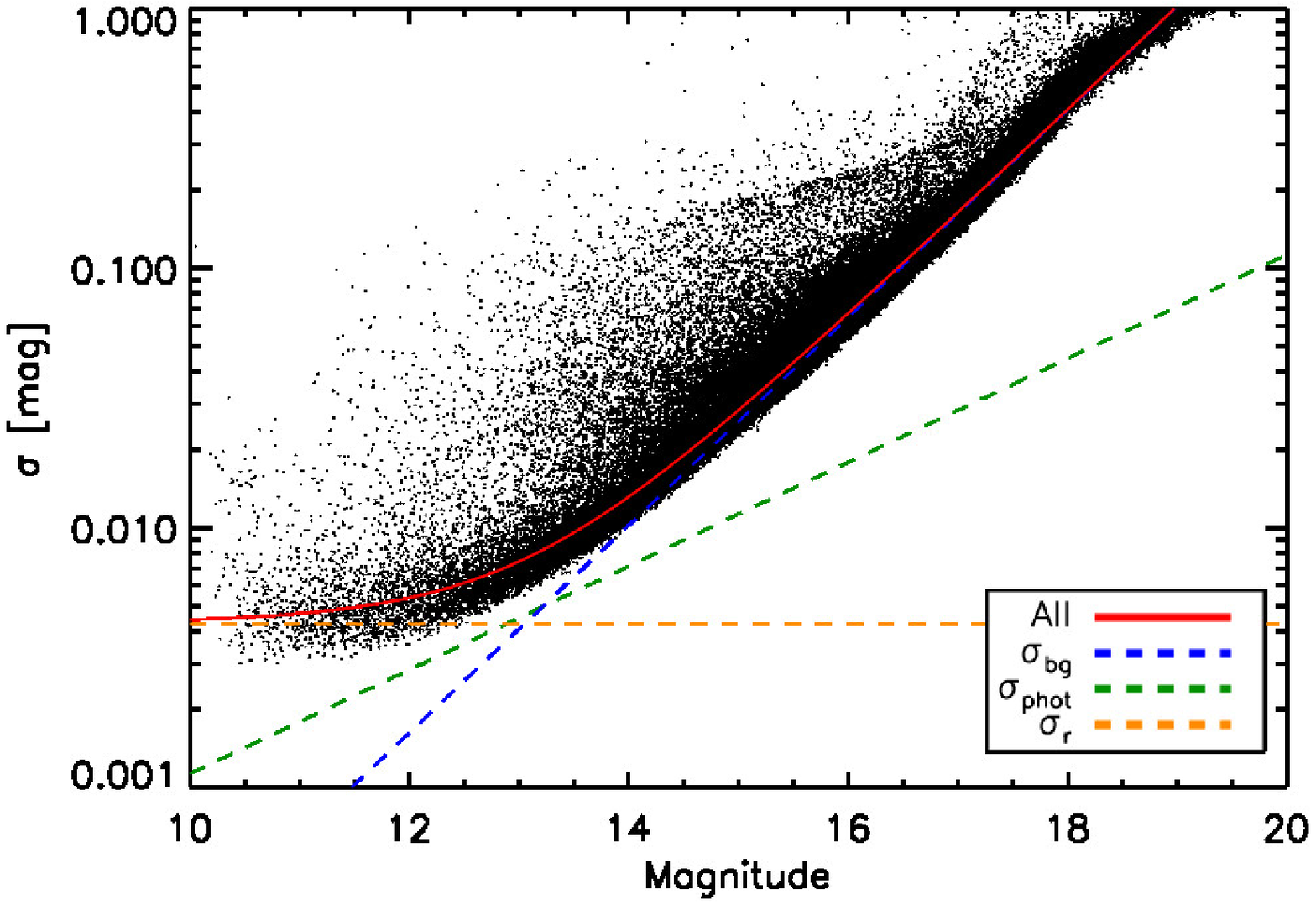}  
  }\\[-.5em]
  \caption{Photometric quality of observations on target fields (a) Exo2 and (b) Exo3 for each site individually. Left plots show median magnitudes and standard deviations for all ASTEP light curves (unbinned), right plots the same for BEST\,II. Lines indicate a fit of Equation~(\ref{eq:sigma-exp}) to the data.
  }
  \label{fig:e23:rmsplots}
\end{figure*}
The noise level of a flux measurement $f$ delimits the transit depth and hence, the planetary radius that can be measured. It is governed by three main factors: First, the signal itself is produced by photons arriving at random and thus follows a Poisson distribution, i.e., the corresponding photon noise $\sigma_\textrm{ph}$ is proportional to $f^{-1/2}$. Second, uncertainties in the background flux level estimation and noise from the calibration process add up to a constant absolute error $\Delta f_\textrm{bg}$ in each pixel, thus yielding a relative noise component $\sigma_\textrm{bg}\propto 1/f$. Third, correlated noise is proportional to $f$ and thus yields a constant relative noise level $\sigma_r$. The combined relative photometric error~$\sigma$ can be described analytically \citep[after][Equation~(12)]{Newberry1991}:
\begin{equation}\label{eq:sigma-exp}
  \sigma\approx 
  \sqrt{
  \sigma_r^2 + \frac{1}{g\cdot f}
  + n^\textrm{ap}_\star\left(1+\frac{1}{n^\textrm{ap}_\textrm{sky}}\right) \cdot \left(\frac{\Delta f_\textrm{bg}}{f}\right)^2
  },
\end{equation}
whereby $\Delta f_\textrm{bg}$ and the stellar flux $f$ are given in ADU, $g$ denotes the CCD gain factor, and $n^\textrm{ap}_\star$ and $n^\textrm{ap}_\textrm{sky}$ specify the number of pixels in the stellar aperture and the background annulus, respectively. This expression is used to determine $\sigma_r$ and $\Delta f_\textrm{bg}$ by fitting to $\left( f,\sigma \right)$ of a large set of light curves (e.g.~Figure~\ref{fig:e23:rmsplots}). 

The parameter $\sigma_r$ quantifies remnant systematic trends in the data that are typically due to multiple factors including both technical issues (e.g., telescope tracking, calibration uncertainties) and environmental constraints (e.g., clouds, scintillation). Since it gives a photometric noise limit that cannot be decreased by increasing the flux $f$ (e.g., through binning or by using a larger telescope), $\sigma_r$ is most interesting for comparing the two observing sites.

\subsection{Transit Detection Yield}\label{sec:detyield} 
This study aims at a quantitative comparison of the expected planet detection yield based on a thorough assessment of the photometric noise budget and observational duty cycle obtained from each site. The model used for that builds on the theoretical framework of \citet{Beatty2008}, but uses some simplifying assumptions; it is described in detail in Appendix~\ref{sec:app:detyield}. By considering a single planetary radius $r_{p0}$ and a small range $\left[p_0,p_1\right]$ of orbital periods $p$, the number of detections is approximated as 
\begin{equation}\label{eq:res:Ndet}
 N_\textrm{det}\cong N_\textrm{S/N}\cdot f_{p0} \cdot f_t,
\end{equation}
whereby $f_{p0}$ gives the fraction of stars that possess a planet in the considered period range with radius $r_{p0}$, and $N_\textrm{S/N}$ denotes the number of light curves with sufficient S/N to detect a transit in a given data set. The value $N_\textrm{S/N}$ is calculated by comparing the observed photometric quality with the expected transit depth in a Monte Carlo simulation, which makes use of the Besan\c{c}on model \citep{Robin2003} to characterize the underlying sample of $N_\star$ stars in each target field. 

The parameter $f_t$ defines a probability that is constrained by timing factors, i.e., it integrates the geometric transit probability $p_g(p)$ together with the observational window function $p_\textrm{win}$ over $\left[p_0,p_1\right]$ (Equation~(\ref{eq:app:detyield:ft})). Since at least three or more transits are usually to be recorded for a detection, the term $p_\textrm{win}$ is approximated by the phase coverage~$p_\textrm{c3}(p)$ of three or more transit events from real observing times.

Since the two ASTEP fields have been monitored for only a relatively short period of time, the detection yield is being investigated for orbits of $p\in\left[1,10\right]$\,days. For this range, the probability for a star to host a Jupiter-sized planet has been estimated by other ground-based surveys. Here, $f_{p0}=0.43\%$ ($M_p=M_J$, $p<11.5$\,days; \citealt{Cumming2008}) is taken as a representative example. It agrees well with the occurance rate as determined by the two transit surveys from space, which both found 0.4\% for hot Jupiters with periods of~$p<10$\,days; CoRoT with $M_p=0.45$--2.5\,$M_J$ \citep{Guenther2012a}, Kepler with~\mbox{$r_p=8$--32\,$r_\oplus$} \citep{Howard2012}.

\section{Photometric Quality}\label{sec:phot}
The photometric quality of Dome~C is being analyzed for the following two science cases. First, Section~\ref{sec:phot:short} evaluates the noise level for time scales of single nights, which is important, for instance, to monitor and characterize individual transit events. Second, Section~\ref{sec:phot:long} analyzes the photometric noise budget of whole time series, which is most relevant for the detection yield of a transit survey.

\subsection{Individual Nights}\label{sec:phot:short} 

\subsubsection{Target Fields Exo2 and Exo3}\label{sec:phot:short:exo23}
\begin{figure}[htpc]\centering
 \includegraphics[width=\linewidth]{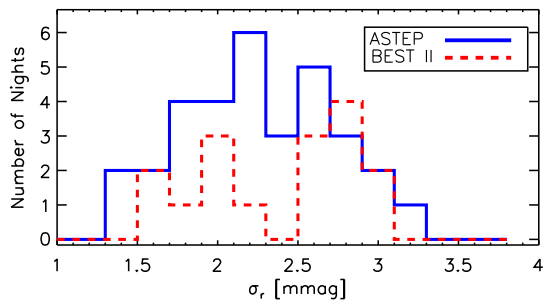}
  \caption{Histogram for systematic noise components $\sigma_r$ in individual nights of Exo2 and Exo3 data. Values have been obtained through fitting as in Figure~\ref{fig:e23:rmsplots}, but for each night individually.}
  \label{fig:e23:sigmar-hist}
\end{figure}

In order to estimate the limiting noise levels of single nights, the value $\sigma_r$ was obtained by fitting Equation~(\ref{eq:sigma-exp}) to individual nights for each large data set and telescope.
The results are summarized in a histogram (Figure~\ref{fig:e23:sigmar-hist}). 
It shows that a rather similar nightly noise component $\sigma_r$ is encountered at each site, typically ranging at excellent values of 1.5--3\,mmag. However, the median of $\sigma_r$ over all nights is 2.28\,mmag for ASTEP and 2.60\,mmag for BEST\,II data, thus indicating a slight advantage for the Antarctic site. 

\subsubsection{WASP-18b}\label{sec:res:wasp18} 
\begin{figure*}[t]\centering
  \includegraphics[width=.85\linewidth]{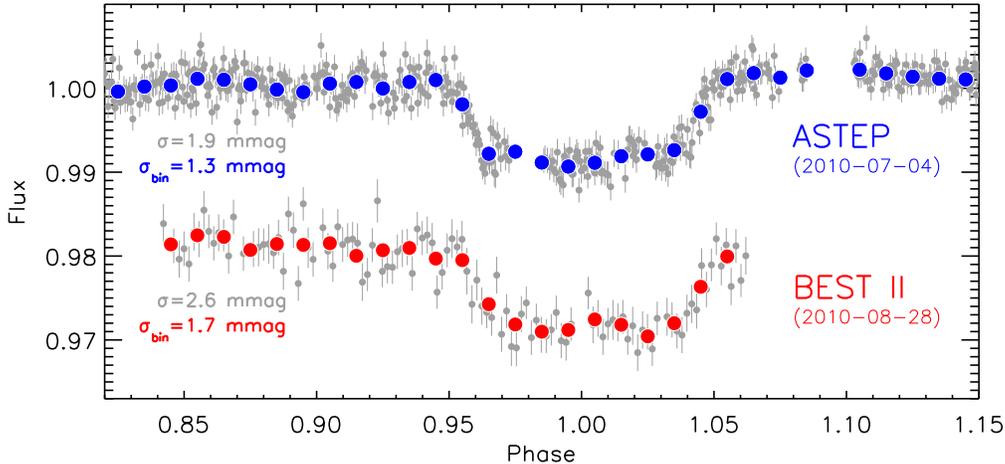}
  \caption{Phase-folded light curves of WASP-18b around transit. Shown are two individual events, one observed with ASTEP (upper line) and one observed with BEST\,II (lower line). In addition to individual measurements (gray), data binned to phase intervals of 0.01 ($\equiv 13.6$\,min) are shown in red (BEST\,II) and blue (ASTEP), respectively. For better visibility, the BEST\,II light curve was shifted in flux by $-0.02$.
  \label{fig:wasp18-single}}
\end{figure*}

Figure~\ref{fig:wasp18-single} shows a comparison of a WASP-18b transit recorded with BEST\,II and ASTEP. While the BEST\,II light curve is the best out of six with a full transit (28th August 2010), the ASTEP night (4th July 2010) was selected without any requirement except for the planet to transit. Note, however, that the statistical significance of the following case study is limited since environmental conditions can vary largely between individual nights.

Table~\ref{tab:obs:wasp18} shows the number of measurements~$N$ and the photometric standard deviation~$\sigma$ for data points inside and outside the transit event, respectively (ephemerides from \citet{Southworth2009}). Out of transit, the noise level of unbinned ASTEP data, 1.87\,mmag, is 27\% lower than unbinned BEST\,II measurements (2.57\,mmag). During the transit, ASTEP yields, with 1.66\,mmag, a 19\% lower noise level than BEST\,II (2.05\,mmag). Assuming only white noise, the transit depth uncertainty~$\sigma_\textrm{dep}$ can be calculated using
\begin{equation}\label{eq:sigmatransit}
  \sigma^2_\textrm{dep} = \sigma^2_\textrm{out}/N_\textrm{out} + \sigma^2_\textrm{in}/N_\textrm{in} .
\end{equation}
This yields an average transit depth of $0.92\%\pm 0.02$\% for ASTEP, and $0.91\%\pm 0.05$\% for BEST\,II, which corresponds to an S/N of 50 and 21, respectively. This difference is mainly determined by the ASTEP mean cadence being four times shorter than for BEST\,II (36\,s and 152\,s, respectively).

A better noise level in the ASTEP time series is also encountered when binning the data to phase intervals of 0.01 ($\equiv$\,13.6\,min; cf.~red and blue points in Figure~\ref{fig:wasp18-single}). However, binning yields significantly larger noise levels than expected from a purely Gaussian noise distribution\footnote{Gaussian noise would yield $\sigma_\textrm{bin}\approx\sigma_\textrm{unbinned}\cdot\sqrt{n_\textrm{bin}}$, whereby $n_\textrm{bin}$ denotes the average number of data points per bin; see also Table~\ref{tab:obs:wasp18}.} and hence, a smaller overall S/N. Using this comparison, both transits are thus considered to contain unfiltered stellar variability and/or remnant systematic noise in the order of $\approx$\,0.3--0.6\,mmag. Consequently, the S/N drops significantly to 24 (by 52\%) for binned ASTEP and to 14 (by 34\%) for binned BEST\,II data.

\begin{table}[t]\centering\footnotesize
\caption{WASP-18b transit. \label{tab:obs:wasp18}}
\begin{tabular}[htbp]{lcccc}
\tableline\tableline
 & \multicolumn{1}{c@{\,}}{ASTEP} & \multicolumn{1}{c}{BEST\,II} & \multicolumn{1}{c@{\,}}{ASTEP} & \multicolumn{1}{c}{BEST\,II} \\
 & \multicolumn{2}{c}{(unbinned)} & \multicolumn{2}{c}{(binned)} \\
\tableline
$N_\textrm{out}$ & 387 & 67 & 24 & 14\\
$N_\textrm{in}$  &  88 & 36 &  8 &  8\\
$\sigma_\textrm{out}$/mmag & 1.87 &  2.57 & 1.27 & 1.65 \\
 &&& (0.47)\tablenotemark{\dagger} & (1.18)\tablenotemark{\dagger} \\
$\sigma_\textrm{in}$/mmag & 1.66 & 2.05 & 0.77 & 1.26 \\
 &&& (0.50)\tablenotemark{\dagger} & (0.97)\tablenotemark{\dagger} \\
 S/N\tablenotemark{\star} & 49.8 & 21.3 & 23.8 & 14.1 \\
\tableline
\end{tabular}\vspace{-3mm}
\tablenotetext{\dagger}{Expected value if corresponding unbinned~$\sigma$ was Poisson\\ noise only (for comparison).}
\tablenotetext{\star}{Using Equation~(\ref{eq:sigmatransit}).}
\end{table}

\subsection{Transit Search}\label{sec:phot:long} 
Since transiting planets are searched in time series spanning multiple nights, the overall light curve noise level determines the detection limit. 

\begin{figure*}[htc]\centering
 \subfigure[Exo2]{\hspace{-3mm}
   \begin{minipage}[b]{0.50\linewidth}
    \includegraphics[width=\linewidth]{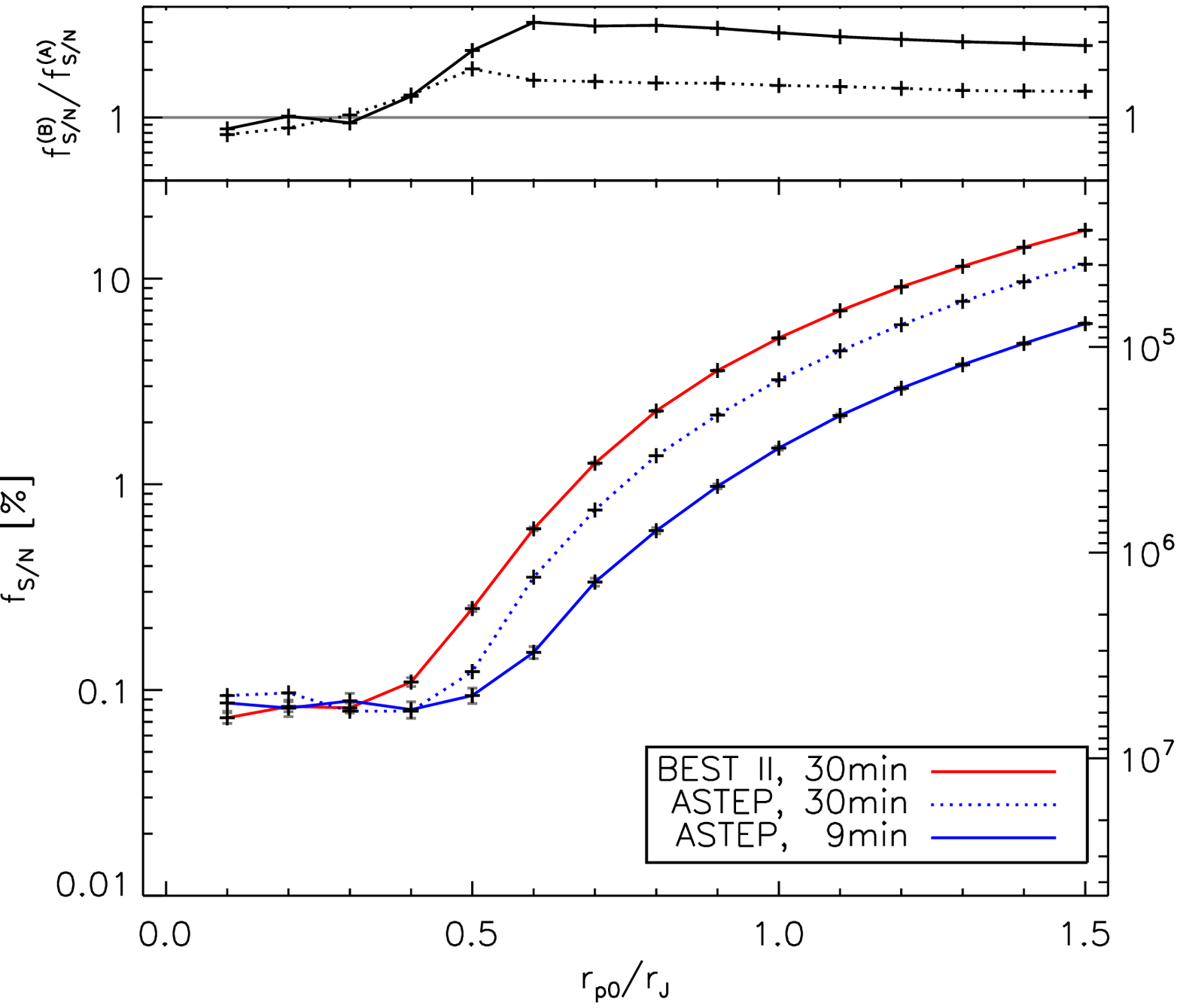}
   \end{minipage}}   
 \subfigure[Exo3]{\hspace{-3mm}
   \begin{minipage}[b]{0.50\linewidth}\hspace*{2.7mm}
    \includegraphics[width=\linewidth]{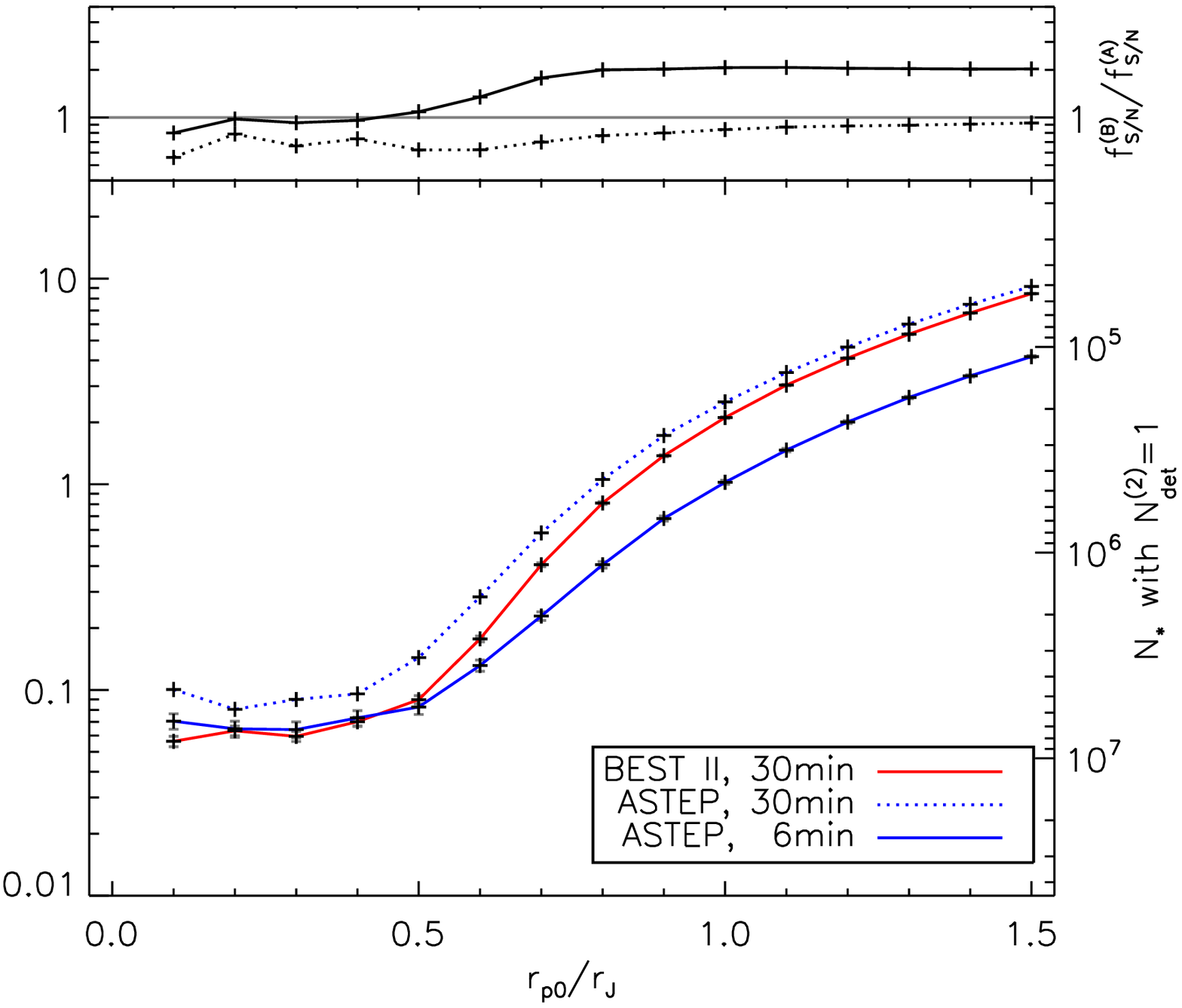}
   \end{minipage}}        
  \vspace*{-2mm}
  \caption{Fraction $f_\textrm{S/N}(r_{p0})$ in fields (a) Exo2 and (b) Exo3 of light curves that provide a photometric noise $\sigma$ sufficient for the detection of a transit signal. The results are calculated as a function of the tested planetary radius $r_{p0}$ and shown for the two binned scenarios, i.e., (1) ASTEP (blue, dotted) and BEST\,II data (red) each binned to 30\,min intervals, and (2) ASTEP data (blue, solid) binned to the same photon noise level $\sigma_\textrm{ph}$ as the corresponding BEST\,II data set (red). The upper panels display the corresponding ratios $f_\textrm{S/N}^B/f_\textrm{S/N}^A$ for a direct comparison. In addition, $f_\textrm{S/N}$ (left $y$-axis) is converted to the total number of stars $N_\star$ (right $y$-axis) that have to be observed for one detection (i.e., using Equation~(\ref{eq:res:Ndet}) with $N_\textrm{det}=1$, an average value of $f_t=0.05$ and the planet fraction $f_{p0}=0.43\%$ \citep{Cumming2008}).
}
\label{fig:astep:detyield:varradius2}
\end{figure*}

In the same way as for single nights, the limiting  noise component~$\sigma_r$ has thus been calculated for each data set (Table~\ref{tab:e23:rmsplotfit}).  
The overall photometric quality of both fields is excellent: As the rms plots in Figure~\ref{fig:e23:rmsplots} show, a precision in the order of $\sim$\,2--4\,mmag is obtained with each telescope at the bright end of the magnitude range. Thus, the very good noise levels of individual nights (Section~\ref{sec:phot:short:exo23}) are largely sustained with both instruments over the whole observing season.

\begin{table}[t]
\footnotesize
\begin{center}
\caption{Photometric noise levels and exposure times $\Delta T$ of joint BEST\,II/ASTEP observations.  \label{tab:e23:rmsplotfit}}
\begin{tabular}{llrcc}
\tableline\tableline
\textsc{Field} & \textsc{Data Set} & $\Delta T$ & $\sigma_r$ & $\Delta f_\textrm{bg}$ \\
\tableline
Exo2 & ASTEP    &  70\,s & 3.4\,mmag & 16.1\,ADU \\
     & BEST\,II & 120\,s & 2.4\,mmag & 12.6\,ADU \\
Exo3 & ASTEP    &  70\,s & 2.8\,mmag & 24.5\,ADU \\
     & BEST\,II &  90\,s & 4.3\,mmag & 15.9\,ADU \\
\tableline
\end{tabular}
\end{center}
\footnotesize
\textsc{Note}.---The limiting noise $\sigma_r$ (unbinned) and the background flux $\Delta f_\textrm{bg}$ are determined by fitting Equation~(\ref{eq:sigma-exp}) to each rms plot $(\overline{m},\sigma)$.
\end{table}

Since the photometric quality obtained in the data sets Exo2 and Exo3 is promising to find transiting planets, both the individual and the combined time series have been searched for transit signals. This yielded a number of planetary candidates that have first been subject to a range of photometric false alarm tests. For candidates that passed them, follow-up observations have been planned. Since they are still ongoing, the results on this transit survey will be presented once it is concluded. The work here, however, aims to compare the two sites in terms of their potential for transit search, since the actual yield is strongly related to the underlying target fields and the two telescopes that obtained them. In this section, the influence of the photometric quality on transit search is thus assessed in a more general context by estimating the detection yield using the model described in Section~\ref{sec:detyield}.

The photometric precision largely depends on the amount of light collected within a period of time. As such, it depends upon the exposure time, telescope aperture, CCD sensitivity, and cadence between two adjacent measurements. The resulting systematic difference is addressed in two scenarios, which are summarized in Table~\ref{tab:astep:binning}. 
\begin{table}[tpc]\centering\footnotesize
\caption{Binning parameters for Exo2 and Exo3.\label{tab:astep:binning}}
\begin{tabular}{lr@{;\ }rr@{;\ }r}
\tableline\tableline
\textsc{Binning}    & \multicolumn{2}{c}{ASTEP} & \multicolumn{2}{c}{BEST\,II}\\
\tableline
\textsc{Scenario 1} & \multicolumn{4}{c}{\dotfill\ same $\sigma_\textrm{ph}$ \dotfill} \\
--- Exo2            & 9.0\,min &  4.5 & 30\,min & 6.1 \\
--- Exo3            & 6.1\,min &  2.7 & 30\,min & 4.8 \\     
\textsc{Scenario 2} & \multicolumn{4}{c}{\dotfill\ same $\Delta t_\textrm{bin}$ \dotfill} \\
--- Exo2            & 30\,min  & 15.1 & 30\,min & 6.1 \\
--- Exo3            & 30\,min  & 10.4 & 30\,min & 4.8 \\            
\tableline
\end{tabular}
\\\begin{flushleft}\footnotesize
\textsc{Note}.---The table gives the binning interval $\Delta t_\textrm{bin}$ and the average number $n_\textrm{bin}$ of measurements per bin for each data set and investigated scenario.
\end{flushleft}
\end{table}
In the first scenario, BEST\,II data are binned into intervals of $\Delta t^\textrm{(1)}_\textrm{bin}=30$\,min in order to keep a reasonable sampling of $\gtrsim 3$ data points during a transit. Since ASTEP collects more light than BEST\,II due to its larger aperture and smaller readout time, a shorter binning interval $\Delta t^\textrm{(2)}_\textrm{bin}$ is used for ASTEP such that it achieves the same photon noise level $\sigma_\textrm{ph}$ as BEST\,II within 30\,min (see Appendix~\ref{sec:app:binning} for details). In a second scenario, ASTEP data are as well binned to intervals of 30\,min in order to compare the actual instrument performance.

The number $N_\textrm{S/N}$ of light curves suitable for transit search is calculated as a function of planetary radius $r_{p0}$ for each scenario. The results are compared based on the fraction $f_\textrm{S/N}=N_\textrm{S/N}/N_\star$, since the stellar count $N_\star$ itself depends on the size of the FOV; as such, it is driven by the project design and not a site characteristic. Furthermore, the two angular resolutions yield a different degree of contamination (see Appendix~\ref{sec:app:datacombi:phot}); however, its influence on the results is largely reduced by excluding stars from the simulation that deviate by more than 0.5\,mag from catalog values (see Appendix~\ref{sec:app:detyield}). 

Figure~\ref{fig:astep:detyield:varradius2} displays the results in each target field and binning scenario. 
For Exo2 (Figure~\ref{fig:astep:detyield:varradius2}a), BEST\,II yields a better fraction~$f_\textrm{S/N}$ than ASTEP for large planets ($r_{p0}\geq 0.6\,r_J$; by 184--299\% larger with the same photon noise level, and by 46--72\% with the same binning interval). ASTEP yields values of $f_\textrm{S/N}$ similar to BEST\,II only for very small planets ($r_{p0}\leq 0.3\,r_J$). For Exo3 (Figure~\ref{fig:astep:detyield:varradius2}b), the ASTEP $f_\textrm{S/N}$ is up to 79\% better than BEST\,II if both data sets are binned to 30\,min. However, if the comparison is made at the same photon noise level, the fraction $f_\textrm{S/N}$ is up to 107\% higher for BEST\,II, whereas ASTEP again yields a larger fraction for very small radii (by 4--25\% larger for $r_{p0}\leq 0.4\,r_J$).
However, as can also be seen in Figure~\ref{fig:astep:detyield:varradius2}, about five million stars would have to be monitored in order to obtain a reasonable detection probability in this regime of small planets, so that either project does not provide a sufficient precision for detecting Neptune-sized planets in these two data sets. 

Overall, the transit detection efficiency of ASTEP and BEST\,II can be considered well comparable; a significant advantage to detect smaller planets from Antarctica is not evident from these two data sets. Environmental conditions that are expected to yield smaller noise levels at Dome~C (such as scintillation) can thus not form the dominant term in the overall noise budget for bright stars. Instead, remnant systematic trends are more likely to have limited the photometric precision at both sites in this data set.

\section{Duty Cycle}\label{sec:dc}
In addition to the photometric precision, the transit detection yield is constrained by the observational duty cycle. This section investigates its influence on the detection yield for the two investigated large data sets. For that, the timing parameter $f_t$ serves as the most important quantity, since it is directly proportional to the detection yield and can readily be computed from the observational phase coverage $p_\textrm{c3}(p)$ (Equation~(\ref{eq:app:detyield:ft})). In the following text, Section~\ref{sec:dc:single} first describes the efficiency for each individual site, while Section~\ref{sec:dc:joint} focuses on the potential advantage of combining data from Antarctica and Chile for transit search.

\subsection{Individual Time Series}\label{sec:dc:single} 
\begin{figure*}[p]\centering
  \subfigure[Exo2]{\begin{minipage}[c]{.48\linewidth}\vspace{0pt}\hspace{-4mm}
   \includegraphics[width=.59\linewidth,angle=90]{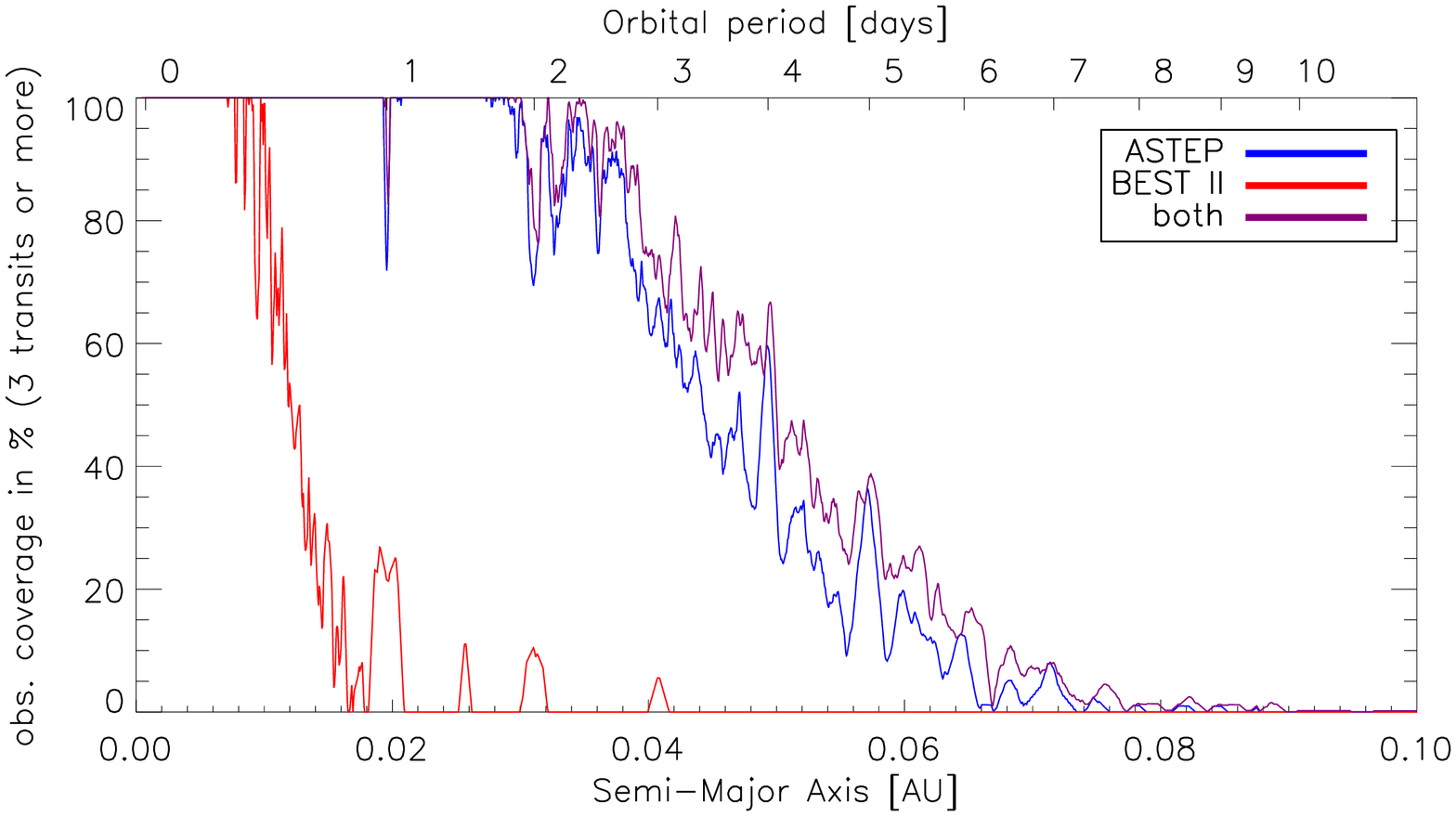}\end{minipage}}
  \subfigure[Exo3]{\begin{minipage}[c]{.48\linewidth}\vspace{0pt}\hspace{0mm}
   \includegraphics[width=.59\linewidth,angle=90]{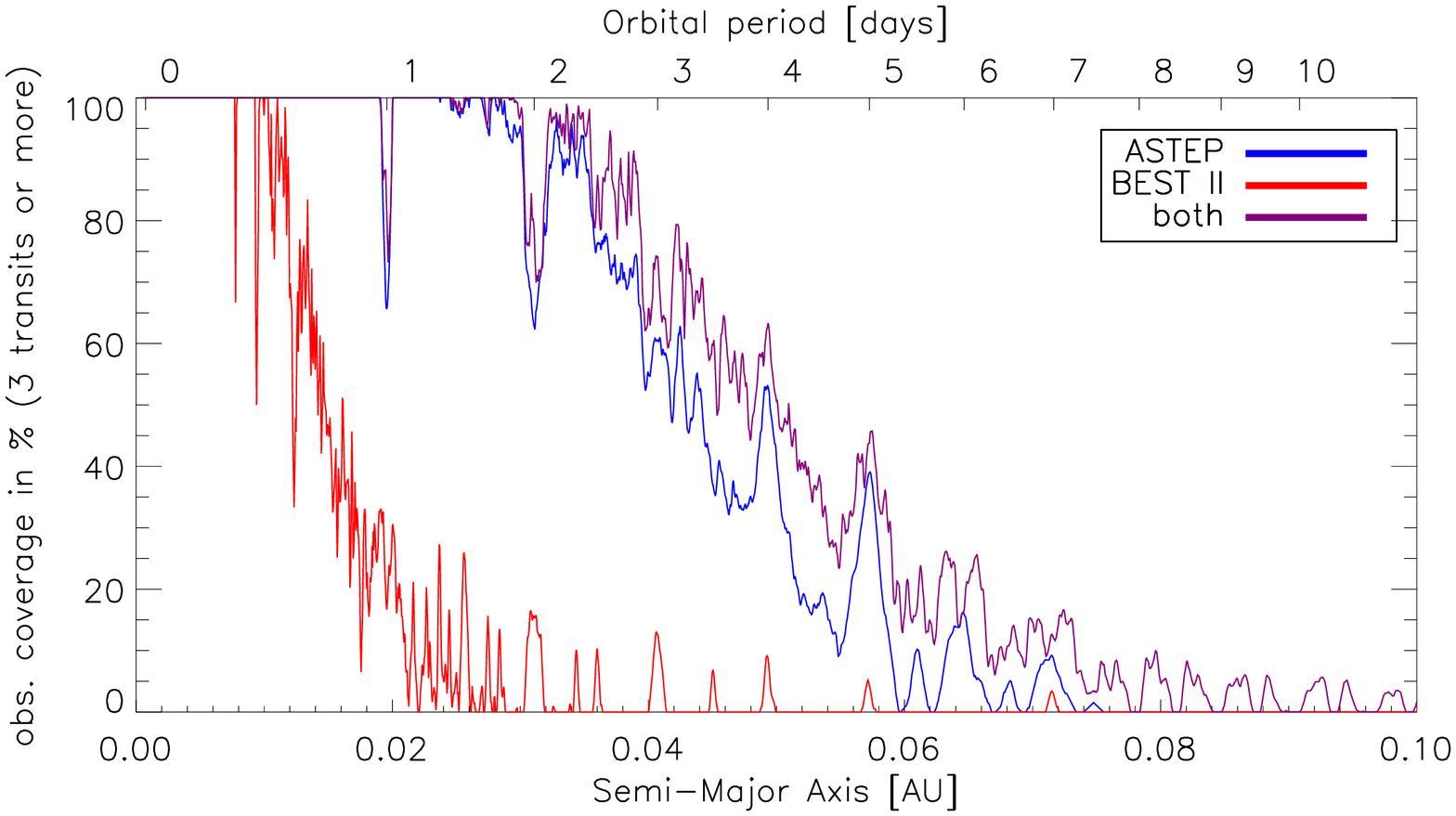}\end{minipage}}
  \vspace*{-2mm}
  \caption{Phase coverage $p_\textrm{c3}(p)$ of three or more transits for the two ASTEP fields as a function of possible orbital periods $p$. The coverage is shown in red for the BEST\,II data, and in blue for ASTEP. The violet line gives the orbital coverage that is obtained using the joint time series. The lower axis gives the corresponding semi-major axis for a solar-mass star. See Figure~\ref{fig:obs:exo23} for a graphical representation of the corresponding observational window functions.}
  \label{fig:detyield1:best2astep-cov}
\end{figure*}
\begin{figure*}[p]\centering
  \subfigure[Short periodic variable (Exo2\_026332)]{
    \includegraphics[width=.85\linewidth]{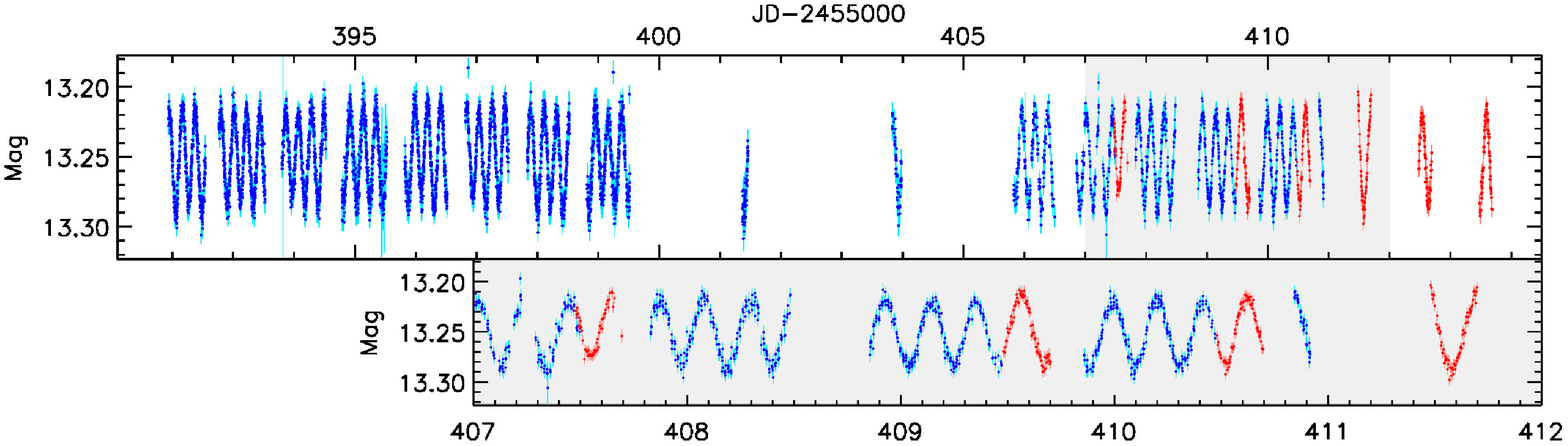}}
  \subfigure[Eclipsing binary (Exo2\_028863)]{  
    \includegraphics[width=.85\linewidth]{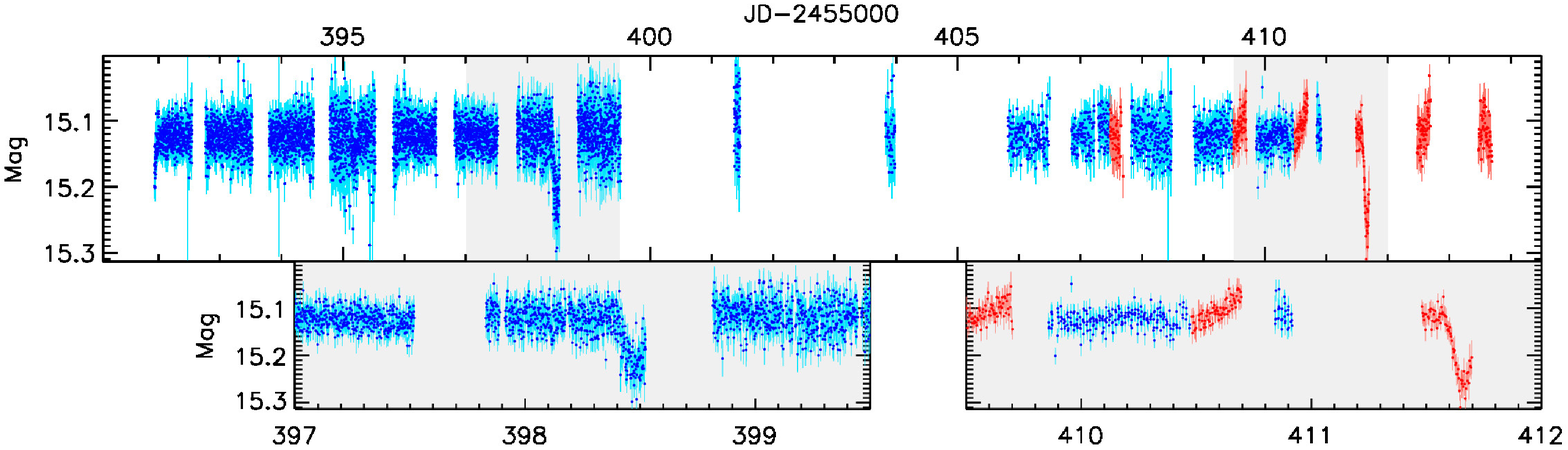}}
  \caption{Examples of two joint BEST\,II/ASTEP light curves. The upper plots show the whole light curve, while the lower plot enlarges interesting sequences (gray shaded). Blue points indicate ASTEP measurements, BEST\,II data appear in red.}
  \label{fig:e23:example}
\end{figure*}Figure~\ref{fig:detyield1:best2astep-cov} displays the observational coverage $p_\textrm{c3}(p)$ of three or more transits for both fields and projects. It shows that ASTEP data cover orbital periods up to 2--3\,days completely, while the additional BEST\,II observations are confined to periods of less than a day. For ASTEP, this yields $f_t=0.0463$ for the field Exo2 and $f_t=0.0438$ for Exo3 (see also Table~\ref{tab:astep:detyield:covsim}), while the BEST\,II observations alone are far too sparse to provide a reasonable phase coverage ($f_t\approx 0.001$). The Antarctic site thus provides a performance within 16 nights that typically requires a whole observing season at a midlatitude site. (BEST\,II typically achieves values of $f_t=0.04$--$0.05$ during a whole campaign covering 30--40 observing nights.)

\subsection{Joint Time Series}\label{sec:dc:joint} 
\subsubsection{Feasibility of Combination}
The combination of ASTEP and BEST\,II data (Section~\ref{sec:datacombi}) yields very homogeneous light curves for bright stars ($m\lesssim 15$\,mag), i.e., the magnitude range that is most interesting for transit search. However, a case-by-case comparison in Appendix~\ref{sec:app:datacombi:phot} shows that significantly different photometric noise levels can be encountered between the two telescopes for fainter targets; as is discussed there in more detail, the difference in angular resolution is considered the dominant cause for this effect.  

Most importantly, a transit search is also feasible in the combined time series: Adding BEST\,II to ASTEP data ($\sim$\,10\% more points) again yields very precise light curves, i.e., the fraction of stars with mmag-precision does not decrease 
(4.4\% for ASTEP alone, and 
4.3\% for matched time series; see Table~\ref{tab:obs:fieldstats}). Detection yield simulations for combined time series also indicate that the high photometric quality is sustained after adding BEST\,II data: They show that the fraction $f_\textrm{S/N}$ of light curves with suitable S/N to detect Jupiter-sized planets remains constant for Exo3, and even increases by 18\% for Exo2. Altogether, the data from the two projects can thus be reasonably combined to extend the observational duty cycle. 

\subsubsection{Examples}\label{sec:detyield:examples} 
Figure~\ref{fig:e23:example} gives two examples of a joint ASTEP\slash BEST\,II light curve. Two variable stars have been selected in order to compare the variation: A short-periodic pulsator ($\delta$~Scuti type, Figure~\ref{fig:e23:example}a) and an eclipsing binary (EA type, Figure~\ref{fig:e23:example}b).

The first case, Figure~\ref{fig:e23:example}a, shows how BEST\,II time series can fill small gaps between two Antarctic nights and thus yield an almost continuous duty cycle. Furthermore, both the amplitude and photometric precision are in very good agreement. The second case, Figure~\ref{fig:e23:example}b, highlights an important and anticipated advantage of joint observations from Antarctica and a midlatitude site: Since only one eclipse event was observed with ASTEP, no period could be derived for this eclipsing binary from ASTEP time series alone. The additional BEST\,II observations, however, uncover a second event, so that the period can be constrained. Again, the amplitudes of both events are in good agreement. 

\subsubsection{Implications for Transit Search}\label{sec:detyield:dutycycle} 
For transit search in general, the phase coverage is extended towards larger orbital periods if ASTEP time series are complemented with BEST\,II data. Since the detection yield is directly proportional to the timing probability $f_t$, this value allows assessment of the effect quantitatively.

\begin{table*}[t]\centering
\caption{Impact of additional BEST\,II observations on the ASTEP planet detection yield.\label{tab:astep:detyield:covsim}}
\begin{tabular}{lc@{\hspace{3.5mm}}r@{\hspace{3.1mm}}c@{\hspace{3.5mm}}c}
\tableline\tableline\\[-.2em]
 & \multicolumn{4}{c}{\textsc{Real ASTEP Time Series}} \\[.2em]
\textsc{Field} & \multicolumn{1}{c}{A$_{\textrm{obs}}$} & \multicolumn{1}{c@{\hspace{3.1mm}}}{A$_{\textrm{obs}}$ + B$_{\textrm{obs}}$} & \multicolumn{1}{c}{A$_{\textrm{obs}}$ + B$_{\textrm{obs}}^{\textrm{shifted}}$} & \multicolumn{1}{c}{A$_{\textrm{obs}}$ + B$_{\textrm{max}}$} 
 \\\cline{2-5}
Exo2 & \textbf{(a$_2$) 0.04630} & (b$_2$) 0.05189 \mbox{\scriptsize(+12\%)} & (c$_2$) 0.05223 \mbox{\scriptsize(+13\%)} & (d$_2$) 0.06144 \mbox{\scriptsize(+33\%)} \\
Exo3 & \textbf{(a$_3$) 0.04381} & (b$_3$) 0.05173 \mbox{\scriptsize(+18\%)} & (c$_3$) 0.05207 \mbox{\scriptsize(+19\%)} & (d$_3$) 0.05527 \mbox{\scriptsize(+26\%)} \\
\\[-.5em]\tableline\tableline\\[-.2em]
 && \multicolumn{2}{c}{\textsc{Maximal ASTEP Time Series}} \\[.2em]
\textsc{Month} && \multicolumn{1}{c@{\hspace{3.1mm}}}{A$_{\textrm{max}}$} & \multicolumn{1}{@{}c}{A$_{\textrm{max}}$ + B$_{\textrm{max}}$} 
 \\\cline{2-5}
{July}   &&  \textbf{(e$_1$) 0.08629}\hspace{4mm}  & (f$_1$) 0.09650 \mbox{\scriptsize(+12\%)} \\
{August} &&  \textbf{(e$_2$) 0.07198}\hspace{4mm}  & (f$_2$) 0.08563 \mbox{\scriptsize(+19\%)} \\
\tableline
\end{tabular}\vspace{-4mm}
\tablecomments{For each case studied, the table gives the timing fraction $f_t$ (Equation~(\ref{eq:app:detyield:ft})) that is proportional to the planet detection yield (Equation~(\ref{eq:res:Ndet})). Cases include real ASTEP (A$_{\textrm{obs}}$) and BEST\,II (B$_{\textrm{obs}}$) time series, shifted time series (B$_{\textrm{obs}}^{\textrm{shifted}}$), as well as hypothetical scenarios (A$_{\textrm{max}}$, B$_{\textrm{max}}$); see the text for a detailed description. In addition, the table shows the relative improvement of $f_t$ compared to the reference case (bold) of ASTEP observations alone.}
\end{table*}
\begin{figure*}[t]\centering\footnotesize
 \begin{minipage}[b]{0.25\linewidth}\centering
   \begin{minipage}[b]{0.45\linewidth}\centering\includegraphics[width=\linewidth]{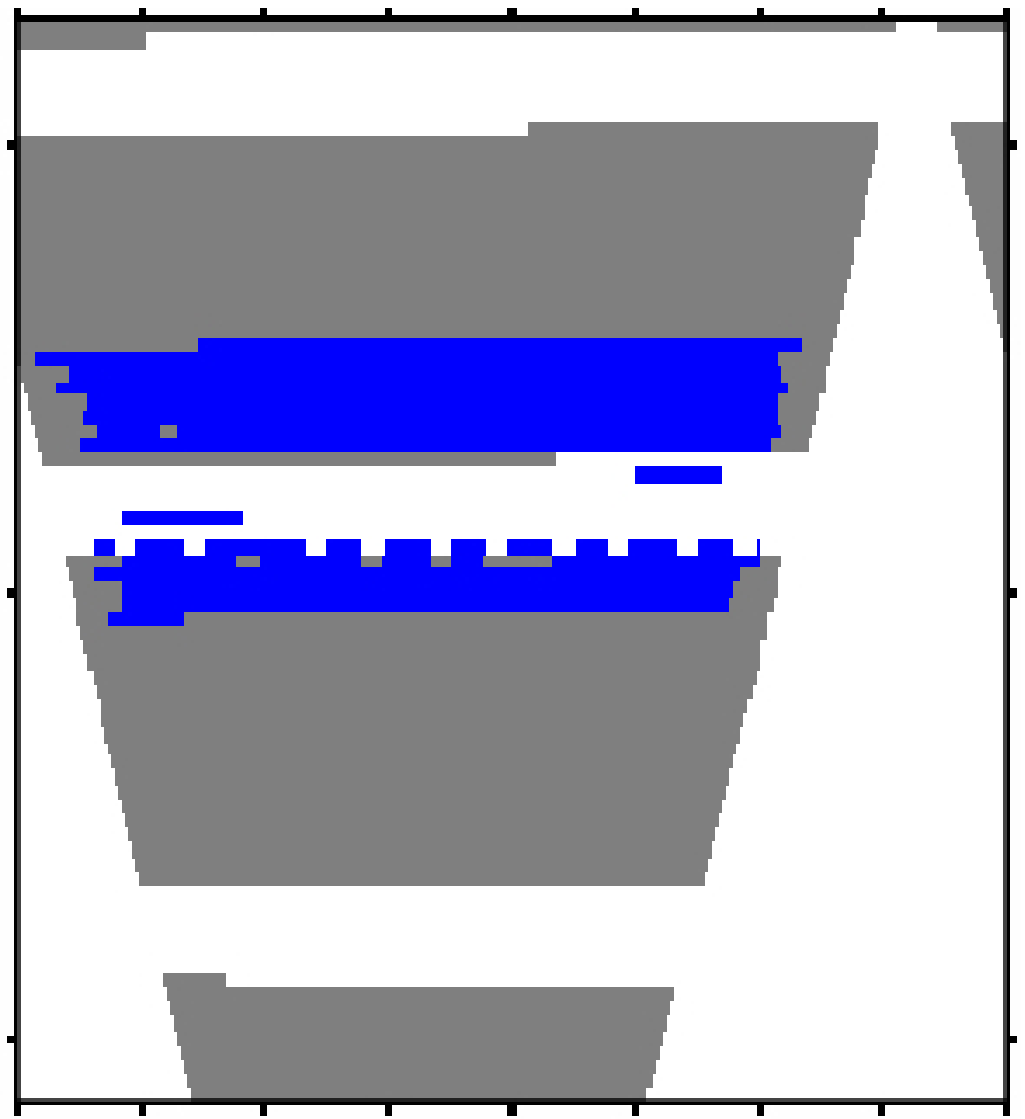}    \\(a$_2$)\vspace{0mm}\end{minipage}
   \begin{minipage}[b]{0.45\linewidth}\centering\includegraphics[width=\linewidth]{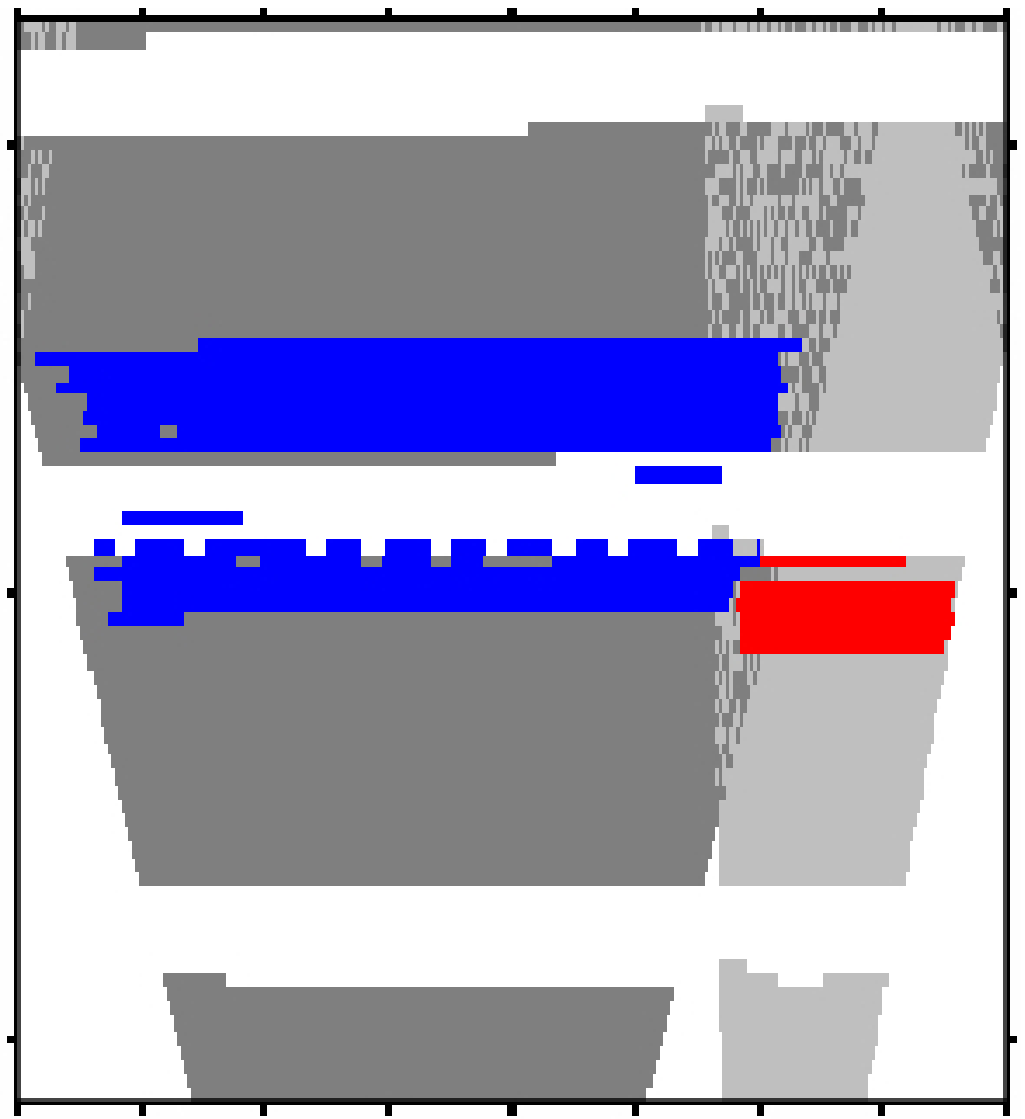}    \\(b$_2$)\vspace{0mm}\end{minipage}
   \begin{minipage}[b]{0.45\linewidth}\centering\includegraphics[width=\linewidth]{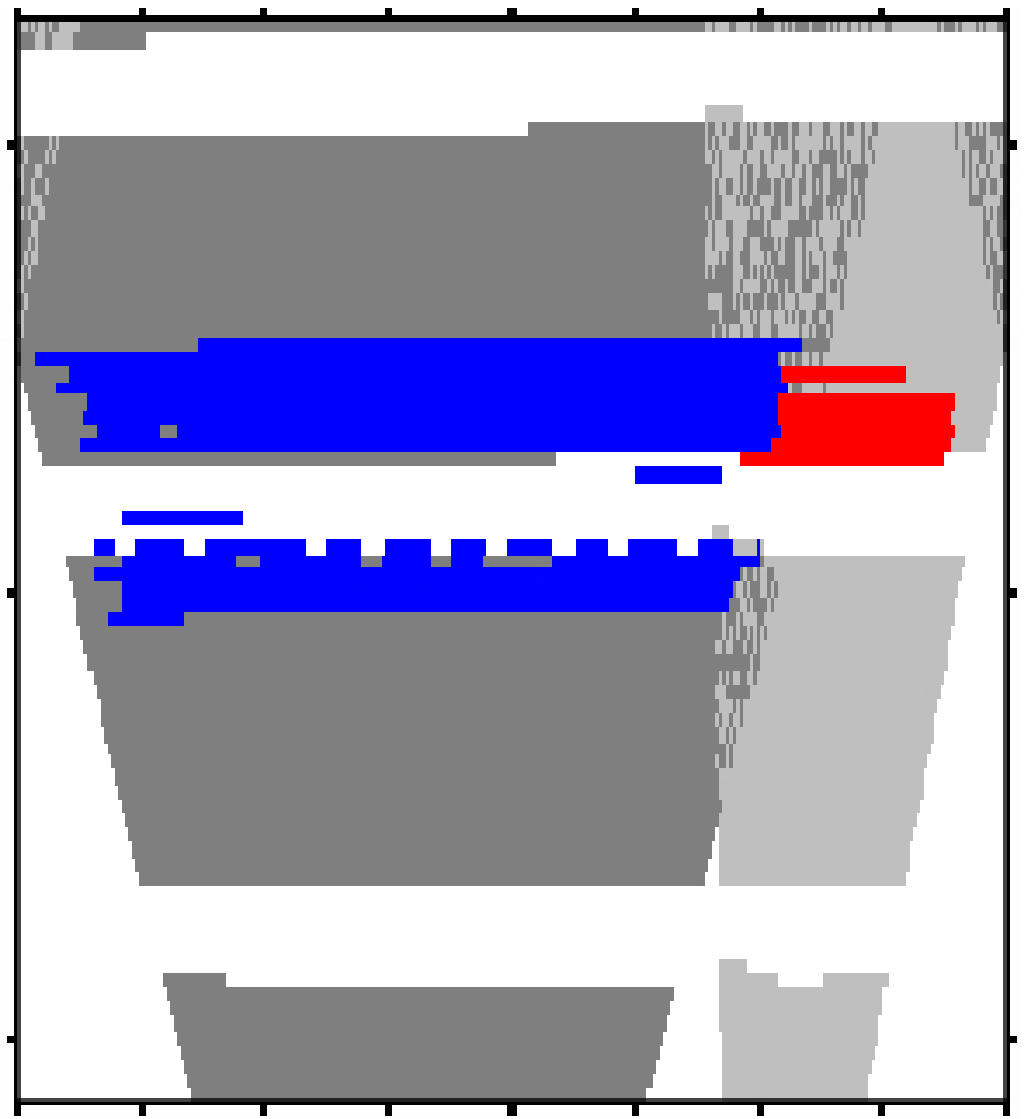}\\(c$_2$)\vspace{0mm}\end{minipage}
   \begin{minipage}[b]{0.45\linewidth}\centering\includegraphics[width=\linewidth]{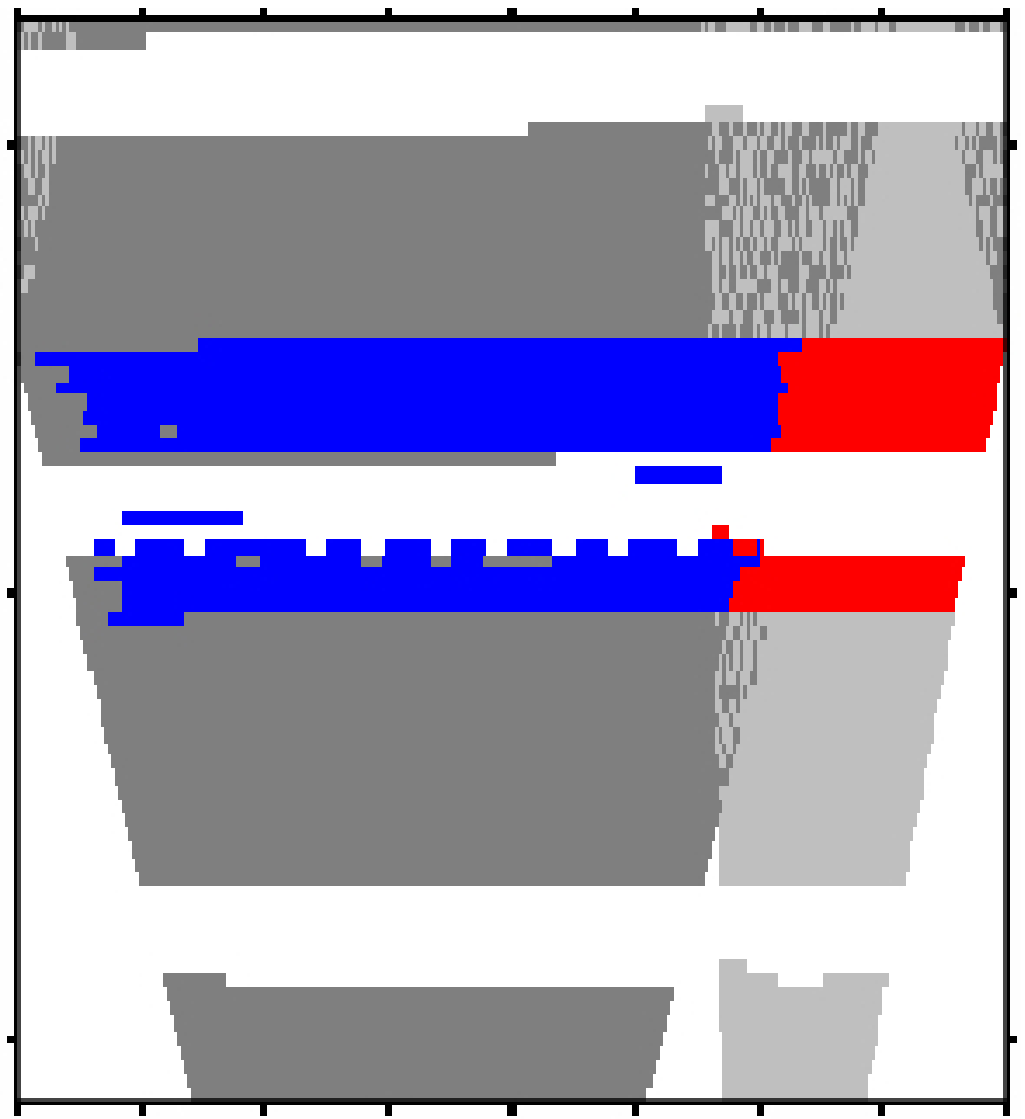}    \\(d$_2$)\vspace{0mm}\end{minipage}
 \textbf{Exo2}
 \end{minipage}\hspace{5mm}
 \begin{minipage}[b]{0.25\linewidth}\centering
   \begin{minipage}[b]{0.45\linewidth}\centering\includegraphics[width=\linewidth]{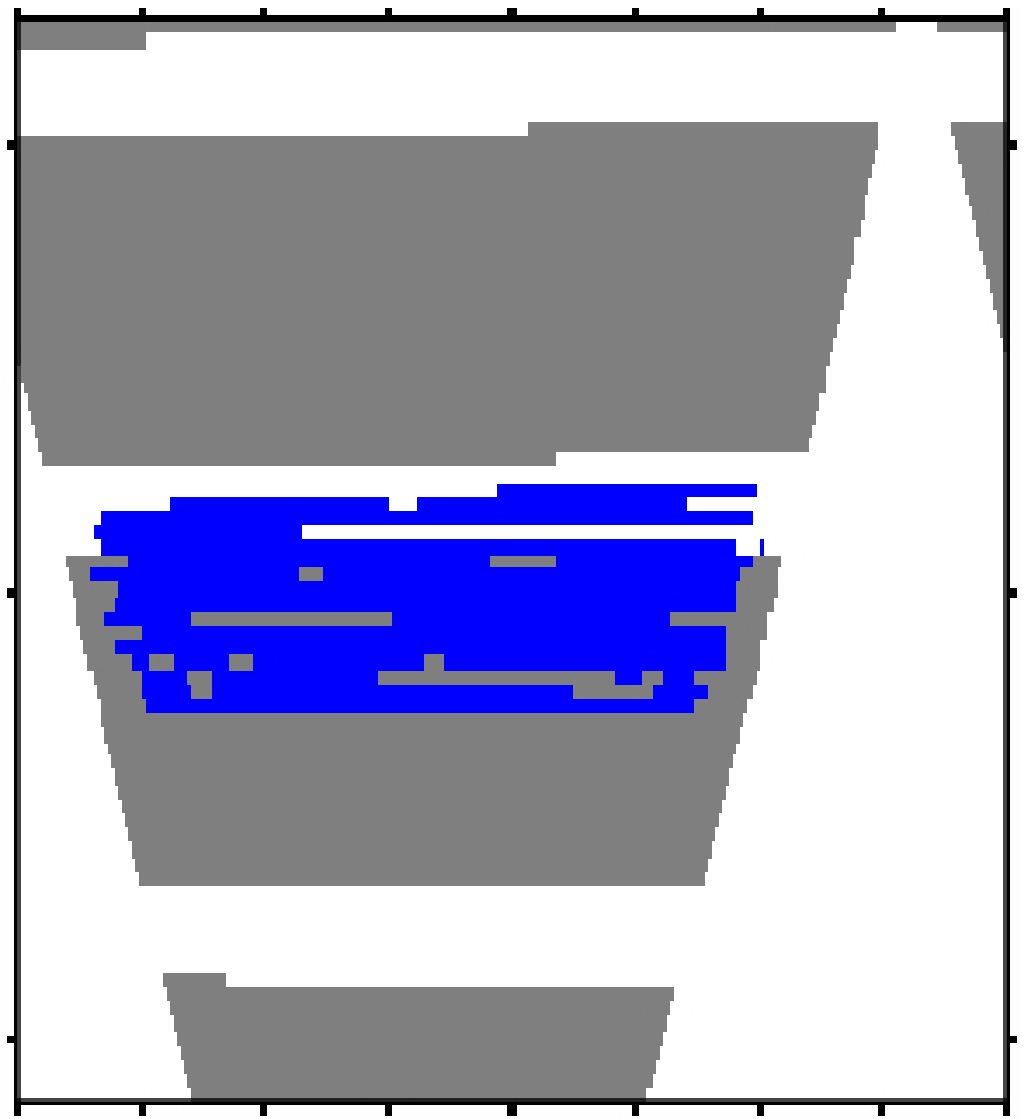}    \\(a$_3$)\vspace{0mm}\end{minipage}
   \begin{minipage}[b]{0.45\linewidth}\centering\includegraphics[width=\linewidth]{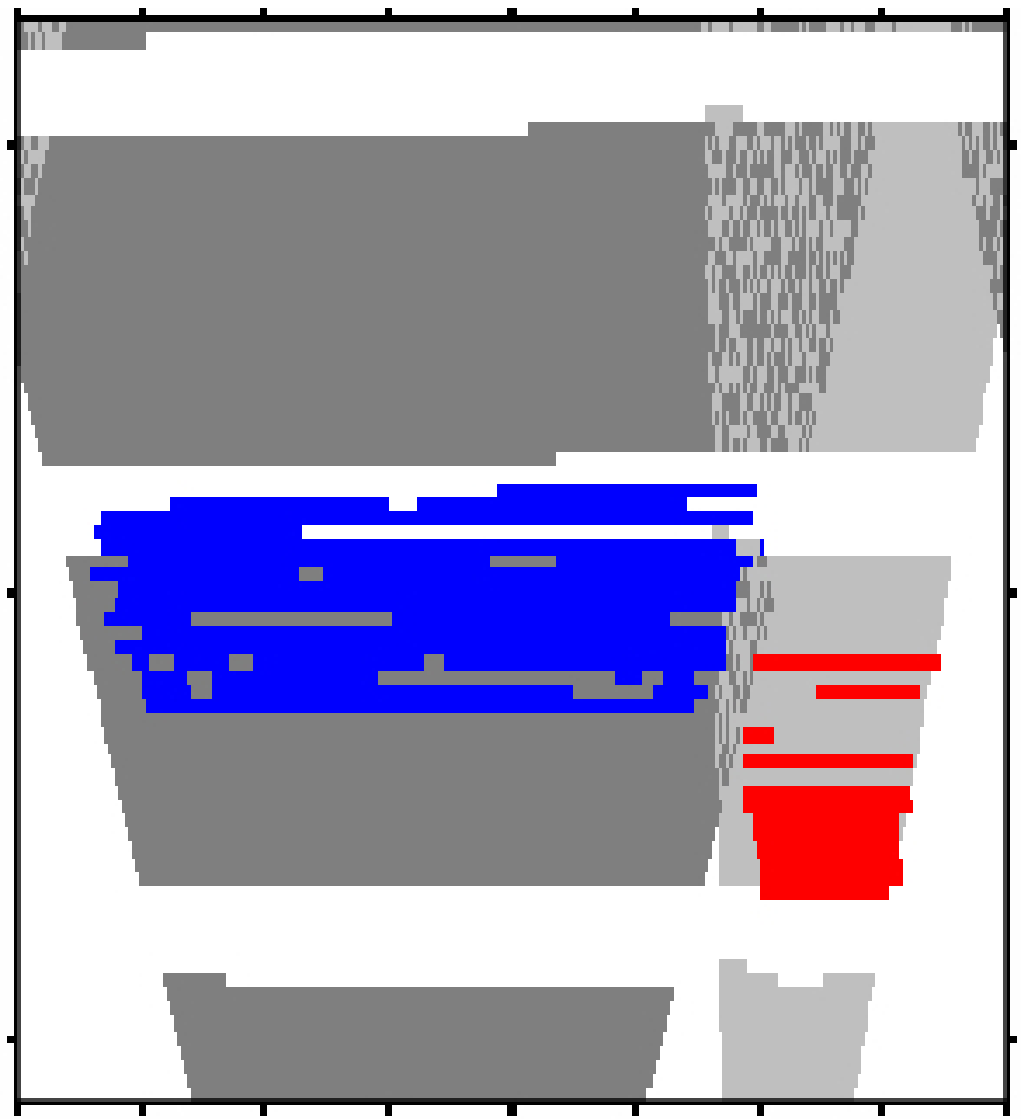}    \\(b$_3$)\vspace{0mm}\end{minipage}
   \begin{minipage}[b]{0.45\linewidth}\centering\includegraphics[width=\linewidth]{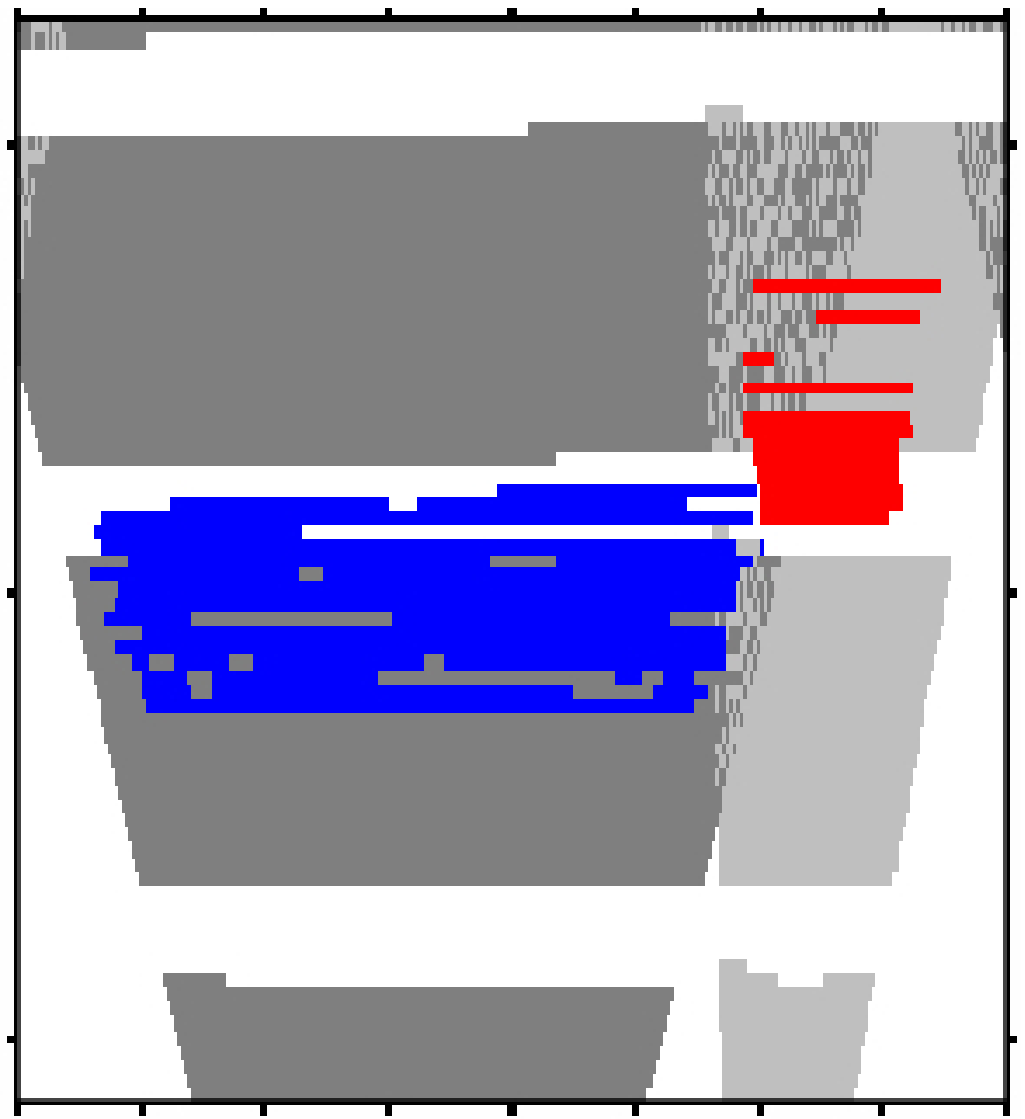}\\(c$_3$)\vspace{0mm}\end{minipage}
   \begin{minipage}[b]{0.45\linewidth}\centering\includegraphics[width=\linewidth]{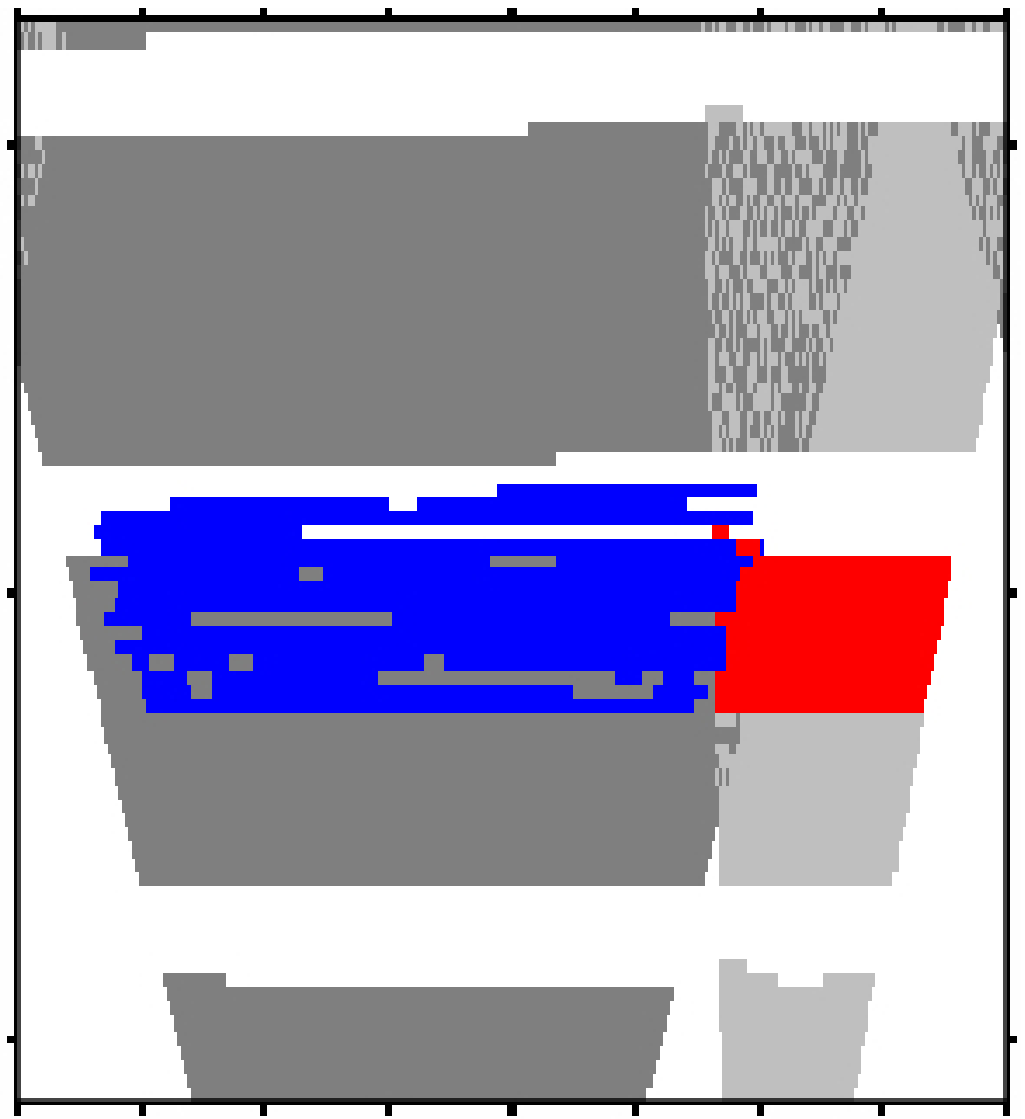}    \\(d$_3$)\vspace{0mm}\end{minipage}
   \textbf{Exo3}
 \end{minipage}\hspace{5mm}
 \begin{minipage}[b]{0.25\linewidth}\centering
   \begin{minipage}[b]{0.45\linewidth}\centering\includegraphics[width=\linewidth]{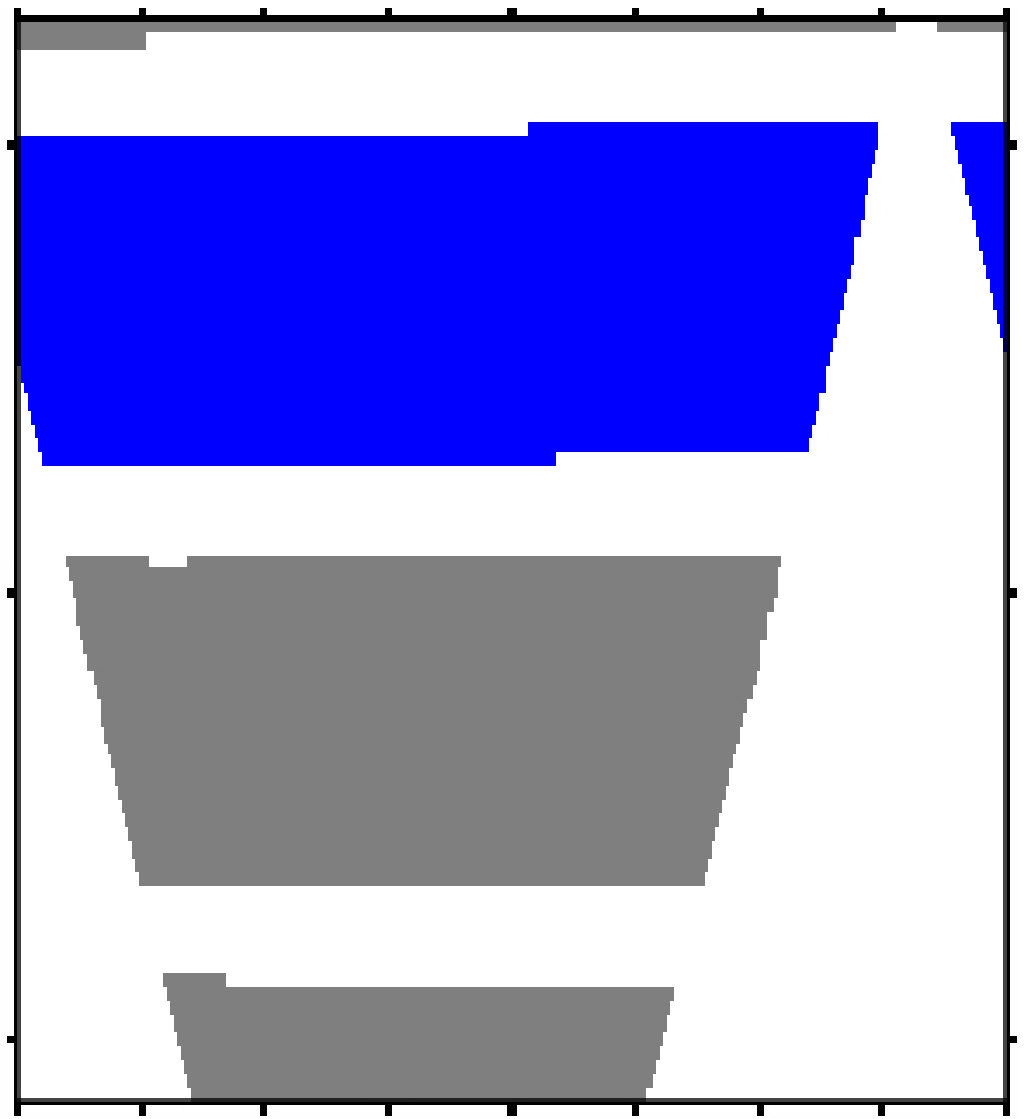}\\(e$_1$)\vspace{0mm}\end{minipage}
   \begin{minipage}[b]{0.45\linewidth}\centering\includegraphics[width=\linewidth]{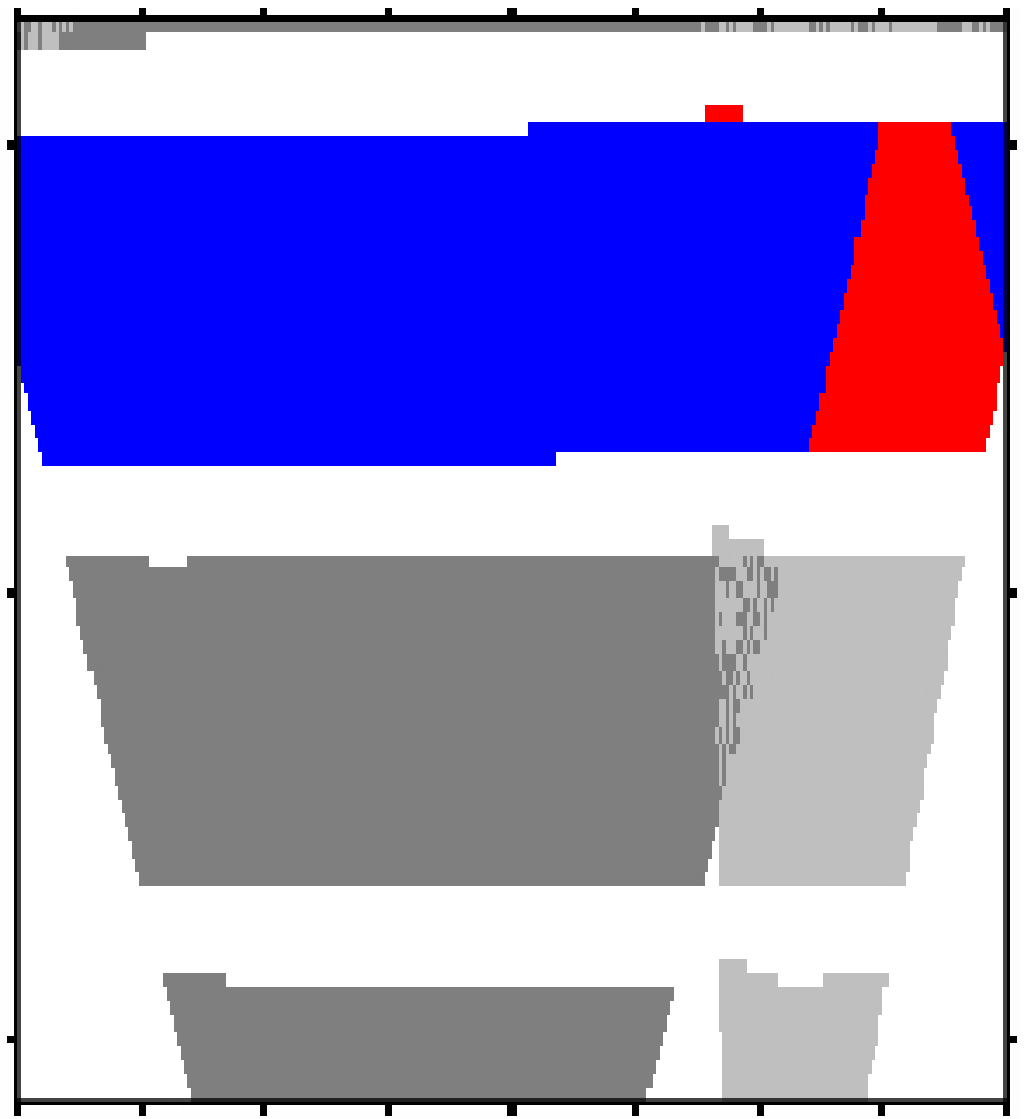}\\(f$_1$)\vspace{0mm}\end{minipage}
   \begin{minipage}[b]{0.45\linewidth}\centering\includegraphics[width=\linewidth]{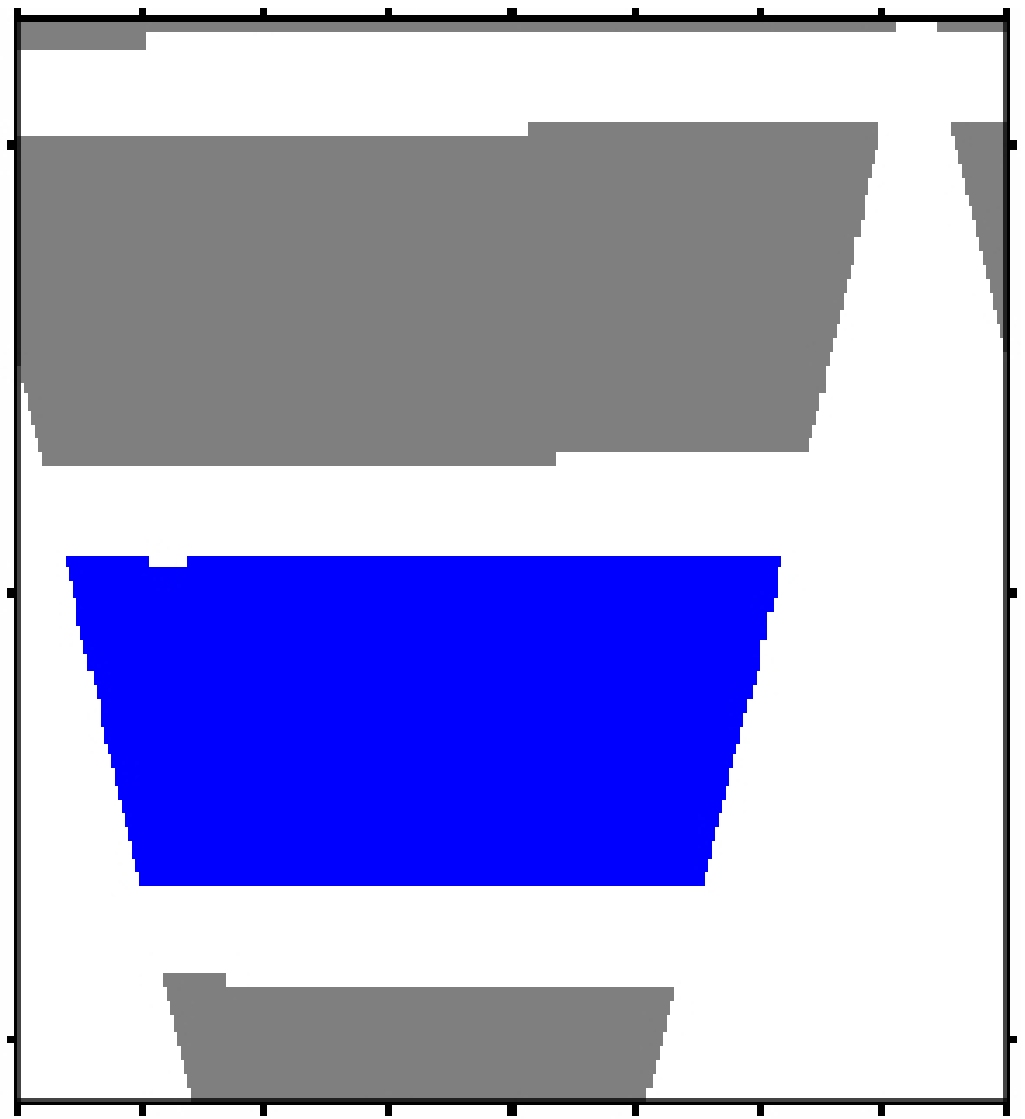}\\(e$_2$)\vspace{0mm}\end{minipage}
   \begin{minipage}[b]{0.45\linewidth}\centering\includegraphics[width=\linewidth]{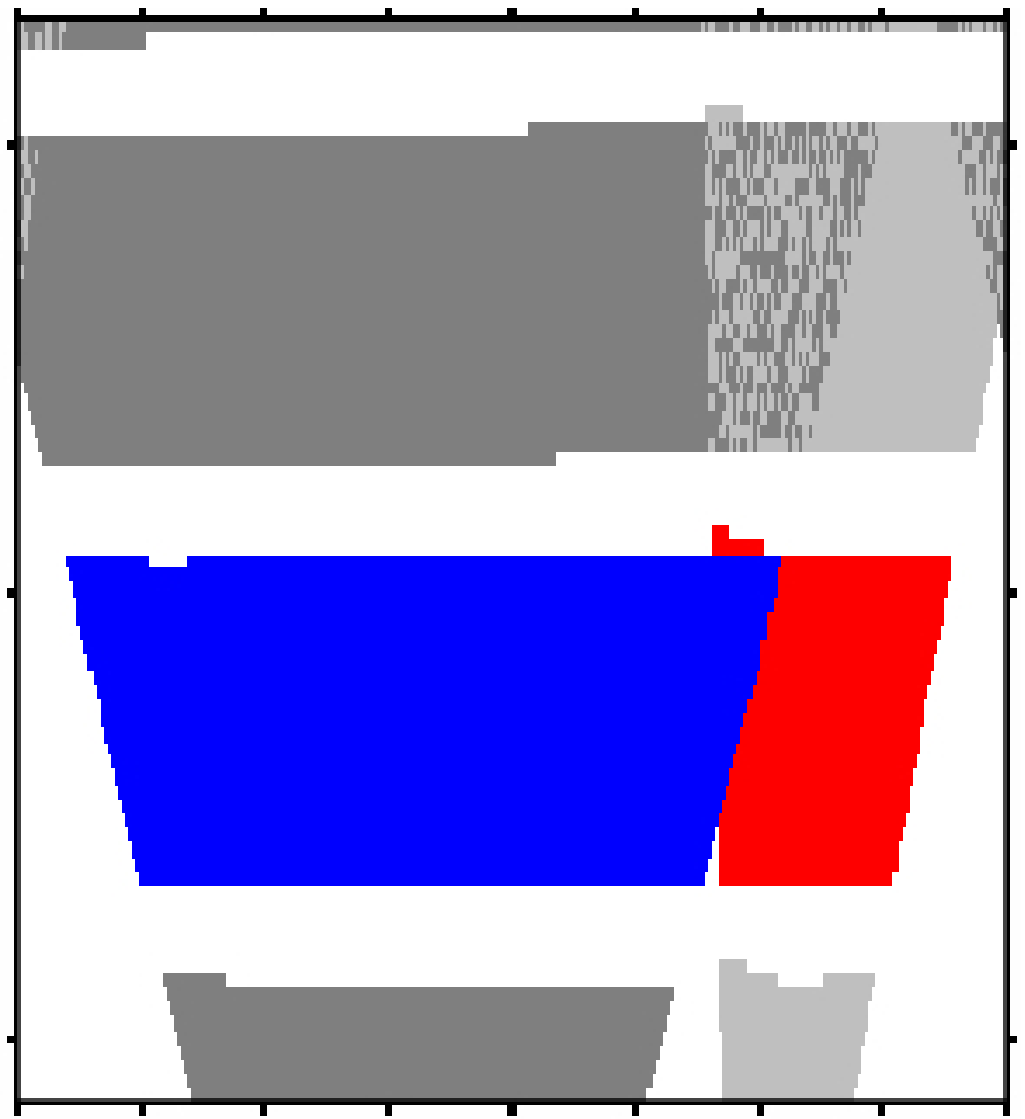}\\(f$_2$)\vspace{0mm}\end{minipage}
   \textbf{Full month runs}
 \end{minipage}   
  \vspace*{-2mm}
  \caption{Case study for joint observations of ASTEP and BEST\,II, based on actual observations of field Exo2 (a$_2$--d$_2$), Exo3 (a$_3$--d$_3$), and fully hypothetical time series~\mbox{(e--f)}. Axes and colors as in Figure~\ref{fig:obs:exo23}. For case descriptions, see the text.}
  \label{fig:detyield2:covsim}
\end{figure*}

In addition to the phase coverage of the actual time series, the prospects of joint observations are examined further in a case study; its aim is to suggest an optimized observing strategy for future campaigns. 
The cases studied are shown in Figure~\ref{fig:detyield2:covsim} and indicated with letters both in the figure and the text. They include four scenarios that are based on actual ASTEP observing dates of target field Exo2 (a$_2$--d$_2$), four on those of Exo3 (a$_3$--d$_3$), and four fully hypothetical scenarios (e$_1$, e$_2$, f$_1$, f$_2$). A brief description of all cases is given in the following list.
\begin{itemize}
	\item[\textbf{(a)}] First, they include the time series as obtained by ASTEP in the 2010 season for the fields Exo2 (a$_2$) and Exo3 (a$_3$), which are used as a minimum reference for comparison.
	\item[\textbf{(b)}] Second, joint observations of ASTEP and BEST\,II are reviewed.
	\item[\textbf{(c)}] Third, BEST\,II time series have been shifted back in time by 1, 2, $\ldots$ days in order to investigate how much could be gained from an improved scheduling of joint observations; the case with the largest value $f_t$ is shown for each field.
	\item[\textbf{(d)}] Fourth, the maximum increase that BEST\,II observations could possibly yield is evaluated by assuming an optimal duty cycle in Chile during nights with observations from Antarctica. 
	\item[\textbf{(e--f)}] Finally, ASTEP observations are also replaced by an optimal time series that could be obtained between two full Moons, i.e., about twice as long as the actual observations. Cases cover the months of July~\mbox{(e$_1$, f$_1$)} and August 2010 (e$_2$, f$_2$), whereby the two cases (e) only include observations from Antarctica, and (f) complement these with an optimal duty cycle from Chile.
\end{itemize}
The impact of these different scenarios on the planet detection yield is quantified in Table~\ref{tab:astep:detyield:covsim}. For each case, it gives the timing factor $f_t$ and the relative increase of the ASTEP detection yield due to additional BEST\,II observations.

The actual ASTEP observations of fields Exo2 and Exo3 alone yield timing factors $f_t$ of~0.044--0.046 (a). If BEST\,II data are added~(b), the factor~$f_t$ increases by~12--18\% to~0.052 for both fields. The slight offset between ASTEP and BEST\,II time series seems to have a negligible impact, as a shift would at most have increased $f_t$ by 0.00034~(c). However, if more observations were taken from Chile and scheduled optimally with ASTEP~(d), $f_t$ could be raised significantly up to~0.055--0.061, i.e., the detection yield could be increased by~26--33\% compared to ASTEP alone.

A relative increase of~12--19\% due to additional observations from Chile would even be encountered if ASTEP and BEST\,II observed the two fields continuously during one month~(cases~e and~f): the factor $f_t$ raises from 0.086 to~0.097 for July, and from~0.072 to~0.086 for August, respectively. As expected, the impact of complementary BEST\,II observations increases with the length of observing interruptions experienced during noon in Antarctica.

\section{Summary and Discussion}\label{sec:summary}
This study presents first joint observations from Antarctica and Chile: The known transiting planet WASP-18b and two target fields were monitored together with ASTEP and BEST\,II in 2010. For the two target fields, ASTEP measurements span 25 nights and include 94{,}965 stars, while BEST\,II obtained 224{,}552 light curves during 18 nights; joint observations are available for 58{,}822 stars. Their comparison aims at a first quantitative evaluation of the potential for transit search at Dome~C that is solely based on real photometric time series. Particular attention was paid to the photometric precision and observational phase coverage, which are both expected to yield advantageous conditions for transit searches in Antarctica. 

For a single transit of WASP-18b, ASTEP yields an unbinned, out-of-transit standard deviation of 1.9\,mmag. The data from Antarctica thus show a smaller noise level compared to 2.6\,mmag achieved with BEST\,II. However, the difference is not in the order of a factor~2--4 as expected from a smaller scintillation noise \citep{Kenyon2006b}, indicating that unfiltered stellar variability and/or systematic effects still contribute significantly to the noise budget of either light curve.
An analysis of the two large data sets from both telescopes showed that their photometric quality as well is excellent, reaching a precision of 2--4\,mmag for bright stars from both Antarctica and Chile over each observing campaign. However, the photometric precision is very similar: A simulation shows that BEST\,II and ASTEP overall yield a well comparable detection yield for a range of planetary radii.
An advantage to find significantly smaller planets from Antarctica is thus not evident from these first data. Whether the limiting noise component is an inherent site characteristic or can be further decreased will become more apparent while the experience with the ASTEP\,400 system, its unique environment, and the data reduction grows through continued operation. In this respect, it would also be most interesting to directly compare the photometric quality with another Antarctic site, e.g., with data from one of the Antarctic Survey Telescopes \citep[AST3;][]{Li2013} at Dome~A.

In contrast to the photometric quality, the long polar night yields a clear advantage for transit search in Antarctica. Within two weeks of observations, ASTEP yields a detection for planets at short periods that can only be achieved during a whole season from Chile. For the first time, light curves from Antarctica and Chile were combined in order to extend the observational duty cycle. A case-by-case comparison showed that the photometric systems of ASTEP and BEST\,II compare well and such a combination is practicable to the precision needed for transit search. If BEST\,II observations are added, the yield increases by 12--18\%. Furthermore, a case study has shown that a similar relative increase is even encountered if the duty cycle of ASTEP was extended further, and could be increased up to 26--33\% if BEST\,II observations could be obtained in parallel to each night with observations from Antarctica. 

Thus, such network observations can indeed increase the detection yield significantly compared to time series obtained from Antarctica alone. Note that a key advantage of the combination Chile-Dome~C is its large longitudinal separation (166$\degr$), and slightly less favorable conditions are expected from combining a Chilean site with another Antarctic observatory such as Dome~A~(148$\degr$) or Dome~F~(110$\degr$).

\acknowledgments
\textbf{Acknowledgments.} This work was funded by Deutsches Zentrum f\"ur Luft- und Raumfahrt and partly by the Nordrhein-Westf\"alische Akademie der Wissenschaften. The ASTEP project is supervised by the Observatoire de la C\^ote d'Azur, and installed at the Concordia station, Antarctica, with support from the Institut Paul \'Emile Victor. Our research made use of the UCAC3, USNO-B1.0, 2MASS (Two Micron All Sky Survey), and GSC2.2 catalogs, and the Besan\c{c}on model of the Galaxy.

\bibliographystyle{apj}
\bibliography{ms}

\begin{thebibliography}{55}
\expandafter\ifx\csname natexlab\endcsname\relax\def\natexlab#1{#1}\fi

\bibitem[{{Abe} {et~al.}(2013{\natexlab{a}}){Abe}, {Epchtein}, {Ansorge},
  {Argentini}, {Bryson}, {Carbillet}, {Dalton}, {David}, {Esau}, {Genthon},
  {Langlois}, {Le Bertre}, {Lemrani}, {Le Roux}, {Marchiori}, {M{\'e}karnia},
  {Montnacher}, {Moretto}, {Prugniel}, {Rivet}, {Ruch}, {Tao}, {Tilquin}, \&
  {Vauglin}}]{Abe2013a}
{Abe}, L., {Epchtein}, N., {Ansorge}, W., {et~al.} 2013{\natexlab{a}}, in IAU
  Symposium, Vol. 288, IAU Symposium, ed. M.~G. {Burton}, X.~{Cui}, \& N.~F.~H.
  {Tothill}, 243--250

\bibitem[{{Abe} {et~al.}(2013{\natexlab{b}}){Abe}, {Gon{\c c}alves}, {Agabi},
  {Alapini}, {Guillot}, {M{\'e}karnia}, {Rivet}, {Schmider}, {Crouzet},
  {Fortney}, {Pont}, {Barbieri}, {Daban}, {Fante{\"i}-Caujolle}, {Gouvret},
  {Bresson}, {Roussel}, {Bonhomme}, {Robini}, {Dugu{\'e}}, {Bondoux},
  {P{\'e}ron}, {Petit}, {Szul{\'a}gyi}, {Fruth}, {Erikson}, {Rauer}, {Fressin},
  {Valbousquet}, {Blanc}, {Le van Suu}, \& {Aigrain}}]{Abe2013}
{Abe}, L., {Gon{\c c}alves}, I., {Agabi}, A., {et~al.} 2013{\natexlab{b}},
  \aap, 553, A49

\bibitem[{{Alard}(2000)}]{Alard2000}
{Alard}, C. 2000, \aaps, 144, 363

\bibitem[{{Alard} \& {Lupton}(1998)}]{Alard1998}
{Alard}, C., \& {Lupton}, R.~H. 1998, \apj, 503, 325

\bibitem[{{Baglin} {et~al.}(2006){Baglin}, {Auvergne}, {Barge}, {Deleuil},
  {Catala}, {Michel}, {Weiss}, \& {COROT Team}}]{Baglin2006}
{Baglin}, A., {Auvergne}, M., {Barge}, P., {et~al.} 2006, in {The COROT Mission
  Pre-Launch Status}, ed. M.~{Fridlund}, A.~{Baglin}, J.~{Lochard}, \&
  L.~{Conroy}, ESA SP-1306, 33--37

\bibitem[{{Barnes}(2007)}]{Barnes2007}
{Barnes}, J.~W. 2007, \pasp, 119, 986

\bibitem[{{Bayliss} \& {Sackett}(2011)}]{Bayliss2011}
{Bayliss}, D.~D.~R., \& {Sackett}, P.~D. 2011, \apj, 743, 103

\bibitem[{{Beatty} \& {Gaudi}(2008)}]{Beatty2008}
{Beatty}, T.~G., \& {Gaudi}, B.~S. 2008, \apj, 686, 1302

\bibitem[{{Burton}(2010)}]{Burton2010}
{Burton}, M.~G. 2010, \aapr, 18, 417

\bibitem[{{Burton} {et~al.}(2005){Burton}, {Lawrence}, {Ashley}, {Bailey},
  {Blake}, {Bedding}, {Bland-Hawthorn}, {Bond}, {Glazebrook}, {Hidas}, {Lewis},
  {Longmore}, {Maddison}, {Mattila}, {Minier}, {Ryder}, {Sharp}, {Smith},
  {Storey}, {Tinney}, {Tuthill}, {Walsh}, {Walsh}, {Whiting}, {Wong}, {Woods},
  \& {Yock}}]{Burton2005}
{Burton}, M.~G., {Lawrence}, J.~S., {Ashley}, M.~C.~B., {et~al.} 2005, \pasa,
  22, 199

\bibitem[{{Caldwell} {et~al.}(2004){Caldwell}, {Borucki}, {Showen}, {Jenkins},
  {Doyle}, {Ninkov}, \& {Ashley}}]{Caldwell2004}
{Caldwell}, D.~A., {Borucki}, W.~J., {Showen}, R.~L., {et~al.} 2004, in
  {Proceedings of the International Astronomical Union: Bioastronomy 2002 --
  Life Among the Stars (IAU S213)}, ed. R.~{Norris} \& F.~{Stootman}
  (Astronomical Society of the Pacific), 93--96

\bibitem[{Cox(2000)}]{Cox2000}
Cox, A. 2000, {Allen's Astrophysical Quantities} (Springer)

\bibitem[{{Crouzet} {et~al.}(2010){Crouzet}, {Guillot}, {Agabi}, {Rivet},
  {Bondoux}, {Challita}, {Fante{\"i}-Caujolle}, {Fressin}, {M{\'e}karnia},
  {Schmider}, {Valbousquet}, {Blazit}, {Bonhomme}, {Abe}, {Daban}, {Gouvret},
  {Fruth}, {Rauer}, {Erikson}, {Barbieri}, {Aigrain}, \& {Pont}}]{Crouzet2010}
{Crouzet}, N., {Guillot}, T., {Agabi}, A., {et~al.} 2010, \aap, 511, A36

\bibitem[{{Crouzet} {et~al.}(2011){Crouzet}, {Guillot}, {Agabi}, {Daban},
  {Abe}, {Mekarnia}, {Rivet}, {Fante{\"i}-Caujolle}, {Fressin}, {Gouvret},
  {Schmider}, {Valbousquet}, {Blazit}, {Rauer}, {Erikson}, {Fruth}, {Aigrain},
  {Pont}, \& {Barbieri}}]{Crouzet2011}
{Crouzet}, N., {Guillot}, T., {Agabi}, K., {et~al.} 2011, in {Detection and
  Dynamics of Transiting Exoplanets}, ed. F.~{Bouchy}, R.~{D{\'{\i}}az}, \&
  C.~{Moutou}, EPJ Web of Conferences 11, id.06001

\bibitem[{{Csizmadia} {et~al.}(2011){Csizmadia}, {Moutou}, {Deleuil},
  {Cabrera}, {Fridlund}, {Gandolfi}, {Aigrain}, {Alonso}, {Almenara},
  {Auvergne}, {Baglin}, {Barge}, {Bonomo}, {Bord{\'e}}, {Bouchy}, {Bruntt},
  {Carone}, {Carpano}, {Cavarroc}, {Cochran}, {Deeg}, {D{\'{\i}}az}, {Dvorak},
  {Endl}, {Erikson}, {Ferraz-Mello}, {Fruth}, {Gazzano}, {Gillon}, {Guenther},
  {Guillot}, {Hatzes}, {Havel}, {H{\'e}brard}, {Jehin}, {Jorda}, {L{\'e}ger},
  {Llebaria}, {Lammer}, {Lovis}, {MacQueen}, {Mazeh}, {Ollivier},
  {P{\"a}tzold}, {Queloz}, {Rauer}, {Rouan}, {Santerne}, {Schneider},
  {Tingley}, {Titz-Weider}, \& {Wuchterl}}]{Csizmadia2011}
{Csizmadia}, {\relax Sz}., {Moutou}, C., {Deleuil}, M., {et~al.} 2011, \aap,
  531, A41

\bibitem[{{Cui}(2010)}]{Cui2010}
{Cui}, X. 2010, Highlights of Astronomy, 15, 639

\bibitem[{{Cumming} {et~al.}(2008){Cumming}, {Butler}, {Marcy}, {Vogt},
  {Wright}, \& {Fischer}}]{Cumming2008}
{Cumming}, A., {Butler}, R.~P., {Marcy}, G.~W., {et~al.} 2008, \pasp, 120, 531

\bibitem[{{Daban} {et~al.}(2010){Daban}, {Gouvret}, {Guillot}, {Agabi},
  {Crouzet}, {Rivet}, {Mekarnia}, {Abe}, {Bondoux}, {Fante{\"i}-Caujolle},
  {Fressin}, {Schmider}, {Valbousquet}, {Blanc}, {Le van Suu}, {Rauer},
  {Erikson}, {Pont}, \& {Aigrain}}]{Daban2010}
{Daban}, J.-B., {Gouvret}, C., {Guillot}, T., {et~al.} 2010, in {Ground-based
  and Airborne Telescopes III}, ed. L.~M. {Stepp}, R.~{Gilmozzi}, \& H.~J.
  {Hall}, Proc.~SPIE 7733, 77334T(1--9)

\bibitem[{{Deeg} {et~al.}(2009){Deeg}, {Gillon}, {Shporer}, {Rouan},
  {Stecklum}, {Aigrain}, {Alapini}, {Almenara}, {Alonso}, {Barbieri}, {Bouchy},
  {Eisl{\"o}ffel}, {Erikson}, {Fridlund}, {Eigm{\"u}ller}, {Handler}, {Hatzes},
  {Kabath}, {Lendl}, {Mazeh}, {Moutou}, {Queloz}, {Rauer}, {Rabus}, {Tingley},
  \& {Titz}}]{Deeg2009}
{Deeg}, H.~J., {Gillon}, M., {Shporer}, A., {et~al.} 2009, \aap, 506, 343

\bibitem[{{Fossat} {et~al.}(2010){Fossat}, {Aristidi}, {Agabi}, {Bondoux},
  {Challita}, {Jeanneaux}, \& {M{\'e}karnia}}]{Fossat2010}
{Fossat}, E., {Aristidi}, E., {Agabi}, A., {et~al.} 2010, \aap, 517, A69

\bibitem[{{Fressin} {et~al.}(2006){Fressin}, {Guillot}, {Schmider}, {Agabi},
  {Moutou}, {Bouchy}, {Boer}, {Pont}, {Erikson}, {Rauer}, \& {The Astep
  Team}}]{Fressin2006}
{Fressin}, F., {Guillot}, T., {Schmider}, F.~X., {et~al.} 2006, in {The COROT
  Mission Pre-Launch Status}, ed. M.~{Fridlund}, A.~{Baglin}, J.~{Lochard}, \&
  L.~{Conroy}, ESA SP-1306, 513--516

\bibitem[{{Fruth} {et~al.}(2012){Fruth}, {Kabath}, {Cabrera}, {Chini},
  {Csizmadia}, {Eigm\"uller}, {Erikson}, {Kirste}, {Lemke}, {Murphy},
  {Pasternacki}, {Rauer}, \& {Titz-Weider}}]{Fruth2012}
{Fruth}, T., {Kabath}, P., {Cabrera}, J., {et~al.} 2012, \aj, 143, 140

\bibitem[{{Fruth} {et~al.}(2013){Fruth}, {Kabath}, {Cabrera}, {Chini},
  {Csizmadia}, {Eigm\"uller}, {Erikson}, {Kirste}, {Lemke}, {Murphy},
  {Pasternacki}, {Rauer}, \& {Titz-Weider}}]{Fruth2013}
---. 2013, \aj, 146, 136

\bibitem[{{Guenther} {et~al.}(2012){Guenther}, {Gandolfi}, {Sebastian},
  {Deleuil}, {Moutou}, \& {Cusano}}]{Guenther2012a}
{Guenther}, E.~W., {Gandolfi}, D., {Sebastian}, D., {et~al.} 2012, \aap, 543,
  A125

\bibitem[{{Hellier} {et~al.}(2009){Hellier}, {Anderson}, {Cameron}, {Gillon},
  {Hebb}, {Maxted}, {Queloz}, {Smalley}, {Triaud}, {West}, {Wilson}, {Bentley},
  {Enoch}, {Horne}, {Irwin}, {Lister}, {Mayor}, {Parley}, {Pepe}, {Pollacco},
  {Segransan}, {Udry}, \& {Wheatley}}]{Hellier2009}
{Hellier}, C., {Anderson}, D.~R., {Cameron}, A.~C., {et~al.} 2009, \nat, 460,
  1098

\bibitem[{{Howard} {et~al.}(2012){Howard}, {Marcy}, {Bryson}, {Jenkins},
  {Rowe}, {Batalha}, {Borucki}, {Koch}, {Dunham}, {Gautier}, {Van Cleve},
  {Cochran}, {Latham}, {Lissauer}, {Torres}, {Brown}, {Gilliland}, {Buchhave},
  {Caldwell}, {Christensen-Dalsgaard}, {Ciardi}, {Fressin}, {Haas}, {Howell},
  {Kjeldsen}, {Seager}, {Rogers}, {Sasselov}, {Steffen}, {Basri},
  {Charbonneau}, {Christiansen}, {Clarke}, {Dupree}, {Fabrycky}, {Fischer},
  {Ford}, {Fortney}, {Tarter}, {Girouard}, {Holman}, {Johnson}, {Klaus},
  {Machalek}, {Moorhead}, {Morehead}, {Ragozzine}, {Tenenbaum}, {Twicken},
  {Quinn}, {Isaacson}, {Shporer}, {Lucas}, {Walkowicz}, {Welsh}, {Boss},
  {Devore}, {Gould}, {Smith}, {Morris}, {Prsa}, {Morton}, {Still}, {Thompson},
  {Mullally}, {Endl}, \& {MacQueen}}]{Howard2012}
{Howard}, A.~W., {Marcy}, G.~W., {Bryson}, S.~T., {et~al.} 2012, \apjs, 201, 15

\bibitem[{{Ichikawa}(2010)}]{Ichikawa2010}
{Ichikawa}, T. 2010, Highlights of Astronomy, 15, 632

\bibitem[{{Indermuehle} {et~al.}(2005){Indermuehle}, {Burton}, \&
  {Maddison}}]{Indermuehle2005}
{Indermuehle}, B.~T., {Burton}, M.~G., \& {Maddison}, S.~T. 2005, \pasa, 22, 73

\bibitem[{{Kabath} {et~al.}(2007){Kabath}, {Eigm{\"u}ller}, {Erikson},
  {Hedelt}, {Rauer}, {Titz}, {Wiese}, \& {Karoff}}]{Kabath2007}
{Kabath}, P., {Eigm{\"u}ller}, P., {Erikson}, A., {et~al.} 2007, \aj, 134, 1560

\bibitem[{{Kabath} {et~al.}(2008){Kabath}, {Eigm{\"u}ller}, {Erikson},
  {Hedelt}, {von Paris}, {Rauer}, {Renner}, {Titz}, \& {Karoff}}]{Kabath2008}
---. 2008, \aj, 136, 654

\bibitem[{{Kabath} {et~al.}(2009{\natexlab{a}}){Kabath}, {Fruth}, {Rauer},
  {Erikson}, {Murphy}, {Chini}, {Lemke}, {Csizmadia}, {Eigm{\"u}ller},
  {Pasternacki}, \& {Titz}}]{Kabath2009}
{Kabath}, P., {Fruth}, T., {Rauer}, H., {et~al.} 2009{\natexlab{a}}, \aj, 137,
  3911

\bibitem[{{Kabath} {et~al.}(2009{\natexlab{b}}){Kabath}, {Erikson}, {Rauer},
  {Pasternacki}, {Csizmadia}, {Chini}, {Lemke}, {Murphy}, {Fruth}, {Titz}, \&
  {Eigm{\"u}ller}}]{Kabath2009a}
{Kabath}, P., {Erikson}, A., {Rauer}, H., {et~al.} 2009{\natexlab{b}}, \aap,
  506, 569

\bibitem[{{Karoff} {et~al.}(2007){Karoff}, {Rauer}, {Erikson}, {Voss},
  {Kabath}, {Wiese}, {Deleuil}, {Moutou}, {Meunier}, \& {Deeg}}]{Karoff2007}
{Karoff}, C., {Rauer}, H., {Erikson}, A., {et~al.} 2007, \aj, 134, 766

\bibitem[{{Kenyon} {et~al.}(2006){Kenyon}, {Lawrence}, {Ashley}, {Storey},
  {Tokovinin}, \& {Fossat}}]{Kenyon2006b}
{Kenyon}, S.~L., {Lawrence}, J.~S., {Ashley}, M.~C.~B., {et~al.} 2006, \pasp,
  118, 924

\bibitem[{{Kenyon} \& {Storey}(2006)}]{Kenyon2006a}
{Kenyon}, S.~L., \& {Storey}, J.~W.~V. 2006, \pasp, 118, 489

\bibitem[{{Klagyivik} {et~al.}(2013){Klagyivik}, {Csizmadia}, {Pasternacki},
  {Fruth}, {Erikson}, {Cabrera}, {Chini}, {Eigm{\"u}ller}, {Kabath}, {Kirste},
  {Lemke}, {Murphy}, {Rauer}, \& {Titz-Weider}}]{Klagyivik2013}
{Klagyivik}, P., {Csizmadia}, {\relax Sz}., {Pasternacki}, T., {et~al.} 2013,
  \apj, 773, 54

\bibitem[{{Kov{\'a}cs} {et~al.}(2002){Kov{\'a}cs}, {Zucker}, \&
  {Mazeh}}]{Kovacs2002}
{Kov{\'a}cs}, G., {Zucker}, S., \& {Mazeh}, T. 2002, \aap, 391, 369

\bibitem[{{Lasker} {et~al.}(2008){Lasker}, {Lattanzi}, {McLean}, {Bucciarelli},
  {Drimmel}, {Garcia}, {Greene}, {Guglielmetti}, {Hanley}, {Hawkins},
  {Laidler}, {Loomis}, {Meakes}, {Mignani}, {Morbidelli}, {Morrison},
  {Pannunzio}, {Rosenberg}, {Sarasso}, {Smart}, {Spagna}, {Sturch},
  {Volpicelli}, {White}, {Wolfe}, \& {Zacchei}}]{Lasker2008}
{Lasker}, B.~M., {Lattanzi}, M.~G., {McLean}, B.~J., {et~al.} 2008, \aj, 136,
  735

\bibitem[{{Li} \& {Wang}(2013)}]{Li2013}
{Li}, X., \& {Wang}, D. 2013, in IAU Symposium, Vol. 288, IAU Symposium, ed.
  M.~G. {Burton}, X.~{Cui}, \& N.~F.~H. {Tothill}, 329--330

\bibitem[{{Monet} {et~al.}(2003){Monet}, {Levine}, {Canzian}, {Ables}, {Bird},
  {Dahn}, {Guetter}, {Harris}, {Henden}, {Leggett}, {Levison}, {Luginbuhl},
  {Martini}, {Monet}, {Munn}, {Pier}, {Rhodes}, {Riepe}, {Sell}, {Stone},
  {Vrba}, {Walker}, {Westerhout}, {Brucato}, {Reid}, {Schoening}, {Hartley},
  {Read}, \& {Tritton}}]{Monet2003}
{Monet}, D.~G., {Levine}, S.~E., {Canzian}, B., {et~al.} 2003, \aj, 125, 984

\bibitem[{{Newberry}(1991)}]{Newberry1991}
{Newberry}, M.~V. 1991, \pasp, 103, 122

\bibitem[{{P{\'a}l} \& {Bakos}(2006)}]{Pal2006}
{P{\'a}l}, A., \& {Bakos}, G.~{\'A}. 2006, \pasp, 118, 1474

\bibitem[{{Pasternacki} {et~al.}(2011){Pasternacki}, {Csizmadia}, {Cabrera},
  {Eigm\"uller}, {Erikson}, {Fruth}, {von Paris}, {Rauer}, {Titz},
  {Eisl\"offel}, {Hatzes}, {Boer}, {Tournois}, {Kabath}, {Hedelt}, \&
  {Voss}}]{Pasternacki2011}
{Pasternacki}, T., {Csizmadia}, {\relax Sz}., {Cabrera}, J., {et~al.} 2011,
  \aj, 142, 114

\bibitem[{{Pont} \& {Bouchy}(2005)}]{Pont2005a}
{Pont}, F., \& {Bouchy}, F. 2005, in EAS Publications Series, Vol.~14, {Dome C
  Astronomy and Astrophysics Meeting}, ed. M.~{Giard}, F.~{Casoli}, \&
  F.~{Paletou}, 155--160

\bibitem[{{Rauer} \& {Deeg}(2010)}]{Rauer2010a}
{Rauer}, H., \& {Deeg}, H. 2010, in EAS Publications Series, Vol.~40, {3rd
  ARENA Conference: An Astronomical Observatory at Concordia (Dome C,
  Antarctica)}, ed. L.~{Spinoglio} \& N.~{Epchtein}, 349--360

\bibitem[{{Rauer} {et~al.}(2008){Rauer}, {Fruth}, \& {Erikson}}]{Rauer2008}
{Rauer}, H., {Fruth}, T., \& {Erikson}, A. 2008, \pasp, 120, 852

\bibitem[{{Rauer} {et~al.}(2010){Rauer}, {Erikson}, {Kabath}, {Hedelt}, {Boer},
  {Carone}, {Csizmadia}, {Eigm{\"u}ller}, {Paris}, {Renner}, {Tournois},
  {Titz}, \& {Voss}}]{Rauer2010}
{Rauer}, H., {Erikson}, A., {Kabath}, P., {et~al.} 2010, \aj, 139, 53

\bibitem[{{Reyl{\'e}} {et~al.}(2010){Reyl{\'e}}, {Robin}, {Schultheis}, \&
  {Marshall}}]{Reyle2010}
{Reyl{\'e}}, C., {Robin}, A.~C., {Schultheis}, M., \& {Marshall}, D.~J. 2010,
  in SF2A-2010: Proceedings of the Annual meeting of the French Society of
  Astronomy and Astrophysics, ed. S.~{Boissier}, M.~{Heydari-Malayeri},
  R.~{Samadi}, \& D.~{Valls-Gabaud}, 51--54

\bibitem[{{Robin} {et~al.}(2003){Robin}, {Reyl{\'e}}, {Derri{\`e}re}, \&
  {Picaud}}]{Robin2003}
{Robin}, A.~C., {Reyl{\'e}}, C., {Derri{\`e}re}, S., \& {Picaud}, S. 2003,
  \aap, 409, 523

\bibitem[{{Saunders} {et~al.}(2009){Saunders}, {Lawrence}, {Storey}, {Ashley},
  {Kato}, {Minnis}, {Winker}, {Liu}, \& {Kulesa}}]{Saunders2009}
{Saunders}, W., {Lawrence}, J.~S., {Storey}, J.~W.~V., {et~al.} 2009, \pasp,
  121, 976

\bibitem[{{Skrutskie} {et~al.}(2006){Skrutskie}, {Cutri}, {Stiening},
  {Weinberg}, {Schneider}, {Carpenter}, {Beichman}, {Capps}, {Chester},
  {Elias}, {Huchra}, {Liebert}, {Lonsdale}, {Monet}, {Price}, {Seitzer},
  {Jarrett}, {Kirkpatrick}, {Gizis}, {Howard}, {Evans}, {Fowler}, {Fullmer},
  {Hurt}, {Light}, {Kopan}, {Marsh}, {McCallon}, {Tam}, {Van Dyk}, \&
  {Wheelock}}]{Skrutskie2006}
{Skrutskie}, M.~F., {Cutri}, R.~M., {Stiening}, R., {et~al.} 2006, \aj, 131,
  1163

\bibitem[{{Southworth} {et~al.}(2009){Southworth}, {Hinse}, {Dominik},
  {Glitrup}, {J{\o}rgensen}, {Liebig}, {Mathiasen}, {Anderson}, {Bozza},
  {Browne}, {Burgdorf}, {Calchi Novati}, {Dreizler}, {Finet}, {Harps{\o}e},
  {Hessman}, {Hundertmark}, {Maier}, {Mancini}, {Maxted}, {Rahvar}, {Ricci},
  {Scarpetta}, {Skottfelt}, {Snodgrass}, {Surdej}, \&
  {Zimmer}}]{Southworth2009}
{Southworth}, J., {Hinse}, T.~C., {Dominik}, M., {et~al.} 2009, \apj, 707, 167

\bibitem[{{Tamuz} {et~al.}(2005){Tamuz}, {Mazeh}, \& {Zucker}}]{Tamuz2005}
{Tamuz}, O., {Mazeh}, T., \& {Zucker}, S. 2005, \mnras, 356, 1466

\bibitem[{{Young}(1967)}]{Young1967}
{Young}, A.~T. 1967, \aj, 72, 747

\bibitem[{{Zacharias} {et~al.}(2010){Zacharias}, {Finch}, {Girard}, {Hambly},
  {Wycoff}, {Zacharias}, {Castillo}, {Corbin}, {DiVittorio}, {Dutta}, {Gaume},
  {Gauss}, {Germain}, {Hall}, {Hartkopf}, {Hsu}, {Holdenried}, {Makarov},
  {Martinez}, {Mason}, {Monet}, {Rafferty}, {Rhodes}, {Siemers}, {Smith},
  {Tilleman}, {Urban}, {Wieder}, {Winter}, \& {Young}}]{Zacharias2010}
{Zacharias}, N., {Finch}, C., {Girard}, T., {et~al.} 2010, \aj, 139, 2184

\end{thebibliography}

\appendix

\section{BEST\,II/ASTEP Case-by-Case Comparison}\label{sec:app:datacombi}

\subsection{Photometric Systems}\label{sec:app:photsys}
\begin{figure*}[htpc]\centering\hspace{-8mm}
  \includegraphics[width=.35\linewidth]{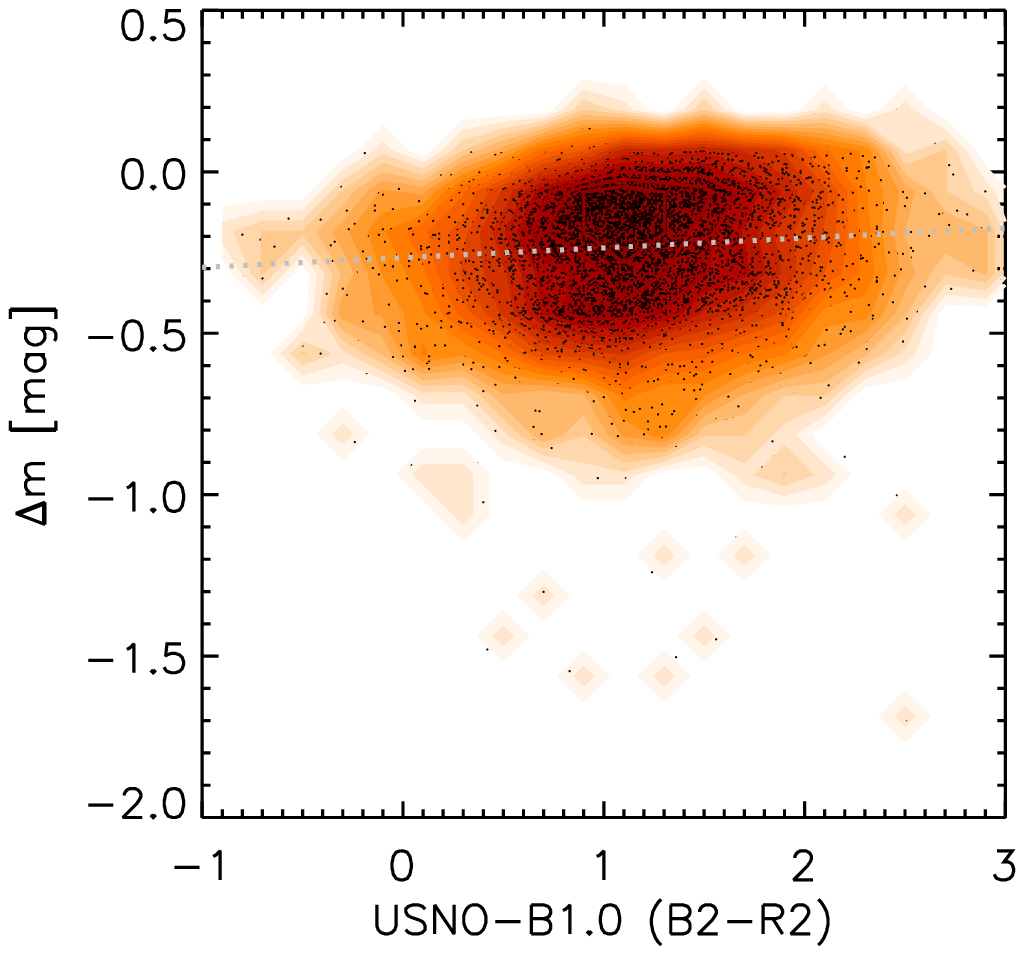}\hspace{8mm}
  \includegraphics[width=.35\linewidth]{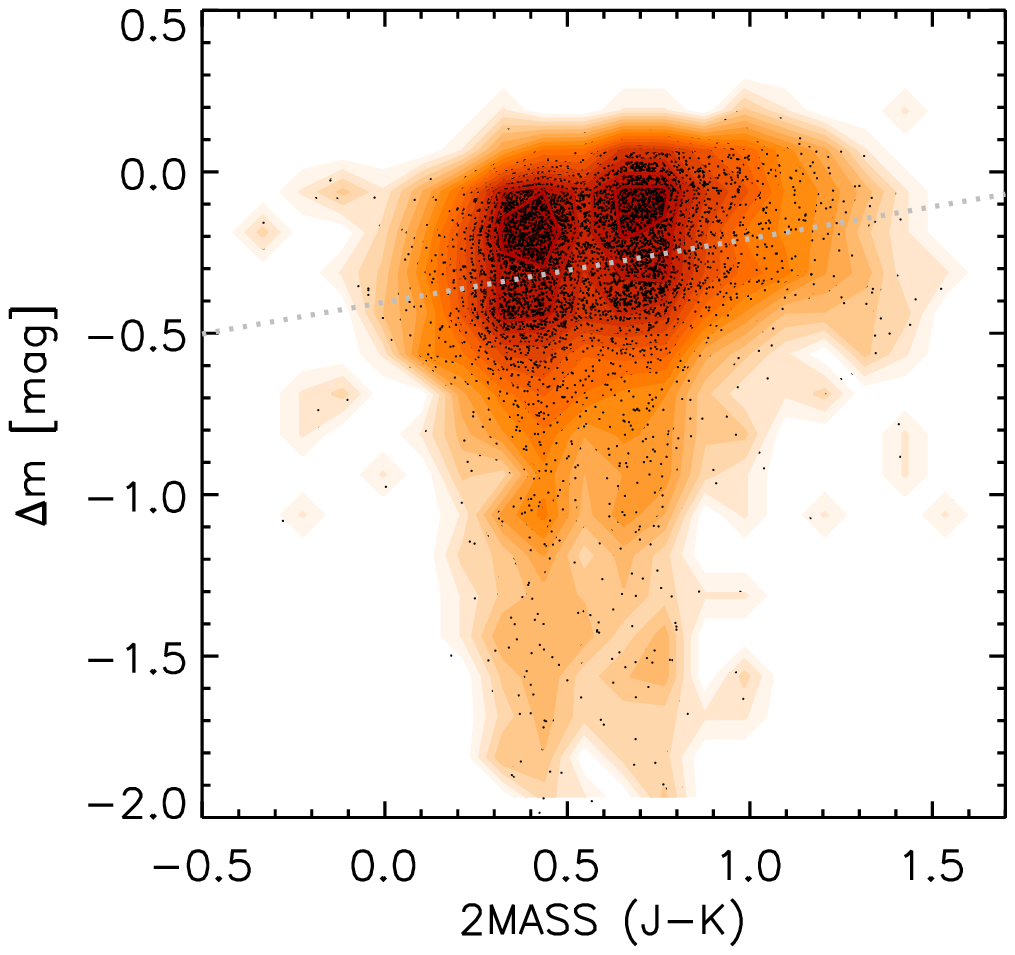}\hspace{-12mm}
  \vspace*{-2mm}
  \caption{Magnitude difference $\Delta m$ between BEST\,II and ASTEP (Equation~(\ref{eq:Deltam})) in field Exo3 as a function of stellar colors. The left plot uses $(B-R)$ from the USNO-B1.0 catalog \citep{Monet2003}, the right plot $(J-K)$ from 2MASS \citep{Skrutskie2006}. Black dots represent individual stars; the background color indicates their number density. A linear fit based on Equation~(\ref{eq:app:colorfit}) is shown by the white dotted line. (Only stars with $\overline{m}_i^A\leq 15$\,mag are shown and used for the fit, since fainter stars exhibit large color uncertainties.)}
  \label{fig:app:best2astep:color-magdiff}
\end{figure*}

In order to evaluate differences in both photometric systems quantitatively, catalog colors have been obtained for matched stars: $(B-R)$ from USNO-B1.0 \citep{Monet2003} is used in the blue, and $(J-K)$ from 2MASS \citep{Skrutskie2006} in the red part of the spectrum. For each star, they are compared with~$\Delta m$ (Equation~(\ref{eq:Deltam})), which is expected to increase as a function of both. Figure~\ref{fig:app:best2astep:color-magdiff} shows the analysis for the stars in target field Exo3. A linear fit to the data yields
\begin{equation}\label{eq:app:colorfit}
\begin{array}{rcl}
  \Delta m \left[(B-R)\right] &=& \left(-0.266 \pm 0.006 \right)\,\mbox{mag} + (B-R) \cdot \left( 0.030 \pm 0.005\right) \\
  \Delta m \left[(J-K)\right] &=& \left(-0.404 \pm 0.011 \right)\,\mbox{mag} + (J-K) \cdot \left( 0.20 \pm 0.02\right) \ .
\end{array}
\end{equation}
The smaller number of matched stars in Exo2 show a very similar relation. 

\subsection{Magnitudes and Photometric Precision}\label{sec:app:datacombi:phot}
\begin{figure*}[htc]\centering
  \subfigure[Exo2]{\begin{minipage}[c]{\linewidth}\vspace{0pt}
   \includegraphics[width=.31\linewidth]{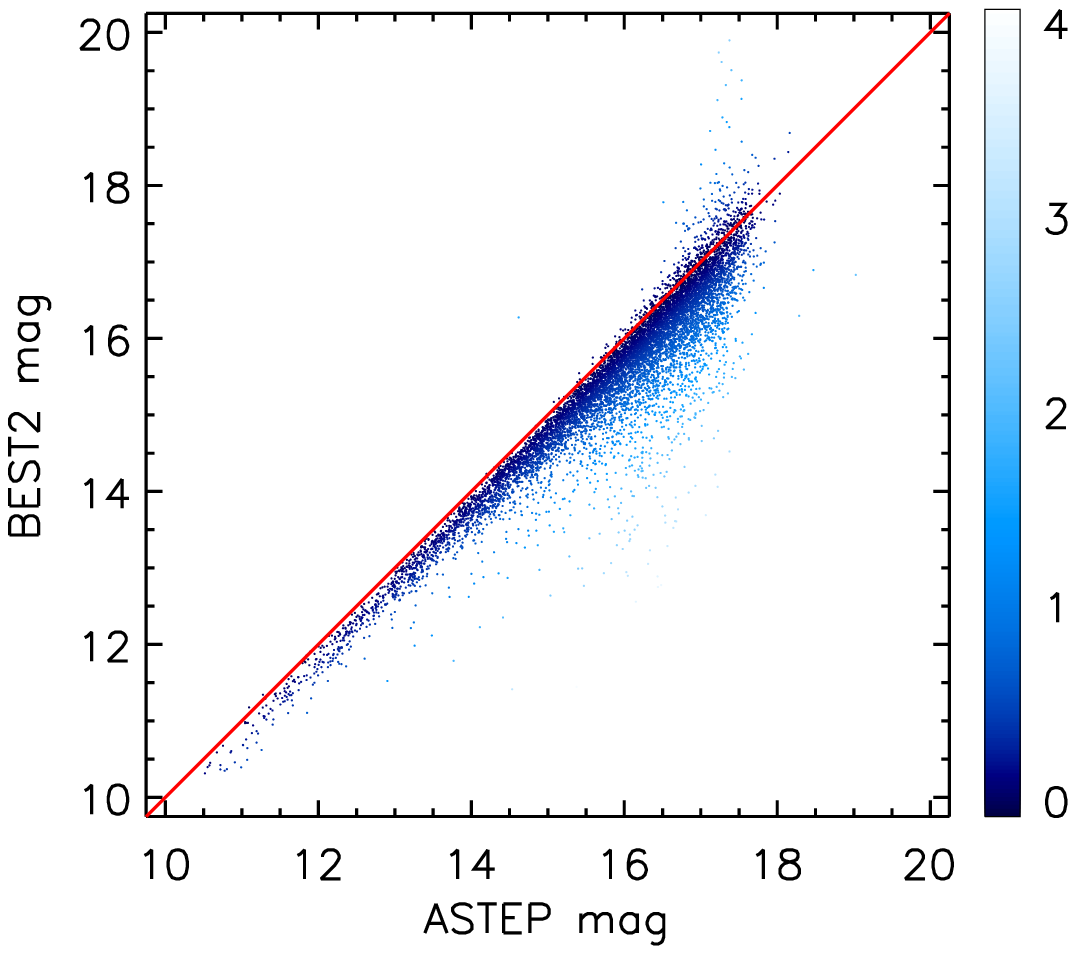}
   \includegraphics[width=.31\linewidth]{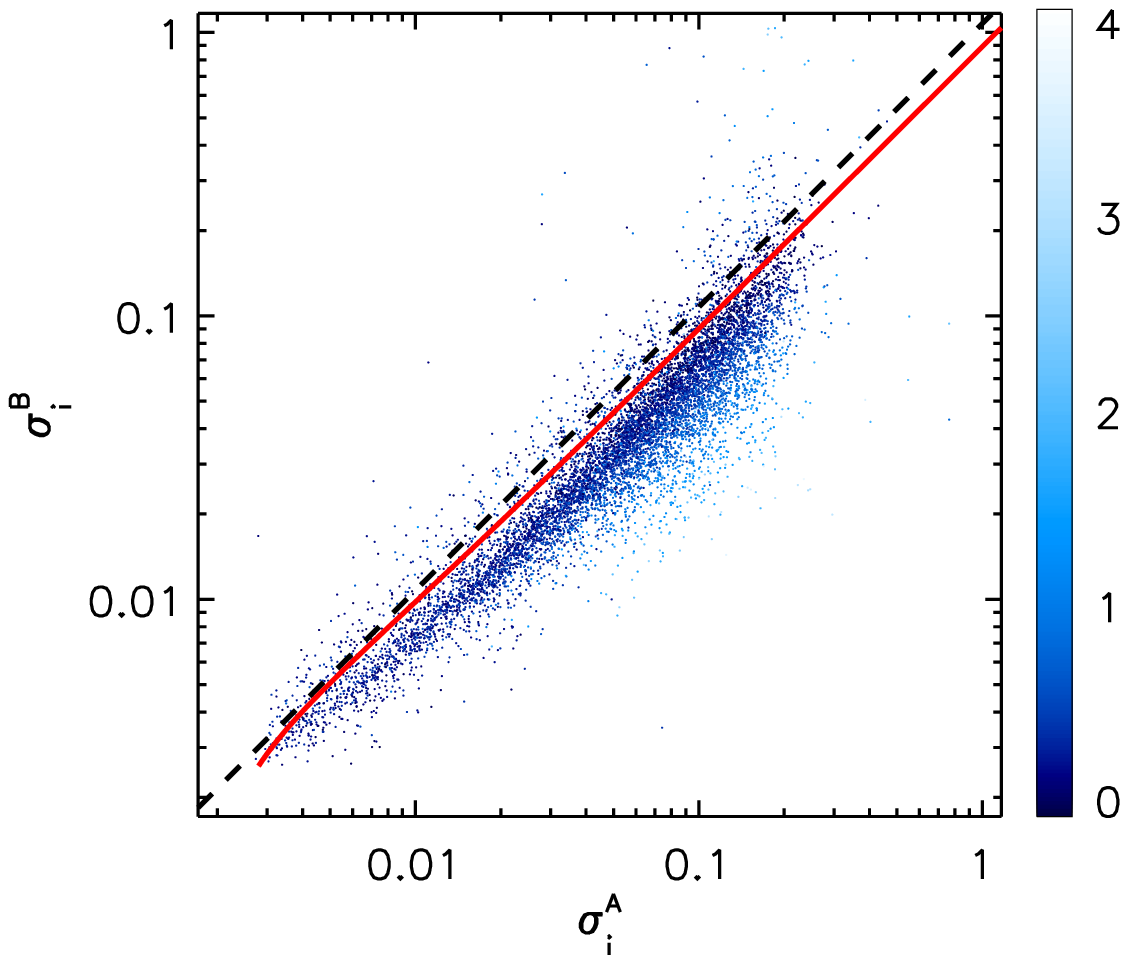}
   \includegraphics[width=.31\linewidth]{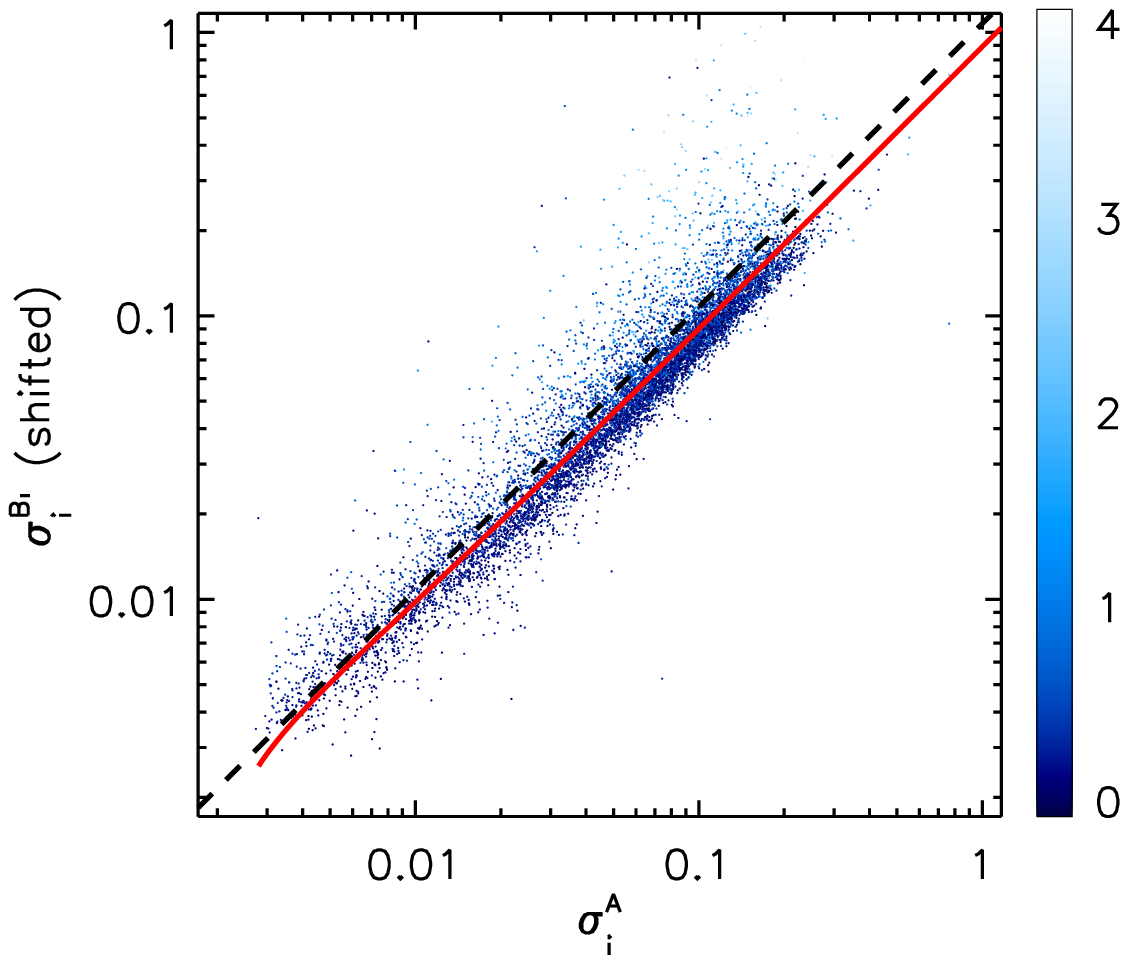}   
  \end{minipage}}
  \subfigure[Exo3]{\begin{minipage}[c]{\linewidth}\vspace{0pt}
   \includegraphics[width=.31\linewidth]{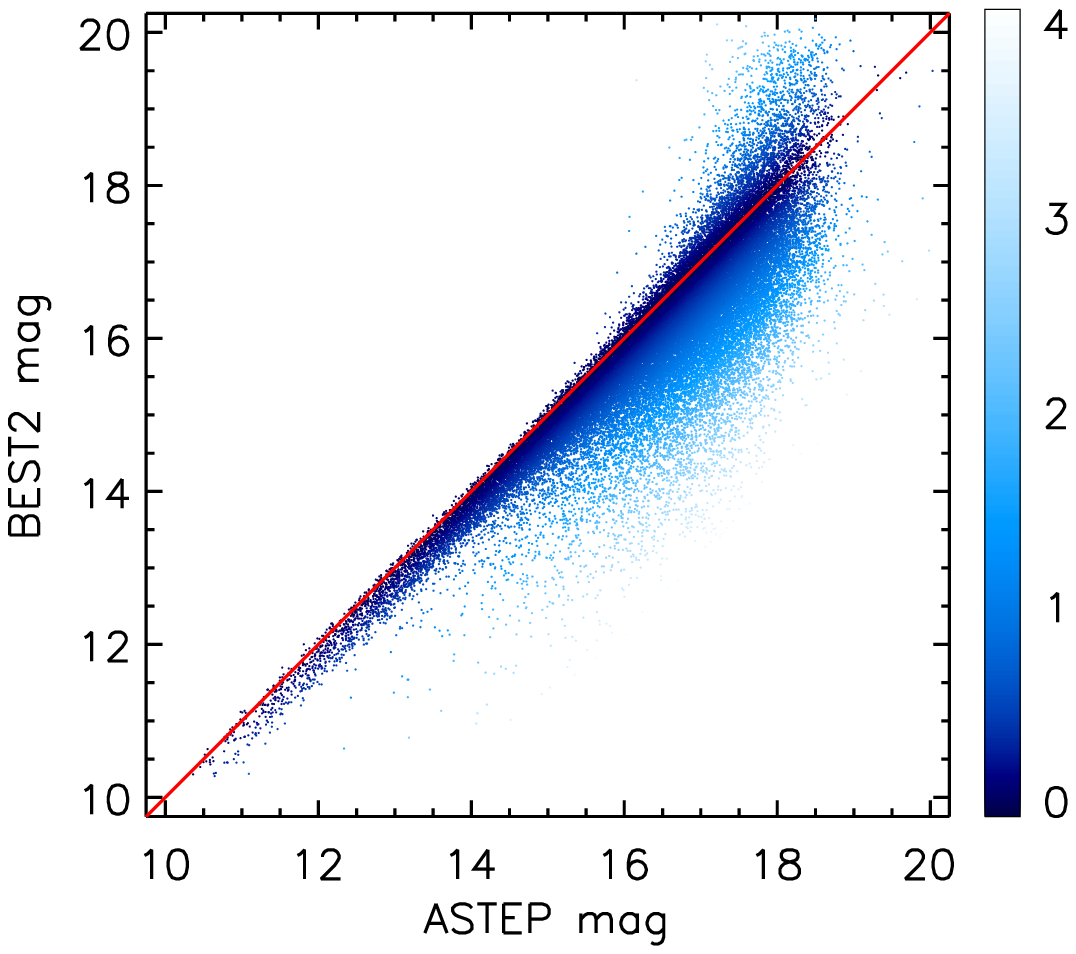}  
   \includegraphics[width=.31\linewidth]{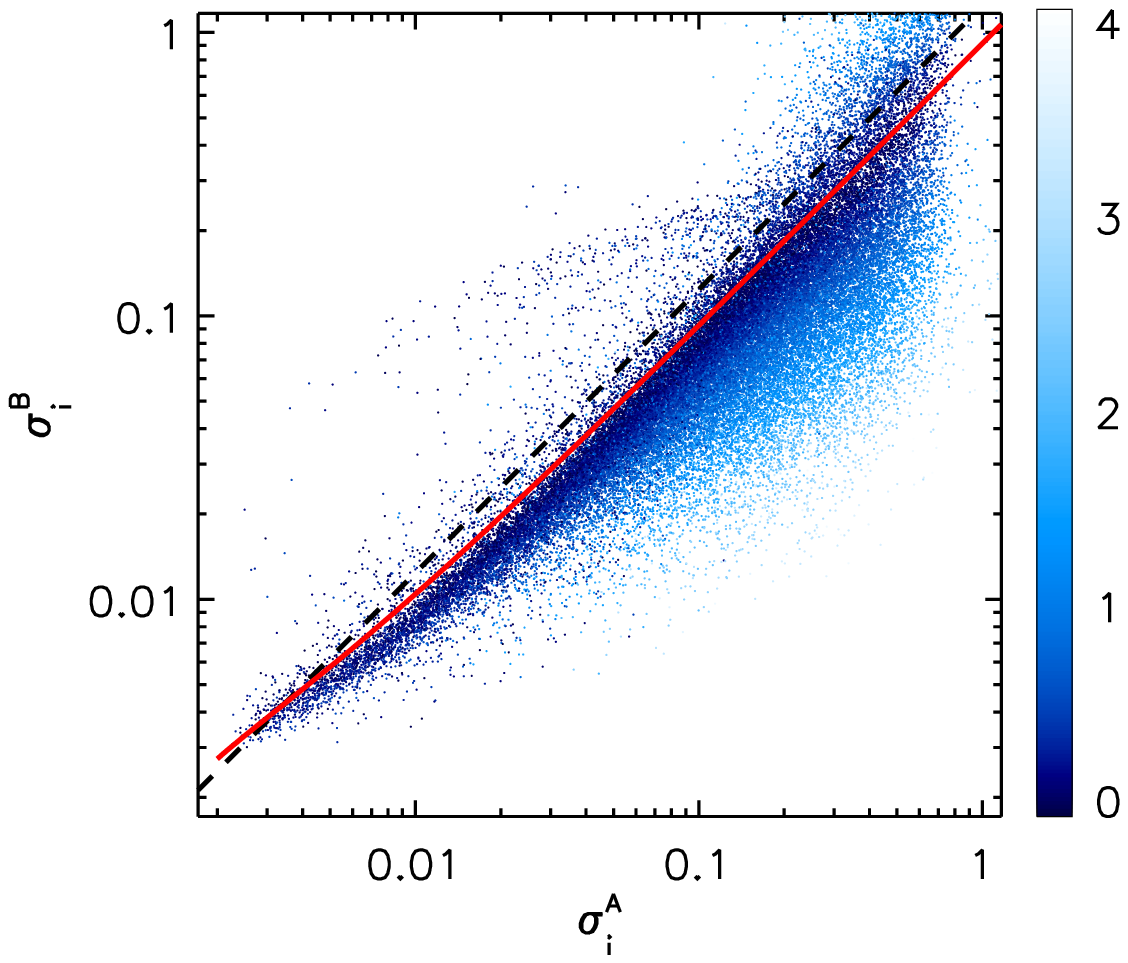}
   \includegraphics[width=.31\linewidth]{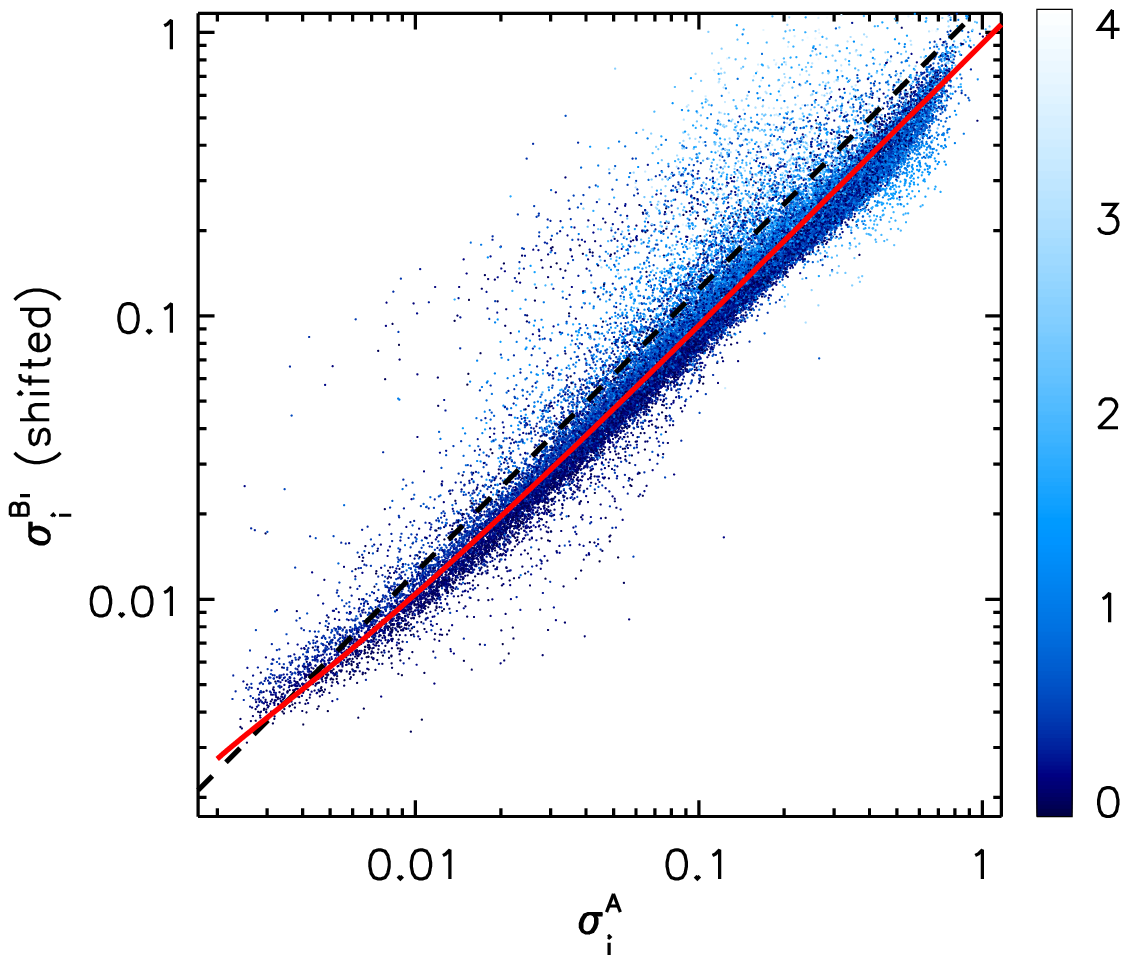}   
  \end{minipage}}  
  \vspace*{-2mm}
  \caption{Direct comparison of ASTEP and BEST\,II photometry for each matched star~$i$ in data sets (a)~Exo2 and (b)~Exo3. Left panels show mean magnitudes $(\overline{m}^A_i,\overline{m}^B_i)$; the red line denotes $\overline{m}^A=\overline{m}^B$. Light curve standard deviations $(\sigma^A_i,\sigma^B_i)$ are shown in the middle, while right panels compare the shifted BEST\,II noise level ${\sigma^B_i}'=\sigma^B_i\cdot 10^{-0.4\cdot \Delta m_i}$ with $\sigma^A_i$. In the middle and right plots, the red line denotes the expected relationship between $\sigma^A$ and $\sigma^B$ for~$\overline{m}^A=\overline{m}^B$ using the rms fits $\sigma^A(\overline{m}^A)$ and $\sigma^B(\overline{m}^B)$ (red lines in Figure~\ref{fig:e23:rmsplots}). For comparison, the black dashed line shows the expected dependency from photon noise only (Equation~(\ref{eq:app:noiseratio})). The absolute magnitude difference $\left|\Delta m\right|$ (Equation~(\ref{eq:Deltam})) is shown in blue as indicated in the color bar.}
  \label{fig:app:best2astep:mag-rms}
\end{figure*}
The light curve combination allows a comparison of ASTEP (indicated with ``A'' in the following) and BEST\,II (``B'') photometry for each matched star individually: The left panel of Figure~\ref{fig:app:best2astep:mag-rms} shows the mean magnitudes $\overline{m}^A$ and $\overline{m}^B$ that are used for the base level adjustment (Equation~(\ref{eq:Deltam})). The middle panel shows both standard deviations, i.e., the pairs $(\sigma_i^A,\sigma_i^B)$, and compares them to the respective $\sigma$-fits of Figure~\ref{fig:e23:rmsplots}. 

When $\sigma_i^A$ and $\sigma_i^B$ are being compared without accounting for systematic differences between the two instruments, the majority of stars shows lower overall photometric noise levels in BEST\,II data compared to the respective ASTEP measurements. However, the deviation increases with the magnitude difference $\left|\Delta m\right|$ between BEST\,II and ASTEP, as the color coding of Figure~\ref{fig:app:best2astep:mag-rms} indicates; potential interpretations of this effect are being discussed in the following text.

Different exposure times~$\Delta T_{A/B}$ (Table~\ref{tab:e23:rmsplotfit}) and telescope apertures $D_{A/B}$ alone cannot explain it, since that would yield
\begin{equation}\label{eq:app:noiseratio}
  \frac{\sigma_i^B}{\sigma_i^A} = \frac{D_A}{D_B} \sqrt{ \frac{g_B}{g_A} \cdot \frac{\Delta T_A}{\Delta T_B}} \ \ \left( = 1.4065\cdot\sqrt{\frac{\Delta T_A}{\Delta T_B}} \ \mbox{for BEST\,II/ASTEP}\right)
\end{equation}
for identical photometric systems (with gain factors $g_{A/B}$), i.e., a noise ratio that is independent of the stellar magnitude. Therefore, two alternative hypotheses have been investigated: Differences between the photometric systems or the angular resolution of both telescopes.

First, the magnitude difference $\Delta m$ between both systems shows a slight color dependence. Hence, the noise level is also expected to vary with the stellar color: With the assumption of photon noise only, i.e., $\sigma=1.0857\cdot\delta f\propto f^{-1/2}$, it follows that~\mbox{$\sigma^B_i/\sigma^A_i\propto 10^{0.2\Delta m_i}$}. However, the derived color dependency of $\Delta m$ (Equation~(\ref{eq:app:colorfit})) is too small to explain differences up to an order of magnitude between~$\sigma^B_i$ and $\sigma^A_i$: With $\Delta m\gtrsim -0.5$\,mag, it follows that only $\sigma^B_i\gtrsim 0.8\cdot\sigma^A_i$. 

Second, the BEST\,II pixel scale is larger ($1.5''$/Px compared to $0.9''$/Px for ASTEP\,400). Therefore, BEST\,II fields are more affected by crowding, i.e., more stellar apertures overlap each other than in the ASTEP data. In turn, this yields a systematic magnitude difference $m_B<m_A$ and hence, negative differences $\Delta m$ for contaminated stars. In addition, overlapping apertures yield an underestimation of brightness variations such that
\begin{equation}
  \delta m = \delta m_t \cdot 10^{0.4(\overline{m}-\overline{m}_t)},
\label{eq:app:deltam2}
\end{equation}
whereby $\overline{m}_t$ and $\delta m_t$ refer to the target's mean magnitude and variation, respectively, and $\overline{m}$ and $\delta m$ denote the corresponding values for the whole aperture, i.e., including the target and contamination. To test the relevance of crowding in the two data sets, a shifted value ${\sigma^B_i}'$ was calculated from the initial photometric noise~$\sigma^B_i$ of BEST\,II using Equation~(\ref{eq:app:deltam2}) and the assumption that magnitude differences~$\Delta m_i$ are solely caused by contamination. The right panel of Figure~\ref{fig:app:best2astep:mag-rms} compares ${\sigma^B_i}'$ with the corresponding (unshifted) ASTEP noise $\sigma^A_i$. It shows that the large majority of BEST\,II light curves with initial values of $\sigma^B_i\ll \sigma^A_i$ distribute smoothly around the noise dependency that is expected from the photometric quality of both data sets.

Thus, the analysis suggests that crowding introduces significant systematic differences to absolute and relative brightness measurements, although the pixel scale of BEST\,II is only 1.7~times larger compared to ASTEP. This is particularly important for fainter stars ($m\gtrsim 15$\,mag) in both fields, which are, however, less interesting for transit search. Also note that this bias is taken into account for the calculation of the detection yield by excluding stars with mean magnitudes that deviate significantly from their respective catalog value (Appendix~\ref{sec:app:detyield}). For bright stars, the photometric precision of both instruments compares well (see Figures~\ref{fig:e23:rmsplots} and~\ref{fig:app:best2astep:mag-rms}).

\section{Detection Yield Calculation}\label{sec:app:detyield}
The calculation of the estimated detection yield $N_\textrm{det}$ in this study is based on the general description 
\begin{equation}\label{eq:app:detyield:dNdet}
\frac{d^6 N_\textrm{det}}{dr_p\ dp\ dM_\star\ dr\ dl\ db} = \rho_\star(r,l,b)\ r^2 \cos b\ \frac{dn}{dM_\star}\ \frac{df(r_p,p)}{dr_p\ dp} \cdot\ p_\textrm{det}(M_\star,r,r_p,p),
\end{equation}
given by \citet{Beatty2008}. Therein, $r$, $l$, and $b$ specify Galactic coordinates, $\frac{dn}{dM_\star}$ the present day mass function, and $\frac{df(r_p,p)}{dr_p\ dp}$ the probability that a star will possess a planet of radius $r_p$ and orbital period $p$. The function $p_\textrm{det}$ describes the probability that a planetary system around a star of mass $M_\star$ at distance $r$ presents a detectable transit. It comprises the probability $p_g$ for a transit geometry \citep[see, e.g.,][Equation~(8)]{Barnes2007}, observational coverage $p_\textrm{win}$ of the transit, and the probability $p_\textrm{S/N}$ of a sufficient signal-to-noise ratio (S/N), i.e.,
\begin{equation}\label{eq:app:detyield:Pdet}
p_\textrm{det}(M_\star,r,r_p,p) = p_\textrm{S/N}(M_\star,r,r_p,p) \cdot p_g(p) \cdot p_\textrm{win}(p)  .
\end{equation}
In order to estimate $N_\textrm{det}$ for a given data set by using Equation~(\ref{eq:app:detyield:dNdet}), we made some reasonable assumptions. 

First, the calculation can be simplified by assuming a single planetary radius of interest ($r_p\equiv r_{p0}$), and by neglecting the period dependency of the probability $p_\textrm{S/N}$ for a system to show a detectable transit, i.e., 
\begin{equation}\label{eq:app:detyield:pSN}
 p_\textrm{S/N}(M_\star,r,r_p,p) \approx p_\textrm{S/N}(M_\star,r) \cdot \delta(r_{p0}) .
\end{equation}
The latter assumption can be made since the number of transits covered with ASTEP/BEST\,II observations is typically constrained to a small range of $\sim$\,2--4 events, so that averaging individual transits does not yield a significant difference in the detection sensitivity within the relevant period range. 

Second, the probability $df(p)/dp$ that a star possesses a planet of radius $r_{p0}$ is approximately constant if the period range $\left[p_0,p_1\right]$ is small (e.g., for hot Jupiters). Thus, it can be estimated with a mean value $f_{p0}$, i.e.,
\begin{equation}
 f_{p0}= \int_{p_0}^{p_1}{\frac{df(p)}{dp}\ dp} \approx (p_1-p_0) \cdot \frac{df(p)}{dp}\ .
\end{equation}

Finally, integration of Equation~(\ref{eq:app:detyield:dNdet}) yields Equation~(\ref{eq:res:Ndet}), whereby 
\begin{equation}\label{eq:app:detyield:Nsn}
 N_\textrm{S/N} = \iiiint \rho_\star(r,l,b)\ r^2 \cos b\ \frac{dn}{dM_\star} \cdot p_\textrm{S/N}(M_\star,r)\ dr\ dl\ db\ dM_\star
\end{equation}
describes the number of stars in the field with a sufficient S/N to detect a transit, and the timing factor
\begin{equation}\label{eq:app:detyield:ft}
 f_t = \frac{1}{p_1-p_0}\int_{p_0}^{p_1}{p_g(p) \cdot p_\textrm{win}(p) \ dp} 
\end{equation}
encompasses the observational coverage $p_\textrm{win}(p)$ folded with the geometric probability~$p_g(p)$. The latter is calculated using Kepler's third law, and the approximations $e=0$, $r_\star=r_\odot$, $M_\star\gg M_p$, and $r_\star\gg r_p$, thus yielding
\begin{equation}\label{eq:app:detyield:pg}
 p_g(p) \approx \frac{r_\odot+r_{p0}}{p^{2/3}} 
\end{equation}
(with $p$ in years, $r_\odot$ and $r_{p0}$ in astronomical units).

The calculation of $N_\textrm{S/N}$ requires more knowledge than is usually available a priori: Since the stellar radii determine the transit depth, any comparison between the achieved precision and the precision that is actually necessary for transit detection requires information about the stellar population in a given field. The modeling of stellar fields and its link to observational data now presented follows the approach of \citet{Bayliss2011}. 

First, the Besan\c{c}on model of the Galaxy \citep{Robin2003} is used to assess the stellar content of target fields. For each pointing, stars are simulated within the magnitude range $R\in\left[10,17\right]$, and the results are compared and adjusted to catalog data. For the target fields of this study, the total number difference between model and star catalog is minimal for a low interstellar extinction (parameter $a_v=0.0$--0.1\,mag/kpc), and the $R$ band of the GSC2.2 catalog \citep{Lasker2008} showed the best agreement with the Besan\c{c}on results from a group of several catalogs tested (e.g., UCAC3, USNO-A2, NOMAD; for a direct comparison with GSC2.2, see also \citet{Reyle2010}; \citet{Crouzet2010} use the same combination in a very similar context).

Second, each ASTEP/BEST\,II star is matched with the GSC2.2 catalog (within $2''$ radius), and the median difference $\delta m$ between instrumental magnitudes $\overline{m}_i$ and the GSC2.2 $R$ band is calculated. Since the same catalog as in the first step is used, shifted magnitudes $\overline{m}_i+\delta m$ match the Besan\c{c}on catalog reasonably well, and the stellar content can be compared homogeneously.

Third, each simulated star $i$ is assigned a photometric noise level $\sigma_i^\textrm{sim}$ that is typical for its magnitude $m_i^\textrm{sim}$ in the given data set. For that, each light curve is binned to a typical transit time scale of 30~minutes, and the corresponding standard deviation $\sigma^\textrm{bin}$ is calculated. The noise $\sigma_i^\textrm{sim}$ is then determined as a random value following the $\sigma^\textrm{bin}$-distribution of all light curves with similar brightness, i.e., having $\left|\overline{m}_i+\delta m-m_i^\textrm{sim}\right|\leq 0.04$\,mag. In order to limit the effect of crowding and other systematic factors, only stars which differ by less than 0.5\,mag from their corresponding catalog magnitude (i.e., with $\left|\overline{m}_i+\delta m-m_i^\textrm{cat}\right|<0.5$\,mag) are included in the determination of $\sigma_i^\textrm{sim}$ (see also Appendix~\ref{sec:app:datacombi:phot}).

With $\sigma_i^\textrm{sim}$ at hand, it is possible to estimate if the photometric precision of star $i$ allows the detection of a transiting planet with radius $r_{p0}$. In the following, 
\begin{equation}\label{eq:app:detyield:detcriterion}
  \sigma_i^\textrm{sim} \leq c \cdot \delta F \ \ \mbox{with} \ \ \delta F \equiv \left(\frac{r_{p0}}{r_\star}\right)^2
\end{equation}
is used as a simple criterion for sufficient S/N. As in Equation~(\ref{eq:app:detyield:pSN}), the period dependency on the detection threshold is considered small. Tests to detect artificial transit signals in BEST\,II data using the Box-fitting Least Squares algorithm \citep[BLS;][]{Kovacs2002} yielded an approximation of $c\approx 0.64$. Note that since both $\sigma_i^\textrm{sim}$ and $c$ refer to photometric data binned to 30~minutes, the simulation also enables comparing the detection yield between data sets that are obtained with a different time sampling.

The stellar radii $r_\star$ for $\delta F$ in Equation~(\ref{eq:app:detyield:detcriterion}) are derived from the masses $M_\star$ (Besan\c{c}on output) using the power law \citep{Cox2000}
\begin{equation}
  \log_{10} \left(r_\star/r_\odot\right) = 0.917 \log_{10}\left( M_\star/M_\odot \right) - 0.020
\end{equation}
for the main sequence. Other luminosity classes are disregarded for possible detections, as planetary transits for these are generally well below the threshold of the surveys investigated in this work. The simulation is repeated five times for each star and yields $N_\textrm{S/N}=f_\textrm{S/N} \cdot N_\star$, i.e., the number of dwarf stars for which Equation~(\ref{eq:app:detyield:detcriterion}) holds, whereby $N_\star$ denotes the total count of simulated Besan\c{c}on stars.

\section{Photon Noise Adjustment Through Binning}\label{sec:app:binning}
This section describes how photon noise $\sigma_\textrm{ph}$ can artifically be equalized between photometric time series that have been obtained under different instrumental conditions. If $n_\textrm{bin}$ photometric measurements are binned, the number of detected electrons is
\begin{equation}
 N_e = n_\textrm{bin} \cdot f \cdot g ,
\end{equation}
whereby $g$ denotes the gain factor (in e$^-$/ADU) and $f$ the flux of a single exposure measured in ADU. The value $N_e$ determines the photon noise and is directly accessible from the photometry. Using the magnitude definition and $\sigma_\textrm{ph}=N_e^{-1/2}$, the condition $\sigma_\textrm{ph}^A=\sigma_\textrm{ph}^B$ is obtained, if
\begin{equation}\label{eq:app:binning:nn}
 n_\textrm{bin}^A = \frac{g_B}{g_A} \cdot 10^{0.4(\delta m_B-\delta m_A)} \cdot n_\textrm{bin}^B \, .
\end{equation}
The index A/B identifies ASTEP/BEST\,II parameters, and $\delta m=\textrm{median}(m_i^\textrm{cat}-\overline{m}_i)$ denotes the magnitude shift applied during data reduction ($m_{ij}' = m_{ij} + \delta m$ for each observation $j$) in order to adjust the mean instrumental magnitudes $\left(\overline{m}_i\right)$ to the corresponding catalog values $\left(m_i^\textrm{cat}\right)$. 
For the two large data sets of this study, Equation~(\ref{eq:app:binning:nn}) yields $n_\textrm{bin}^A=0.74\,n_\textrm{bin}^B$ for field Exo2, and $n_\textrm{bin}^A=0.55\,n_\textrm{bin}^B$ for Exo3, respectively. With $\Delta t^\textrm{(1)}_\textrm{bin}=30$\,min for BEST\,II data, these ratios are realized using an ASTEP binning interval of 9.0\,min for Exo2, and 6.1\,min for Exo3, respectively.

An alternative way to equalize the photon noise levels is to take the different characteristics (telescope size, exposure time, CCD gain) directly into account by using Equation~(\ref{eq:app:noiseratio}). This approach yields comparable results ($n_\textrm{bin}^A=0.67\,n_\textrm{bin}^B$ for field Exo2, and $n_\textrm{bin}^A=0.50\,n_\textrm{bin}^B$ for Exo3, respectively). However, it does not reflect any additional instrumental effects (e.g., due to differences in the optics), whereas the chosen approach directly relates to the photometric signal and is thus considered to yield a more accurate comparison.

\end{document}